\documentclass[]{article}

\usepackage[affil-it]{authblk}
\usepackage{a4wide}
\usepackage{amsmath}
\usepackage{multirow}
\usepackage{booktabs}
\usepackage[ruled,vlined]{algorithm2e}
\usepackage{array}
\usepackage{color}
\usepackage[normalem]{ulem}
\usepackage{hyperref}
\usepackage{amssymb}
\usepackage{graphicx}
\usepackage{amsfonts}
\usepackage[english]{babel}
\usepackage[sort&compress,square,comma,authoryear]{natbib}

\providecommand{\keywords}[1]{\textbf{Keywords ---} #1}

\newcommand{\SNR}{\mathrm{SNR}}
\newcommand{\SNRT}{\mathrm{SNR}_\mathrm{T}}
\newcommand{\RSNR}{\mathrm{RSNR}}
\newcommand{\RSNRmin}{\mathrm{RSNR}_{\mathrm{min}}}
\newcommand{\bRSNRmin}{\overline{\mathrm{RSNR}_{\mathrm{min}}}}
\newcommand{\RSNRmax}{\mathrm{RSNR}_{\mathrm{max}}}
\newcommand{\bRSNRmax}{\overline{\mathrm{RSNR}_{\mathrm{max}}}}
\newcommand{\RSNRmm}{\mathrm{RSNR}_{\mathrm{max/min}}}
\newcommand{\Lmax}{L_\mathrm{max}}
\newcommand{\Ls}{L_\mathrm{s}}
\newcommand{\Lmi}{L_\mathrm{m}^i}
\newcommand{\sll}{\Psi_\mathrm{l}}
\newcommand{\slw}{\Psi_\mathrm{w}}

\definecolor{Red}{RGB}{212,28,48}
\definecolor{Green}{RGB}{81,153,74}
\definecolor{Blue}{RGB}{0,128,255}
\definecolor{Yellow}{RGB}{200,200,200}

\providecommand{\Yellow}[1]{}

\begin{document}
\bibliographystyle{apalike}

\title{Single Pulse Detection Algorithms for Real-time Fast Radio Burst Searches using GPUs}

\author[1]{Karel Ad\'{a}mek}
\author[1]{Wesley~Armour \thanks{E-mail address: \texttt{wes.armour@oerc.ox.ac.uk}} }
\affil[1]{Oxford e-Research Centre, Department of Engineering Sciences, University of Oxford, 7 Keble road, OX1 3QG, Oxford, United Kingdom}

\maketitle

\begin{abstract}
The detection of non-repeating or irregular events in time-domain radio astronomy has gained importance over the last decade due to the discovery of fast radio bursts. Existing or upcoming radio telescopes are gathering more and more data and consequently the software, which is an important part of these telescopes, must process large data volumes at high data rates. Data has to be searched through to detect new and interesting events, often in real-time. These requirements necessitate new and fast algorithms which must process data quickly and accurately. In this work we present new algorithms for single pulse detection using boxcar filters. We have quantified the signal loss introduced by single pulse detection algorithms which use boxcar filters and based on these results, we have designed two distinct "lossy" algorithms. Our lossy algorithms use an incomplete set of boxcar filters to accelerate detection at the expense of a small reduction in detected signal power. We present formulae for signal loss, descriptions of our algorithms and their parallel implementation on NVIDIA GPUs using CUDA. We also present tests of correctness, tests on artificial data and the performance achieved. Our implementation can process SKA-MID-like data 266$\times$ faster than real-time on a NVIDIA P100 GPU and 500$\times$ faster than real-time on a NVIDIA Titan V GPU with a mean signal power loss of $7\%$. We conclude with prospects for single pulse detection for beyond SKA era, nanosecond time resolution radio astronomy.
\end{abstract}

\section{Introduction} \label{sec:introduction}
    The discovery of Fast Radio Bursts (FRBs) by  \citet{Lorimeretal:2005:archivalFRB} and Rotating Radio Transients (RRATs) by \citet{RRAT:2006:McLaughlin} has highlighted the importance of single pulse detection in time-domain radio astronomy. FRBs and RRATs are rare, non repeating or irregular events, therefore their accurate detection is of great importance if we are to understand the nature of the objects producing them. Furthermore, on modern radio telescopes the algorithms for single pulse detection are required to process more and more data collected by these instruments. Also, the rise of multiwavelength astronomy which requires real-time or near-real-time (\citet{2017NewAR..79...26M}) detections increases the requirements put on single pulse detection even more. All of this necessitates the need for fast and accurate algorithms designed to detect single isolated pulses. Single pulse detection algorithms are of importance in searches for giant pulses or irregular pulses from nulling pulsars also.
    
    The single pulse search was described by \citet{Cordes-McLaughlin:2003:ApJ:SinglePulse} as an exercise in matched filtering. They employed a set of boxcar filters of widths $2^n$. This technique was used successfully in subsequent works (for example \citet{2006ApJ...637..446C, 2009ApJ...703.2259D, 2010MNRAS.402..855B, 2013MNRAS.428.2857R}). The boxcar filter approach was also adopted by many software packages, like \texttt{Heimdall}\footnote{\url{https://sourceforge.net/projects/heimdall-astro/}}, \texttt{Seek}\footnote{\url{https://github.com/SixByNine/psrsoft}}, \texttt{Destroy}\footnote{\url{https://github.com/evanocathain/destroy_gutted}}. In some cases it is beneficial to use more targeted matched filters as suggested by \citet{Keane:2010:Parker}.
    
    The single pulse detection problem is also being explored using machine learning and deep learning techniques for example \citet{2016PASP..128h4503W, Zhang_2018, 2018AJ....156..256C}. Comparison of these techniques to traditional searches is however outside the scope of this work.

    It is becoming increasingly important to understand the associated sensitivity and signal loss introduced by algorithms and software when considering the overall performance of a radio telescope. As such, any sensitivity loss introduced by algorithms or software must be investigated. This was discussed by \citet{Keane:2015:FRBSensitivity}, where the authors compared the sensitivity of different single pulse detection algorithms. Single pulse detection algorithms can have many different forms all with different sensitivities, and also different computational costs. In the case of single pulse detection by boxcar filters, the decrease in sensitivity can occur when a detected pulse is poorly matched by the filters used to extract it from the noise in which it is embedded.
    
    In this work, we quantify sources of signal loss which occur when single pulses are detected using boxcar filters. Based on these, we have designed two distinct methods to distribute and create a limited set of boxcar filter widths in such a way that the signal loss is controlled. We show that the methods presented significantly increase the computational efficiency of traditional approaches and hence provide a significant reduction in execution time, which is of relevance to real-time processing pipelines. We present formulae for calculating the signal loss introduced by a given set and distribution of boxcar filters for an idealized model which uses rectangular pulses. These formulae allow the user to tune the signal loss of their single pulse detection scheme. We present calculations of the computational complexity and the number of memory accesses associated with each method.

	Based on our two methods we have designed and implemented two single pulse detection algorithms for NVIDIA GPUs. These algorithms rely on reusing partial sums to calculate long boxcar filters required for detection of wide single pulses. This increases computational efficiency and decreases the number of memory accesses. We present optimization techniques used in the implementation of these algorithms on GPUs. We also describe the advantages and disadvantages of these algorithms. Reusing partial sums to increase computational efficiency is well known in the radio astronomy community. It has been successfully used in the fast folding algorithm by \citet{Staelin:1969:FFA} with the most recent implementation by \citet{Parent:2018:FFA_PALFA} (using python). The reuse of partial sums was also successfully used in the tree de-dispersion algorithm by \citet{Zackay:2017:treededisp} who implemented tree de-dispersion using CPU, but also suggested an algorithm suitable for GPUs.
	
	This work is part of the AstroAccelerate software package \cite{AstroAccelerate_2019_2556573}, a GPU optimized time-domain processing pipeline for radio astronomy data. It contains GPU implementations of common signal processing tools used in time-domain radio astronomy, such as de-dispersion by \citet{2012ASPC..461...33A}, a GPU implementation of the Fourier domain acceleration search by \citet{2018ApJS..239...28D}, and also periodicity search with a GPU implementation of the harmonic sum by \citet{2018arXiv181202647A}. Aspects of the work have been used in \citet{10.1093/mnras/stv1306} and \citet{2018ATel11606....1M}.
	
	This paper is structured as follows. First, in section \ref{sec:SPD} we analyze the sensitivity of the single pulse detection method which uses boxcar filters and derive formulae for quantifying the signal loss. In section \ref{sec:SPD_algorithms} we present two distinct algorithms and describe their implementation using the CUDA language extension on NVIDIA GPUs. Our results are presented in section \ref{sec:results}, where we present sensitivity of our algorithm when applied to rectangular, Gaussian and double Gaussian pulse profiles with and without white noise. We prove the correctness of our implementation by comparing the measured signal loss with predicted values. We also present the performance of both algorithms and compare it to Heimdall, a GPU accelerated pipeline by \citet{HeimdallSoft, DDTR:2012:Barsdell}. We conclude the paper in section \ref{sec:conclusions}.

%%%%%%%%%%%%%%%%%%%%%%%%%%%%%%%%%%%%%%%%%%%%%
%%%%%%%%%%%%%%%%%%%%%%%%%%%%%%%%%%%%%%%%%%%%%
\section{Single pulse detection}
\label{sec:SPD}

    The aim of the single pulse search, which represents a whole set of techniques and methods, is to find isolated pulses in input data. In this work we focus on the step responsible for recovering the pulse, located somewhere in the input time-series. We will refer to this step as \textit{single pulse detection} or the \textit{SPD} algorithm.

    Single pulse detection, in general, relies on a match filtering process, which is a convolution of the input time-series $x$ with a response function $h$.
    Convolution in the time domain is given by
    \begin{equation}
    \label{eqa:convolution}
    y[n]=\sum^{T-1}_{i=0}h[i]x[n-i]
    \end{equation}
    where $n$ is the time sample, $y$ is the filtered time-series and $T$ is the length of the response function. The response function is typically designed to detect pulses of a certain or similar shape of width $T$. Significantly longer pulses of the same shape or of different shape require a different response function $h$. A Matched filtering approach thus requires multiple passes through the data in order to cover the desired range of widths and shapes and the convolution is computationally expensive and offers little opportunity to reuse data\footnote{This is because partial sums from which we can construct the output sample $y[n]$ cannot be used to construct sample $y[n+i]$, since the elements of $x$ within these sums are weighted by the matched filter $h$.}.

    Because of the computational expense of matched filtering, we have decided to use boxcar filters for our single pulse detection algorithm. If we assume that the pulse we would like to detect could be located anywhere within the time-series, be of any shape, and have a wide range of widths, the boxcar filter is a viable alternative. Boxcar filters are less sensitive than matched filters, but offer two important advantages over the matched filter approach. They are independent of the pulse shape and they allow us to reuse data and computations, which is critical when producing an algorithm for execution on modern computer architectures, especially accelerator architectures such as GPUs.

    The boxcar filter is a simple running sum which can be expressed as
    \begin{equation}
        \label{eqa:boxcar}
        X_L[n]=\sum^{L-1}_{j=0}x[n+j]\,,
    \end{equation}
    where $L$ is the boxcar width. This formalism allows us to reuse partial sums because we can form a new longer boxcar filter from an appropriate combination of shorter boxcar filters.
    
    We can measure the strength of a sample by calculating the signal-to-noise ratio (SNR). The SNR of the $n$-th sample from a time-series is calculated by the formula
    \begin{equation}
    \label{eqa:SNR}
    \mathrm{SNR}[n]=\frac{x[n]-\mu(x)}{\sigma(x)}\,,
    \end{equation}
    where $\mu(x)$, $\sigma(x)$ is the mean and standard deviation of the underling noise in the initial time-series $x$. 
    By applying a boxcar filter to a time-series we are creating a new time-series with different mean $\mu(X_\mathrm{L})$ and standard deviation $\sigma(X_\mathrm{L})$. The SNR for a pulse from this time-series is then calculated as
    \begin{equation}
    \label{eqa:SNRlong}
    \mathrm{SNR}_\mathrm{L}[n]=\frac{X_\mathrm{L}[n]-\mu(X_\mathrm{L})}{\sigma(X_\mathrm{L})}\,,
    \end{equation}
    where $X_L[n]$ is given by equation \eqref{eqa:boxcar}. The value $X_L[n]$ of the new time-series is a value of the boxcar filter and the bin width, or put another way, the number of the accumulated samples from the initial time-series $x$ is equal to the boxcar width.
    
    We have chosen to adopt the SNR of the pulse as a figure of merit by which we measure how well any SPD algorithm detects individual pulses.
    
    The SNR produced by the algorithm which we call \textit{recovered SNR} (RSNR), to distinguish it from the true SNR of the pulse, is calculated using equation \eqref{eqa:SNRlong}. The value of the RSNR depends not only on the initial SNR of the injected pulse and its shape, but also on its position within the time-series and its width. This is because to localize the pulse and sum the power contained within it, we use an incomplete set of boxcar filters. That is, the boxcar filters do not cover every possible pulse width at every possible time sample. As a consequence a pulse of the same shape could be detected with different $\RSNR$ based on its position in time and its width. Lower $\RSNR$ occurs when the injected pulse is not properly matched by these filters either in width or position in time. The highest $\RSNR$ will be produced by the boxcar with width and position that best fits the unknown signal. 

\subsection{Idealized signal model}
    \label{sec:idealmodel}
    %%%%%%%%%%%%%%%%%%%%% Idealized signal model %%%%%%%%%%%%%%%%%%%%%
    In order to compare algorithms, to evaluate sensitivity loss and to simplify sensitivity analysis, we have introduced an idealized model (simplified toy model) of the input signal.

    The simplest form of signal which fits this role is the signal with a rectangular pulse, without any noise, where all samples except those of the pulse are set to zero. The pulse is described by its position $t_s$ within the time-series, by its width $S$, and by its amplitude $A$. The mean ($\mu$) and standard deviation ($\sigma$) which are required in the calculation of the SNR (eq. \eqref{eqa:SNR}) are set to $\mu(x)=0$ and $\sigma(x)=1$ to simulate the presence of white noise.
    
    To get the value of the mean $\mu(X_\mathrm{L})$ and standard deviation $\sigma(X_\mathrm{L})$ for longer boxcar filters we use the white noise approximation. That is the mean and standard deviation for time-series after application of the boxcar filter of width $L$ is given as
    \begin{equation}
        \label{eqa:whitenoise}
        \begin{aligned}
        \mu(X_\mathrm{L})&=L\mu(x)\,,\\
        \sigma(X_\mathrm{L})&=\sqrt{L}\sigma(x)\,.
        \end{aligned}
    \end{equation}
    For the white noise approximation we get $\mu(X_\mathrm{L})=0$ and $\sigma(X_\mathrm{L})=\sqrt{L}$.
    
    For the purpose of comparing the sensitivity of the SPD algorithm to different pulse widths we normalize the amplitude ($A$) of the rectangular pulse to 
    \begin{equation}
    \label{eqa:amplitudenormalization}
    A(S)=C/\sqrt{S}\,,
    \end{equation}
    where $C$ is a normalization constant. The normalization ensures that the RSNR produced by the boxcar filter which fits the rectangular pulse perfectly is the same regardless of the pulse width $S$.

    %%%%%%%%%%%%%%%%%%%%% Idealized signal model %%%%%%%%%%%%%%%%%%%%%

\subsection{Sensitivity analysis}
    %%%%%%%%%%%%%%%%%%%%% Sensitivity analysis %%%%%%%%%%%%%%%%%%%%%
    We quantify the sensitivity of the SPD algorithm in terms of signal loss.
    The \textit{signal loss} $\Psi$ is given as the fraction of the pulse's true $\SNR$ ($\SNRT$) which was not detected by the algorithm. That is, the difference between RSNR detected by the algorithm and the true value $\SNRT$ of the injected pulse divided by $\SNRT$,
    \begin{equation}
    \label{eqa:signalloss}
        \Psi=\frac{\SNRT-\RSNR}{\SNRT} = 1-\frac{\RSNR(S,L,d_s)}{\SNRT}\,,
    \end{equation}
    where the value of $\RSNR$ depends on the pulse width, width of the boxcar filter used for detection and on the position of the boxcar filter with respect to the pulse's position.
    
\begin{figure}[ht!]
	\centering
	\includegraphics[]{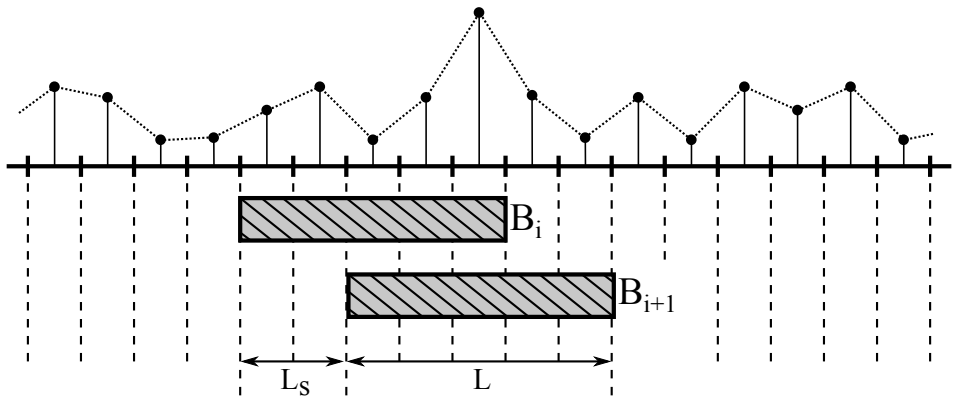}
    \caption{Layout of the boxcar filters covering a time-series $x$. Boxcars ($B_i$, $B_{i+1}$) are shown as hatched boxes. Depending on the value of $L_s$ (here $L_s=2$), multiple boxcars can intersect with themselves. Only two are shown here. \label{fig:boxcarlayout}}
    \end{figure}

    As the position of the signal is unknown we must apply the boxcar filter of width $L$ to the whole time-series $x$, which may or may not by applied to each and every time sample. If we are only interested in the highest $\RSNR$ detected, for fixed boxcar width $L$, then this reduces the sensitivity analysis to a problem of two consecutive boxcar filters separated by $\Ls$ time samples, we call this distance \textit{boxcar separation}. This problem which is depicted graphically in Figure \ref{fig:boxcarlayout} is then repeated throughout the whole time-series $x$. This allows us to avoid studying individual boxcar filters, but still gives enough flexibility to evaluate sensitivity or signal loss introduced by the SPD algorithm for any signal present in the time-series.
    
   In our analysis of the sensitivity of SPD algorithms, for a detailed description see appendix \ref{app:sensitivity}, we have, using the idealized signal model, identified two quantities of the $\RSNR$ which can be used to describe the sensitivity of an SPD algorithm. 

    The first is the lowest RSNR detected $\RSNRmin(S)$ by the SPD algorithm under the worst conditions given the pulse width $S$. This acts as an infimum (greatest lower bound) for $\RSNR$ detected by the SPD algorithm for any pulse of a given width $S$. Along with it, we define \textit{worst case scenario signal loss}
    \begin{equation}
    \Psi_\mathrm{w}(S)=1-\RSNRmin(S)/\SNRT\,,    
    \end{equation}
    that is the greatest signal loss introduced by the SPD algorithm.
    
    The second quantity is the highest possible value of RSNR ($\RSNRmax(S)$) which could be recovered by the SPD algorithm under the best possible circumstances for the given pulse width $S$. This acts as a supremum (least upper bound) for $\RSNR$ which could be detected by the SPD algorithm. To accompany this quantity we have defined the \textit{systematic signal loss} 
    \begin{equation}
    \Psi_\mathrm{l}(S)=1-\RSNRmax(S)/\SNRT\,,
    \end{equation}
    which is always introduced by the SPD algorithm for a pulse of width $S$.

    %%%%%%%%%%%%%%%%%%%%% Sensitivity analysis %%%%%%%%%%%%%%%%%%%%%
    
\subsubsection{Parameters of the SPD algorithm}
    %%%%%%%%%%%%%%%%%%%%% Parameters %%%%%%%%%%%%%%%%%%%%%
    We have also determined two parameters which describe the SPD algorithm and have direct implication on the sensitivity. We characterize the SPD algorithm by the set of boxcar filter widths $\mathfrak{B}=\left\{L_1,L_2,\ldots,\Lmax\right\}$ and by boxcar separation $L_s$ for each boxcar width $L \in \mathfrak{B}$. 
    
    The number of different boxcar widths, that is the sparsity of the set $\mathfrak{B}$ affects the value of $\RSNRmax$. We can increase $\RSNRmax$ (decrease the systematic signal loss $\sll$) by including more boxcar filter widths performed by the SPD algorithm. The set $\mathfrak{B}$ to a lesser degree also affects $\RSNRmin$.
    
    The value of the boxcar separation $\Ls$ is most important for the value of $\RSNRmin$. In order to increase $\RSNRmin$ (decrease the worst case signal loss $\slw$) we have to decrease the boxcar separation $\Ls$.
    
    The SPD algorithm may consist of multiple iterations where both $\mathfrak{B}$ and $L_s$ may differ. We assume that the time-series in which we want to detect pulses is completely covered by boxcar filters of a given width $L$.
    %%%%%%%%%%%%%%%%%%%%% Parameters %%%%%%%%%%%%%%%%%%%%%

\subsubsection{Importance of maximum and minimum RSNR}
\begin{figure}[ht!]
		\centering
		\includegraphics[]{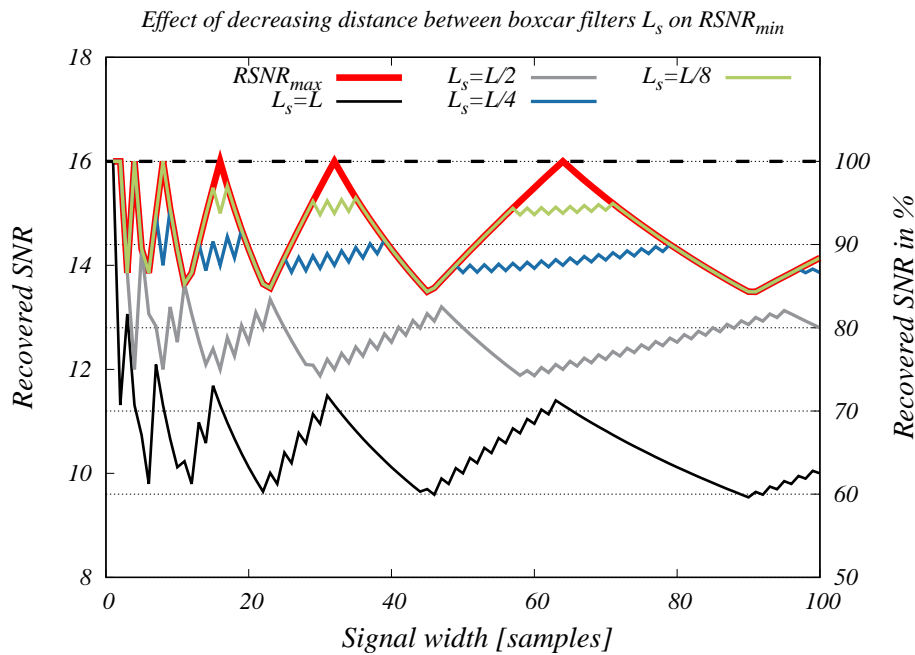}
        \caption{Behavior of $\RSNRmin(S)$ using different boxcar separation $\Ls$. This shows that $\RSNRmin(S)$ is limited by $\RSNRmax(S)$. In order to increase $\RSNRmin(S)$ further, thus increasing the sensitivity of a SPD algorithm, we have to increase $\RSNRmax(S)$ first.
        \label{fig:RSNRdependency}}
    \end{figure}

    %%%%%%%%%%%%%%%%%%%%% Importance of RSNRm/m %%%%%%%%%%%%%%%%%%%%%
    There are several reasons why $\RSNRmm(S)$ is important. Firstly, from surpremum and infimum properties of the $\RSNRmm$ we get $\RSNRmax(S) \geq \RSNRmin(S)$. That is, we cannot increase $\RSNRmin(S)$ above $\RSNRmax(S)$. Since $\RSNRmin(S)$ is mainly improved by decreasing boxcar separation $\Ls$ this means that decreasing $\Ls$ may yield diminishing results. This is shown in Figure \ref{fig:RSNRdependency}.
    
    Secondly, the $\RSNRmm(S)$ gives a hint at how we can increase or decrease the sensitivity of an algorithm, but also how to trade sensitivity for computational performance.

    Lastly, by knowing $\RSNRmm(S)$ for a given SPD algorithm we can compare them and rank them. It can also serve as a verification tool to check if an implementation of a given algorithm works as expected. In this case the algorithm must not produce any RSNR value which would lie outside the limits given by $\RSNRmm(S)$.
    %%%%%%%%%%%%%%%%%%%%% Importance of RSNRm/m %%%%%%%%%%%%%%%%%%%%%

\subsubsection{Error in detected pulse width and time}
    %%%%%%%%%%%%%%%%%%%%% Error in detected width %%%%%%%%%%%%%%%%%%%%%
    The error in the detected width of the pulse for the SPD algorithm can be expressed as
    \begin{equation}
        \label{eqa:errorpulsewidth}
        \epsilon_\mathrm{W} = \frac{\left|S_\mathrm{T}-L\right|}{S_\mathrm{T}}\,
    \end{equation}
    where $S_\mathrm{T}$ is the true pulse width and $L$ is the boxcar width which has detected the pulse, i.e. the detected width. 

    Similarly we can define the error in detected position in time relative to its true position (time localization) as a fraction of its true width $S_\mathrm{T}$ as
    \begin{equation}
        \label{eqa:errorpulsetime}
        \epsilon_\mathrm{t} = \frac{\left|t_\mathrm{T}-t\right|}{S_\mathrm{T}}\,
    \end{equation}
    where $t_\mathrm{T}$ is the true pulse position in time and $t$ is the boxcar time position. 
    %%%%%%%%%%%%%%%%%%%%% Error in detected width %%%%%%%%%%%%%%%%%%%%%

\section{Single pulse detection algorithms}
\label{sec:SPD_algorithms}
%%%%%%%%%%%%%%%%%%%%%%%%%%%%%%%%%%%%%%%%%%%%%%%%%%%%%%%%%%%%%%%%%%%%%%
%%%%%%%%%%%%%%%%%% Single pulse detection algorithm %%%%%%%%%%%%%%%%%%
%%%%%%%%%%%%%%%%%%%%%%%%%%%%%%%%%%%%%%%%%%%%%%%%%%%%%%%%%%%%%%%%%%%%%%

The single pulse detection algorithm may be required to scan for very long pulse widths $S_\mathrm{end}$ with related maximum boxcar filter width $L_\mathrm{end}$. This could lead to unnecessary precision and longer execution time as more boxcar filters than it is necessary would be calculated. This is why both of our proposed SPD algorithms execute in multiple iterations with different parameters and inputs. This allows us to control precision and increase performance. In order to decrease the sensitivity of the SPD algorithm and to lower the amount of data which are processed we use decimation in time (by a factor $D$).

To distinguish quantities from different iterations we decorate them with an index $i$, starting with $i=0$. We assume for simplicity that the decimation factor $D$ does not change during the execution of the algorithm and $D^{(i)}$ means $D$ to the power of $i$ and not an iteration index. Thus we have: 

\begin{itemize}
	\item{$x^i$ input time-series for given iteration ($x^0=x$ is the initial time-series)}
	\item{$\mathfrak{B}^i=\left\{L^i_1,L^i_2,\ldots, \Lmi, \ldots,\Lmax^i\right\}$ a set of boxcar widths used in $i$th iteration}
    \item{$X_L^i [n]$ is the value of the boxcar filter for width $L$ at time sample $n$}
	\item{$X^i [n]=X_{\Lmax}^{i} [n]$ is the value of the boxcar filter at the end of iteration}
	\item{$S_\mathrm{end}$ is the maximum desired pulse width to be searched for}
	\item{$L_\mathrm{end}$ is the maximum boxcar width calculated by the SPD algorithm; $L_\mathrm{end}$ is the boxcar width of the nearest longer boxcar filter to the value of $S_\mathrm{end}$}
\end{itemize}

The output of the SPD algorithm is the highest RSNR value detected by the boxcar filters. We have chosen to separate the output by iteration, that is the output for $i$th iteration is
\begin{equation}
\label{MSD03}
Y^i[n]=\max_{L \in \mathfrak{B}^i}\left(\SNR(X^i_L [n])\right)\,,\\
\end{equation}
where the SNR value is calculated using equation \eqref{eqa:SNRlong}. In addition to the SNR value we assume that the SPD algorithm provides the width of the boxcar which produced the highest SNR value $W^i$.

When designing an algorithm that is suitable for execution on GPUs, it is important to have enough parallelism to ensure high utilization of the GPU hardware. Modern GPUs can process many thousands of threads concurrently. The memory available for a single thread on the GPU is extremely limited, either in the form of registers or the amount of shared memory available. Therefore cooperation between threads within a single threadblock\footnote{Threadblock is a set of threads that can cooperate with each other.} is important. The amount of resources (for example the amount of local memory) consumed by a threadblock and how many threadblocks can be executed concurrently then depends on the number of threads per threadblock.     

\subsection{BoxDIT algorithm}
%%%%%%%%%%%%%%%%%% BoxDIT algorithm %%%%%%%%%%%%%%%%%%%%%%%%%%%
The algorithm we call BoxDIT is based on the ideal SPD algorithm. The ideal SPD algorithm is the algorithm that performs boxcar filters of all widths up to a maximum $\Lmax$ at each element of the initial time-series, thus detects any rectangular pulse with $S<\Lmax$ with its true SNR value. The ideal SPD algorithm has poor performance due to unnecessary high sensitivity. The BoxDIT algorithm trades some sensitivity for performance by introducing the decimation step which reduces the number of boxcar filter widths performed and increases boxcar separation $\Ls$.

\begin{algorithm}
 \SetAlgoLined
	\SetKwFunction{Boxcar}{Boxcars}
	\SetKwFunction{Decimate}{Decimate}	
	\textbf{Input:} $x^0$\;
	\textbf{Output:} $Y$, $W$\;
	$L_s^0=1$\;
	$L^0_\mathrm{MAX}=0$\;
	\For{$i=0$ \KwTo $I$}{
	    \emph{Boxcar separation for this iteration}\;
	    $L_s^{i} = L_s^{i-1}D = D^\mathrm{(i)}$\;
	    \emph{Calculating boxcar filters on decimated data $x^i$}\;
		\Boxcar($Y^i[n]$, $W^i[n]$, $X^{i+1}$, $x^{i}$, $X^i$, $L^i_\mathrm{max}$)\;
		\emph{Decimating data for next iteration}\;
		\Decimate($x^{i+1}$, $x^{i}$)\;
		\emph{Width of the longest boxcar calculated in this iteration}\;
		$L^i_\mathrm{MAX}=L^{i-1}_\mathrm{MAX} + D^\mathrm{(i)}L^i_\mathrm{max}$\;
	}
	\caption{Pseudo-code for the BoxDIT algorithm. The algorithm performs $I$ iterations required to reach desired boxcar width $L_\mathrm{end}$. The quantity $X^i$ represents values of boxcars at the end of the boxcar filter step. These values are used to built up longer boxcar filters in higher iterations. Function \texttt{Boxcars()} calculates boxcar filters up to $\Lmax^i$ and function \texttt{Decimate()} performs decimation in time.}
\label{alg:BOXDIT_host}
\end{algorithm}

The BoxDIT algorithm (algorithm \ref{alg:BOXDIT_host}) is a sequence of iterations, where each iteration has two steps. The first step is to perform all boxcar filters up to $\Lmax^i$ with respect to time-series $x^i$, at every point of the input data $x^i$. From the perspective of the initial time-series $x^0$ the algorithm calculates boxcar filters with a width step $D^{(i)}$ at every $D^{(i)}$ point of the initial time-series. For example, if $L_\mathrm{max}^i=4$ for all $i$ then: boxcar filters of width $L={1,2,3,4}$ are calculated at every point; after decimation, boxcar filters of width $L={6,8,10,12}$ are calculated at every second point and so on. Thus, each subsequent iteration detects wider pulses in the input data. This step uses the previously calculated boxcar filter data to calculate longer boxcar filters. 

\begin{figure}[ht!]
	\centering
	\includegraphics[width=0.80\textwidth]{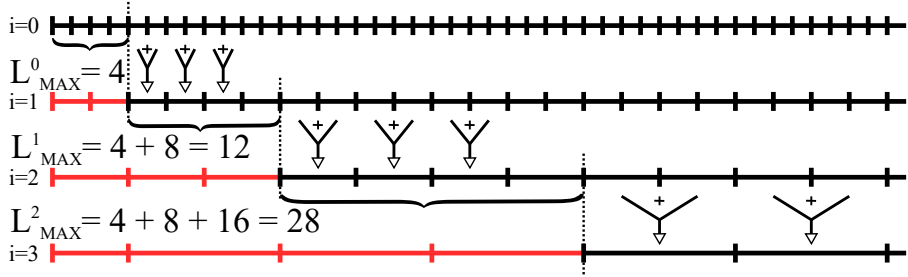}
	\caption{Decimations and maximum widths calculated at each iteration of the BoxDIT algorithm. Here $\Lmax^i=4$ and $D=2$. In the red, we see parts of the time-series which are omitted by the decimations in time, i.e. total shift. \label{fig:boxDITalgoritim}}
\end{figure}

The second step is to decimate the time-series $x^i$ in time by a factor $D$ to create time-series $x^{i+1}$ for the next iteration. This is repeated until $L_\mathrm{end}$ is reached. 
The decimation in time that is used in this paper is given by
\begin{equation}
x^{i+1}[n-\Lmax^{i}/D]=\sum_{j=0}^{D-1} x^i [Dn + j]\,,
\end{equation}
for all $n-\Lmax^{i}/D\geq0$. That is, the $0$th element of the decimated time-series is set to a position of the next time sample which must be added to the partial sum containing $0$th element of the initial time-series $x^0$. This is shown in Figure \ref{fig:boxDITalgoritim}. Such a decimation has the advantage that the maximum width $\Lmax^i$ performed at any iteration of the BoxDIT algorithm has to fulfil only the condition that it is divisible by $D$. During the decimation step we also reduce the number of time samples to $N_i=N_{i-1}/D=N_0/D^{(i)}$, this is important as the higher iterations work with fewer points. Thus each BoxDIT iteration is characterized only by $\Lmax^i$ and $D$.

The width of the boxcar filter calculated by iteration $i$ is given as
\begin{equation}
L^i_\mathrm{M} = \sum_{k=0}^{i-1}D^{(k)}\Lmax^k + D^{(i)}\Lmi\,.
\end{equation}
This could be also used to calculate the maximum boxcar width $L^i_\mathrm{MAX}$ for given iteration $i$.

The value of the boxcar filter at time sample $n$ for iteration $i$ is given as
\begin{equation}
X_{L}[n]=X^{i-1}[n] + \sum_{j=0}^{\Lmi-1} x^i [n_\mathrm{s}^i+j]\,,
\end{equation}
where $n_s^i=n/D^{(i)}$.

The advantage of the BoxDIT algorithm is that the sensitivity of the algorithm is easily adjusted by changing the maximum boxcar width $\Lmax^i$ calculated by the boxcar step of the algorithm. The disadvantage of the BoxDIT algorithm is that it has higher memory bandwidth requirements since in addition to the decimated input data it also needs the values of the longest boxcar filter at every point, which in effect doubles the size of the input data.

\subsubsection{BoxDIT GPU implementation}
%%%%%%%%%%%%%%%%%% BoxDIT algorithm implementation %%%%%%%%%%%%%%%%%%%%%%%
The GPU implementation of the BoxDIT algorithm performs both steps, calculation of boxcar filters and decimation, in one GPU kernel. The implementation must be able to extend the already calculated boxcar filters by using values from the previous iteration. Such an implementation can be used for any iteration of the BoxDIT algorithm without change.

On the input, we have the time-series $x^i$, values for the longest boxcar filter from the previous iteration $X^{i-1}$, and maximum boxcar width $\Lmax^i$ calculated in this iteration. On the output we expect to have the highest RSNR $Y^i$ for every point of the input time-series, associated boxcar width $W^i$, values of the longest boxcar calculated $X^i$ and decimated time-series for next iteration $x^{i+1}$. We have used the decimation factor $D=2$ for the GPU implementation of BoxDIT.

\begin{figure}[ht!]
	\centering
	\includegraphics[width=0.50\textwidth]{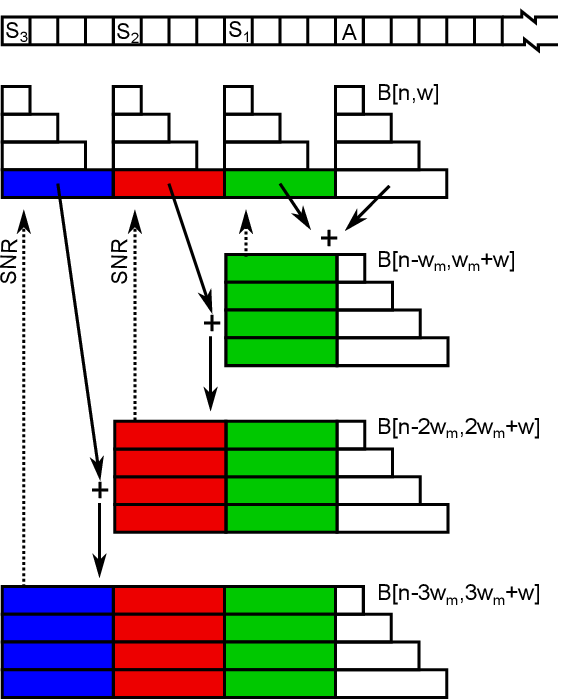}
	\caption{Graphical sketch of the BoxDIT algorithm. Each thread operates on a single sample, threads $A$, $S_1$, $S_2$, and $S_3$ are emphasized. At the beginning, each thread calculates a small scan of size $w_m$ (here $w_m=4$) and stores these values in the thread's registers. This is represented by rectangles of increasing size at below every fourth thread. Each thread then uses sums calculated by threads with smaller index to calculate scan of the required size. The thread $A$ uses largest sums ($w_m$ long) from threads $S_{1,2,3}$ to calculate prefixed sum of length 8, 12, and 16. \label{fig:boxDITGPUimplementation}}
\end{figure}

Calculating all boxcars up to given maximum width $\Lmax^i$ at a given point is equivalent to performing a prefix sum or a scan. In this case, the scan has to be performed at every point of the input time-series. We have found that using the standard prefix sum algorithm like the Hillis-Steel scan \citet{Hillis:1986:DPA:7902.7903} to be slow even when used on cached data. The reason is that series independent scan operations do not cooperate by reusing data enough. Our implementation focuses on increased cooperation between independent prefixed sums within the GPU thread-block.

\begin{algorithm}
 \SetAlgoLined
	\SetKwData{sinput}{s\_input}
	\SetKwData{sSNR}{s\_SNR}
	\SetKwData{staps}{s\_widths}
	
	\SetKwFunction{CalSNR}{Calculate\_SNR}
	\SetKwFunction{ComSNR}{Compare\_SNR}	
	
	\SetKwInOut{Input}{input}
	\SetKwInOut{Output}{output}
	
	\SetKwFunction{SharedMem}{\_\_shared\_\_}
	\SetKwFunction{TDint}{int}
	\SetKwFunction{TDfloat}{float}

	\textbf{Input:} $x^i$, $X^i$, $n$\;
	\textbf{Output:} $x^{f}$, $X^f$, $Y^f$, $Y^f_L$\;
	
	\SharedMem \TDfloat \sinput[nThreads]\;
	\SharedMem \TDfloat \sSNR[nThreads]\;
	\SharedMem \TDint \staps[nThreads]\;
	\TDfloat $B_w$[$w_m$], SNR, temp\;
	\TDint $w$, widths\;
	\TDint t=threadIdx.x\;

	\emph{Calculate initial partial sums (boxcar filters) up to width $w_m$}\;
	\For{$k=0$ \KwTo $w_m$}{
		$B_w[k]=\sum^k_{j=0} x^i[n+t+j]$\;
		\emph{Calculate SNR values corresponding to partial sums $B_w$}\;
		\CalSNR($X^i[n+t]+B_w$, SNR, width)\;
		\emph{Compare SNR values and store highest SNR and its width}\;
		\ComSNR(SNR, width, \sinput, \staps)\;
	}
	\emph{$B_w[w_m]$ is stored into shared memory}\;
	\sinput[t]=$B_w[w_m]$\;
	
	\For{$w=4$ \KwTo $B^i_{max}$}{
		\uIf{$t-w\geq 0$}{
			\emph{We are changing SNR values at position $t-w$, i.e. values $B_w$ are used to build up higher boxcar widths at position $t-w$}\;
			temp=\sinput[$t-w$]\;
			SNR=\sSNR[$t-w$]\;
			width=\staps[$t-w$]\;
			\For{$k=0$ \KwTo $w_m$}{
				$B_w[k]=B_w[k]$ + temp\;
				\CalSNR($X^i[n+t-w]+B_w$, SNR, width)\;
				\ComSNR(SNR, width, \sinput, \staps)\;
			}			
			\sSNR[$t-w$]=SNR\;
			\staps[$t-w$]=width\;			
		}
		$w=w+4$;
	}
	\caption{BoxDIT pseudo code.}
	\label{alg:boxditpseudocode}
\end{algorithm}

The pseudo-code for the parallel GPU implementation of the BoxDIT algorithm is given in Algorithm \ref{alg:boxditpseudocode} and a graphical sketch is presented in Figure \ref{fig:boxDITGPUimplementation}. This figure shows how we divide work among GPU threads and how threads cooperate on a calculation of independent prefixed sums.

As a first step, each thread calculates a much shorter prefixed sum of length $w_m$. That is each thread calculates partial sums $B[n;w]$, where $n$ is a time sample and $w={1,2,\ldots,w_m}$ is the length of the partial sum in the number of time samples. These partial sums $B[n,w]$ are stored into the GPU registers\footnote{GPU registers are the fastest type of GPU memory, but private to an individual thread.} of a given thread. These partial sums serve as accumulators used for calculation of the longer partial sums. The value of the last partial sum $B[n;w_m]$ is also stored into shared memory\footnote{Shared memory is a fast user managed cache on the GPU which all threads in a thread block can access.}, because its value has to be distributed among other threads. The maximum SNR $Y^{i}[n]$ for a given time sample $n$ and boxcar width $W^{i}[n]$ are also stored in the shared memory as they will be modified by other threads.

After this initial phase, each thread reads the partial sum $B[n-w_m;w_m]$ from its appropriate place in shared memory. Then using accumulators stored in its registers (partial sums $B[n;w]$, for $w={1,2,\ldots,w_m}$) each thread is able to produce a prefixed sum up of length $w=2w_m$. That is partial sums $B[n;w]$, where $w={1,2,\ldots,2w_m}$. These partial sums are calculate as
\begin{equation}
B[n-w_m;w_m+w]=B[n-w_m;w_m]+B[n;w]\,.
\end{equation}
The accumulators in the threads registers are then updated to the new values representing partial sums $B[n-w_m;w_m+w]$. The thread calculates SNR value for each accumulator and compares it to the highest detected SNR for a sample $n-w_m$. If the new SNR is higher than the old SNR, the highest SNR $Y^{i}[n]$ is updated together with boxcar width $W^{i}[n]$.

With each following iteration, each thread calculates longer partial sums for time samples which are further from its starting time samples.

The number of iterations required to calculate all required boxcars is $J=\Lmax^i/w_m$. The algorithm used in our implementation can be expressed by equation
\begin{equation}
X^i_{\Lmi}[n]=\sum_{k=0}^{j-1}B[n+kw_m;w_m] + B[n+jw_m;w]\,,
\end{equation}
where $j=\mathrm{ceil}(L^i_m/w_m)$.

At the last step of the algorithm, the maximum values of the SNR $Y^{i}[n]$, the boxcar width $W^{i}[n]$ and the longest accumulated partial sum representing the value of the boxcar filter of width $L_\mathrm{MAX}[n]$ are stored into the device memory. 

The intermediate partial sums produced by the algorithm are not stored to device memory as they are needed only for finding the maximum SNR for every input sample.

\subsection{IGRID algorithm}
%%%%%%%%%%%%%%%%%% IGRID algorithm %%%%%%%%%%%%%%%%%%%%%%%%%%%
Our second algorithm, which we call IGRID, is based on the decimation in time SPD algorithm (DIT algorithm). The DIT algorithm has poor sensitivity, with the average signal loss of $20\%$ and the worst signal loss of $40\%$, but it has high performance. In terms of $\RSNR$ both $\RSNRmm$ are low. The IGRID algorithm decreases signal loss at the expense of the performance.

The DIT algorithm, shown in Algorithm \ref{alg:DIT}, is a sequence of decimations in time applied repeatedly on the input time-series $x^0$, thus in effect producing boxcar filters of width $L^i=2^i$, where $i$ is the number of decimations performed so far. Therefore we have $\mathfrak{B}=\{1,2,4,\ldots,2^I\}$, where $I$ is the total number of iterations performed. The distance between these boxcar filters is the same as the width $\Ls^i=2^i$. Based on our analysis in section \ref{sec:SPD} we can say that large values of $\Ls$ are mostly responsible for low $\RSNRmin$ values, while sparse coverage of boxcar widths $\mathfrak{B}$ are responsible for low $\RSNRmax$ values. The IGRID algorithm corrects for these shortcomings.

\begin{algorithm}
 \SetAlgoLined
	\SetKwData{sinput}{s\_input}
	\SetKwData{sSNR}{s\_SNR}
	\SetKwData{staps}{s\_widths}
	
	\SetKwFunction{CalSNR}{Calculate\_SNR}
	\SetKwFunction{ComSNR}{Compare\_SNR}	
	
	\SetKwInOut{Input}{input}
	\SetKwInOut{Output}{output}
	
	\SetKwFunction{SharedMem}{\_\_shared\_\_}
	\SetKwFunction{TDint}{int}
	\SetKwFunction{TDfloat}{float}

	\textbf{Input:} $x^i$, $n$\;
	\textbf{Output:} $Y[n]$, $W[n]$\;

	\TDfloat SNR, temp\;
	\TDint width\;
	\TDint J \emph{number of iterations performed by the kernel}\;
	
    SNR=0\;
	\For{$j=1$ \KwTo $J$}{
		\CalSNR($x^{j}[n]$, SNR)\;
        width = $2^j$\;
		\ComSNR(SNR, width, $Y[n]$, $W[n]$)\;
		$x^{j+1}[n]=x^{j}[2n]+x^{j}[2n+1]$\;
	}
	\caption{Decimation in time (DIT) algorithm. Function \texttt{Calculate\_SNR} calculates SNR based on boxcar filter value, boxcar width, and mean and standard deviation. Function \texttt{Compare\_SNR} compares SNR values for the same sample and stores highest SNR.}
\label{alg:DIT}
\end{algorithm}

\begin{figure}[ht!]
	\centering
	\includegraphics[width=0.85\textwidth]{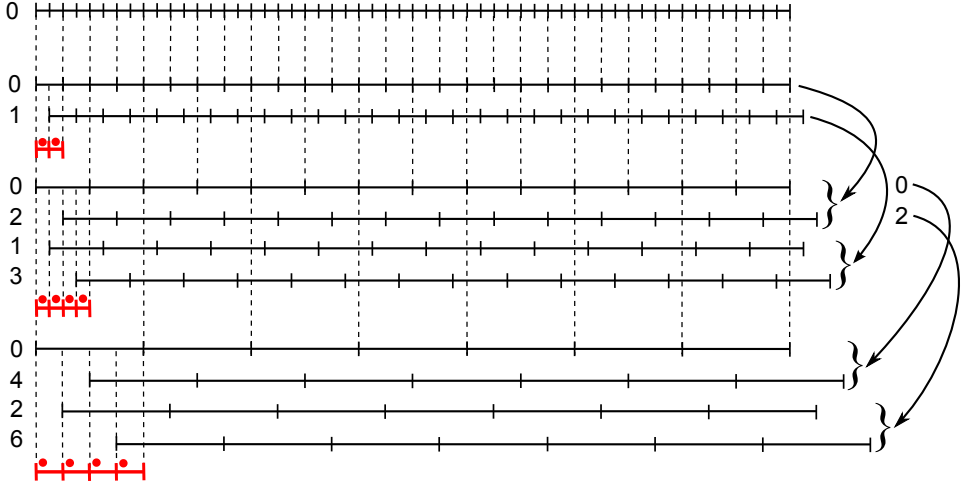}
	\caption{A graphical representation of our IGRID algorithm. The horizontal solid black lines represent the time-series where each slot is one time sample. The DIT algorithm is represented by time-series preceded by zero. Our IGRID algorithm improves the loss in sensitivity of the DIT algorithm by introducing additional layers of boxcar filters which have an additional time shift associated with them. \label{fig:IGRIDalgorithm}}
\end{figure}

One way of decreasing $\Ls$ is to perform an additional DIT operation on the input data $x^i$ producing additional time-series $x^{i+1}$, which is shifted by one sample ($2^i$ samples from the perspective of the initial data). That is for initial time-series we would create one more time-series $x^{1}_1[n]=x^0[2n+1]+x^0[2n+2]$ in addition to the already existing time-series $x^{1}_0[n]=x^0[2n]+x^0[2n+1]$ which results from DIT operation which was not shifted. Let's call these decimated time-series \textit{layers} and decorate them with a subscript indicating their shift in the number of samples with regard to the initial time-series $x^0$. That is $x^i_h$ is a layer in the iteration $i$ with the shift $h$. Let's also introduce \textit{IGRID step} which is a set of layers with different shifts but with the same decimation.

In general if we have a \textit{parent} layer $x^i_h$ which is decimated by $i$ times, that is each sample is a sum of $2^{i}$ samples from the initial time-series $x^0$, we can use it to calculate two new \textit{child} layers, one layer $x^{i+1}_h$ with the same shift $h$ as the parent's and second layer $x^{i+1}_{h+d}$ which is shifted by $d=2^i$ initial time samples compared to the parent. This is shown in Figure \ref{fig:IGRIDalgorithm}, where we have:
\begin{itemize}
	\item{$i=0$ Layer: $0 \rightarrow 0,1$}
	\item{$i=1$ Layer: $0 \rightarrow 0,2$ and $1 \rightarrow 1,3$}
	\item{$i=2$ Layer: $0 \rightarrow 0,4$ and $2 \rightarrow 2,6$ note that layers 1 and 3 are not needed for further iterations}
\end{itemize}
In this example, we have chosen not to use layers $x^2_1$ and $x^2_3$. This means the $\Ls$ increases, but we do not need to keep these layers in memory and process them in further iterations. 

Thus by using $P$ layers we can decrease the distance between boxcar filter to $L_s=2^i/P$. These layers must be shifted by $2^i/P$ samples in order to have a constant step between each boxcar filter. An uneven distribution of boxcar filters leads to increased signal loss at some parts of the input time-series.

\begin{figure}
	\centering
	\includegraphics[width=0.85\textwidth]{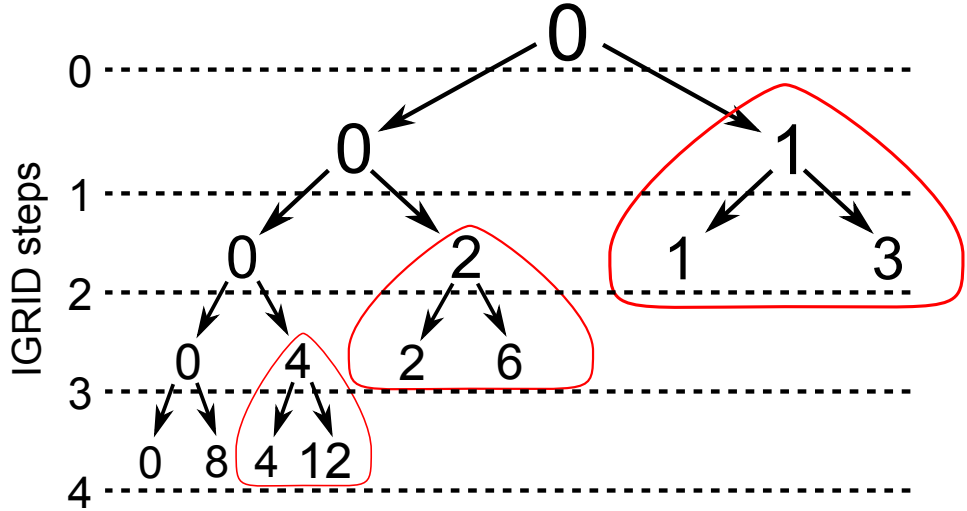}
	\caption{Representation of the IGRID algorithm as a binary tree. Areas which are enclosed in red contains layers which are not needed for the computation of the wider boxcar filters in the following IGRID steps. \label{fig:binarytree}}
\end{figure}

We can look at the algorithm differently using the time shifts alone, which now represent appropriate layers. This is shown in Figure \ref{fig:binarytree}. The layer dependencies (from Figure \ref{fig:IGRIDalgorithm}) have the structure of a binary tree (for decimation factor $D=2$). If we focus on the right branches (marked in red) we see that after a few IGRID steps there are no layers dependent on layers located in these right branches of the binary tree. If we are able to keep and process all layers from these branches in local memory, then all we need to keep in the device memory are layers $x^i_0$.

\begin{figure}[ht!]
	\centering
	\includegraphics[width=0.85\textwidth]{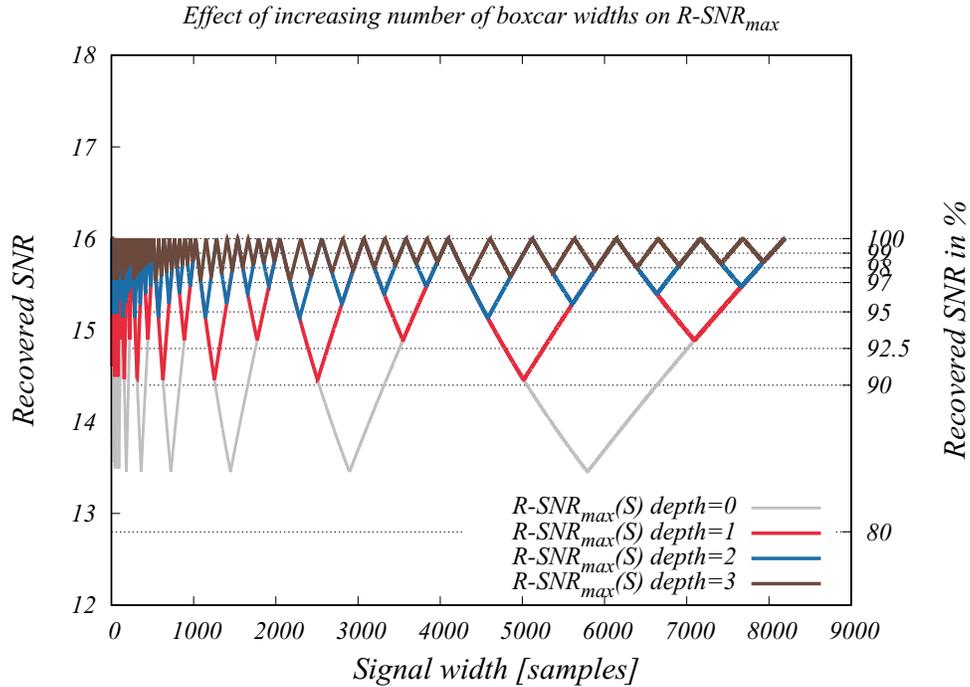}
	\caption{Effect of using the previous IGRID steps to increase the set of boxcar filter widths $\mathfrak{B}$ on $\RSNRmax(S)$.\label{fig:IGRIDrsnrmax}}
\end{figure}

The binary tree structure also hints at how we can deal with the second problem of the DIT algorithm, which is the scarcity of the set of boxcar filter widths $\mathfrak{B}$. By decreasing $\Ls$ we increase $\RSNRmin(S)$ but not $\RSNRmax(S)$, which could result in the situation, where $\RSNRmin(S)$ cannot increase further since $\RSNRmin(S)\leq\RSNRmax(S)$. This is shown in Figure \ref{fig:RSNRdependency}. 

We can increase the number of calculated widths if we use boxcar filters from previous IGRID steps and add them to boxcar filters from the current IGRID step. In other words, we traverse the binary tree in an upward direction toward the root. The IGRID steps contain boxcar filters of width $L=2^i$, thus by using previous IGRID steps we can produce boxcar filters of width
\begin{equation}
    \label{eqa:boxcarsize}
    L=\sum_{k=0}^{d}a_k2^{i-k}\,,
\end{equation}
where $d$ is depth or the number of previous IGRID steps used, and $a_k = \pm1$ are variables determining the width of the resulting boxcar filter. The values of $a_k$ are further restricted to $a_0=1$ for positive $L$ and also $a_1=1$ since smaller boxcars then $L=2^i$ would be calculated in the previous IGRID step. This means that any given IGRID step, depending on depth $d$ can produce any boxcar width $2^i \leq L < 2^{i+1}$. An example of this is shown in table \ref{tab:IGRIDversions} for starting boxcar of width $L=16$. However, by increasing the depth $d$ we increase the number of layers which the IGRID algorithm needs to access and therefore share between IGRID steps. The effect on $\RSNRmax(S)$ caused by increasing the set $\mathfrak{B}$ is shown in Figure \ref{fig:IGRIDrsnrmax}.

\begin{table}[!h]
	\centering
	\caption{Example of possible boxcar widths which could be calculated by traversing the binary tree in an upward direction and adding partial sums from previous iterations using equation \eqref{eqa:boxcarsize}.}
	\begin{tabular}{lccccccccc}
		\toprule
		Depth & \multicolumn{8}{c}{Width} & Increment \\
		 & $a_0$ & $a_3$ & $a_2$ & $a_3$ & $a_1$ & $a_3$ & $a_2$ & $a_3$ & \\
		\hline
		a & 1 & -1 & -1 & 1 & 1 & -1 & 1 & 1 & \\
		
		0 & 16 &    &    &    &    &    &    &    & 0\\
		1 & 16 &    &    &    & 24 &    &    &    & +8\\
		2 & 16 &    & 20 &    & 24 &    & 28 &    & +4\\
		3 & 16 & 18 & 20 & 22 & 24 & 26 & 28 & 30 & +2\\
		\hline
	\end{tabular}
	 \label{tab:IGRIDversions}
\end{table}

The advantage of the IGRID algorithm is that it has low memory bandwidth requirements (for lower precision) as all data required by the algorithm is contained in the decimated input data $x^i$. The disadvantage of the IGRID algorithm is that for higher precision we have to use previous IGRID steps which substantially changes the character of the algorithm (more input data, more complicated calculations of boxcar filters). This makes the IGRID algorithm hard to implement in a flexible way.

\subsubsection{IGRID GPU implementation}
%%%%%%%%%%%%%%%%%% IGRID algorithm implementation %%%%%%%%%%%%%%%%%%%%%%%%%%%
\label{sec:IGRID_GPU_kernel}
We have implemented the IGRID algorithm with three levels of precision, these are abbreviated as $\mathrm{IG}(d)$, where $d$ is the depth. We have used three depths $d=1,2,3$ which corresponds to IGRID algorithms IG(1), IG(2), and  IG(3). Each implementation of the IGRID algorithm also uses a different number of layers.  We have chosen 4 layers for the IG(1) 8 layers for the IG(2) and 16 layers for the IG(3) algorithm per IGRID step. We have implemented three different precision's because it allowed us to optimize each case and get the highest performance.

The main problem of the IGRID algorithm, especially when it is performed with a higher depth parameter, is the amount of data which needs to be shared between IGRID steps. For IG(1) we do not need any additional data since the input layer $x^{i-1}_h$ can also be used to calculate wider boxcar filters. However for IG(2) we need in addition to the input layer $x^{i-1}_h$, also its parent $x^{i-2}_h$, and finally for IG(3) we need the parents of the parents ($x^{i-3}_h$). This means that the size of the input data for IG(2) is twice as big as for IG(1) and the size of the input data for IG(3) is four times bigger than that of IG(1). All of this data needs to be read and written by each IGRID step. This would slow down the code considerably and make it memory bandwidth bound.

\begin{algorithm}
    \SetAlgoLined
	\SetKwData{nChannels}{nChannels}
	\SetKwFunction{TDfloat}{float}
	\SetKwBlock{CudaBlock}{GPU Kernel begin}{end}

	\emph{calculate number of kernel executions}
	$K = I/SpK$\;
	\emph{loop through kernel executions}\;
	\For{$k=0$ \KwTo $K$}{
	    \CudaBlock{
		    \emph{loop through IGRID steps performed by kernel}\;
            \For{$s=0$ \KwTo $SpK$}{
                \emph{loop through A layers}\;
                \For{$a=0$ \KwTo $A$}{
                    \emph{calculations of layers}\;
                }
            }
	    }
	}
 
    \caption{Pseudo-code for IGRID SPD algorithm. Variable $I$ is the number of IGRID steps required for single pulse search, which is calculated on the maximum boxcar width $L_\mathrm{end}$ which is user defined, $SpK$ is the number of IGRID steps performed by the GPU kernel, and $A$ is the number of layers used.}
    \label{alg:GPUIGRID}
\end{algorithm}

In our implementation of the IGRID algorithm we have chosen to process more than one IGRID step per one execution of the GPU kernel. Algorithm \ref{alg:GPUIGRID} describes how the IGRID steps are split into blocks. Processing more IGRID steps per GPU kernel is beneficial in a number of different ways. If we perform more IGRID steps together we not only remove device memory accesses between individual IGRID steps, but we also reduce the amount of data shared between consecutive GPU kernel executions, since each IGRID step reduces the amount of data by half due to decimation in time. The data shared between IGRID steps within one GPU kernel execution are stored in GPU shared memory.

The second benefit comes when we process whole right branches which are marked in Figure \ref{fig:binarytree}. As discussed above, each right branch depends only on the layer with zero shift, but in order to calculate whole IGRID step we might need layers from different branches. That is demonstrated by IGRID with a step of two in Figure \ref{fig:binarytree}. We see that, to calculate a whole step, we need layers with shifts $h=\{0,2,1,3\}$, but layers with shifts $\{2\}$ and $\{1,3\}$ are from different branches. If we do not calculate the entire right branch within one execution we would have to save all or some of the intermediate data to the device memory.

\begin{figure}
	\centering
	\includegraphics[width=0.65\textwidth]{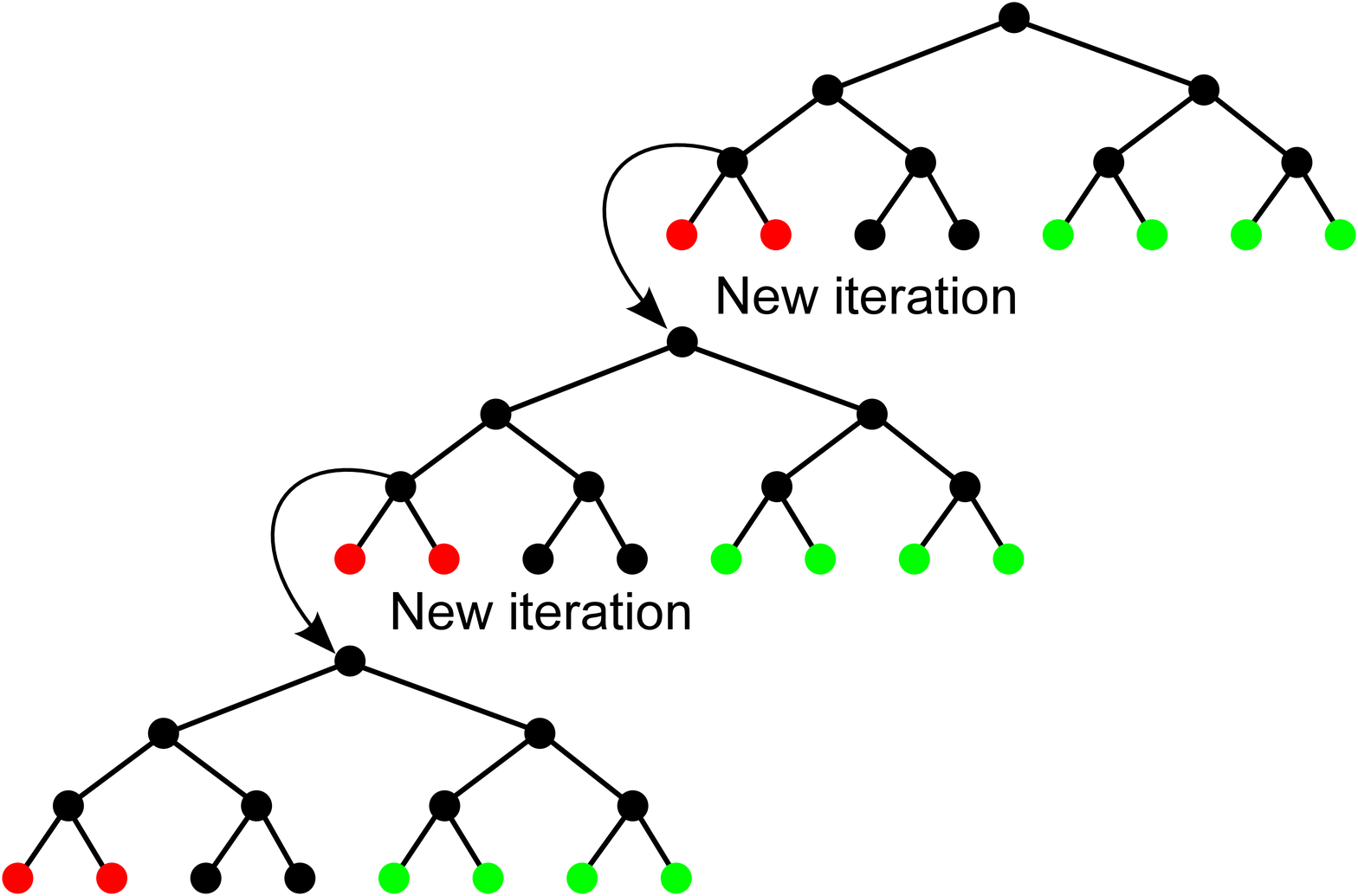}
    \caption{Representation of how IGRID kernels compute layers and progress to the next iteration. Black circles represent layers which we must calculate, in order to reach desired sensitivity. Red circles show layers which are calculated but not written to device memory (thus recalculated later by the next iteration) and Green circles are layers which are calculated but not needed, these are used to increase sensitivity. \label{fig:binarytreeexecution}}
\end{figure}

Therefore we have chosen to calculate whole sub-sections of the binary tree starting always at the layer with zero shift and going down to a chosen depth, even if it means calculating parts of the tree which are later discounted. Recalculating parts of the binary tree allows us to save device memory bandwidth and increases the performance in the process. This is shown in Figure \ref{fig:binarytreeexecution}. The other indirect benefit is that we can achieve higher sensitivity, since we calculate parts of the tree which would not be calculated otherwise.

\begin{figure}[ht!]
	\centering
	\includegraphics[width=0.8\textwidth]{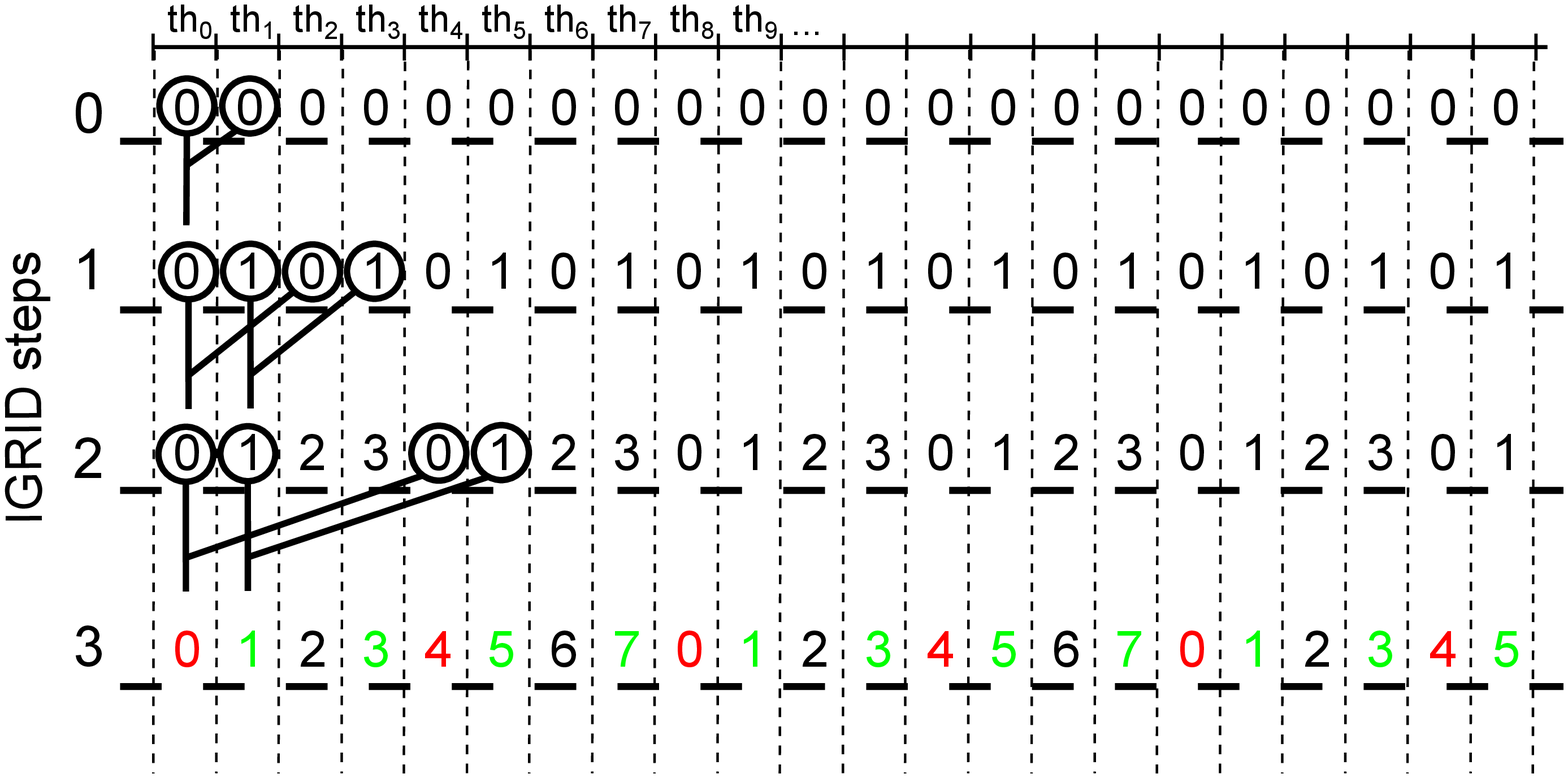}
	\caption{Visualization of IGRID GPU kernel. Iterations of the IGRID algorithm are separated by thick horizontal dashed lines. The number shows to which layer a given time sample belongs. The coloured number uses the same colour coding as in Figure \ref{fig:binarytreeexecution}. Black shows the layer we have to calculate, red a time sample from a layer which is calculated but not used, and green a layer which is calculated but it is not necessary to achieve the desired sensitivity. \label{fig:IGRIDalgorithmGPU}}
\end{figure}

The calculation of the binary tree itself inside the GPU kernel exploits the property of decimation in time, where the output decimated  time-series is half the size of the input. After decimation we are able to fit two decimated layers into the memory space instead of one. The process of calculating these two child layers from the parent layer can be performed in one operation. If we use a running sum of size two (a boxcar filter of width two) we can calculate the two child layers in-place but they will be mixed together, one in even samples, the second in odd samples. Figure \ref{fig:IGRIDalgorithmGPU} depicts how this calculation of a sub-section of the binary tree is performed. As we perform more decimations we can fit more layers into the same memory space but also the individual layers are more intertwined as we see in Figure \ref{fig:IGRIDalgorithmGPU} for later IGRID steps.

\begin{algorithm}
 \SetAlgoLined
	\SetKwData{sBTone}{s\_BT1}
	\SetKwData{sBT2}{s\_BT2}
	\SetKwData{sBT3}{s\_BT3}
	\SetKwData{sBT3}{s\_BT3_temp}
	
	\SetKwFunction{CalSNR}{Calculate\_SNR}
	\SetKwFunction{ComSNR}{Compare\_SNR}	
	
	\SetKwInOut{Input}{input}
	\SetKwInOut{Output}{output}
	
	\SetKwFunction{SharedMem}{\_\_shared\_\_}
	\SetKwFunction{TDint}{int}
	\SetKwFunction{TDfloat}{float}

	\textbf{Input:} $x^{i}_{h}$, $x^{i-1}$, $n$, $i$\;
	\textbf{Output:} $x^{i+1}_{h}$, $x^{i+1}_{h+2^{i}}$, $Y^i$, $Y^i$\;
	
	\emph{B0 and Bd represent boxcar filters}\;
	\TDfloat SNR, B0, Bd\;
	\TDint width\;

		\emph{Calculate two child layers}\;
        $d=2^i$\;
        width = $2d$\;
        B0 = $x^{i}_{h}[2n]$ + $x^{i}_{h}[2n+1]$\;
        Bd = $x^{i}_{h}[2n+1]$ + $x^{i}_{h}[2n+2]$\;
        \emph{highest SNR is stored in initial data coordinates}\;
        SNR = \CalSNR(B0, width)\;
		\ComSNR(SNR, width, $Y^i[2dn]$, $Y^i_L[2dn]$)\;
        SNR = \CalSNR(Bd, width)\;
		\ComSNR(SNR, width, $Y^i[2dn+d]$, $Y^i_L[2dn+d]$)\;
        
        \emph{storing to memory}\;
        $x^{i+1}_{h}[n]$ = B0\;
        $x^{i+1}_{h+d}[n]$ = Bd\;
        
        \emph{Traversing binary tree first depth}\;
        width = width + $d$\;
        B0 = B0 + $x^{i}_{h}[2n+2]$\;
        Bd = Bd + $x^{i}_{h}[2n+3]$\;
        \CalSNR($\ldots$)\;
        \ComSNR($\ldots$)\;
        
        \emph{Traversing binary tree second depth}\;
        width = width + $d$\;
        B0 = B0 + $x^{i}_{h}[2n+2]$ + $x^{i-1}_{h}[4n+6]$\;
        B0 = B0 + $x^{i-1}_{h}[4n+4]$\;
        Bd = Bd + $x^{i}_{h}[2n+3]$ + $x^{i-1}_{h}[4n+8]$\;
        Bd = Bd + $x^{i-1}_{h}[6n+6]$\;
        \CalSNR($\ldots$)\;
        \ComSNR($\ldots$)\;
	
	\caption{The IGRID IG(2) algorithm. Function \texttt{Calculate\_SNR} calculates SNR and \texttt{Compare\_SNR} compares SNR for the same sample.}
\label{alg:IGRID}
\end{algorithm}

The GPU kernel uses shared memory to calculate these decimations. A thread which performs the boxcar filter of width two has to add the sample with increasing step, in order to use samples from the correct layer. Each thread always writes into the same memory address. Thus during the calculation a thread calculates time samples from different layers as is shown in Figure \ref{fig:IGRIDalgorithm}, where, for example, the 7th thread calculates 8th time sample from the zero shifted layer, then the 4th time sample from the layer which is shifted by one sample and so on. Accesses to shared memory are without bank conflicts. The algorithm for IG(2) is shown in Algorithm \ref{alg:IGRID}.

\subsection{Computational and memory complexity}
The number of operations and number of memory accesses of one iteration for both proposed algorithms is summarised in table \ref{tab:algorithms}. 

\begin{table}[!h]
	\centering
	\caption{Number of operations and number of memory accesses of selected SPD algorithms, where $B$ is the number of boxcar filters performed in one iteration.}
	\begin{tabular}{lcc}
		\toprule
		\multirow{2}{*}[-2pt]{Algorithm} & \# floating & \# global \\
		 & point operations & memory accesses \\
		\hline
		Ideal boxcar filter & $N^2$ & $N^2 $ \\
		%\hline
		Decimation in time & $N\log_2N$ & $4N\log_2N$ \\
		%\hline
		BoxDIT & $BN\log_2\left(\frac{N}{B}+1\right)$ & $6N\log_2\left(\frac{N}{B}+1\right)$ \\
		%\hline
		IGRID(IG(1)) & $N\log_2N$ & $4N\log_2N$ \\
		\hline
	\end{tabular}
	\label{tab:algorithms}
\end{table}

\section{Results}
%%%%%%%%%%%%%%%%%%%%%%%%%%%%%%%%%%%%%%%%%%%%%
%%%%%%%%%%%%% RESULTS %%%%%%%%%%%%%%%%%%%%%%%
%%%%%%%%%%%%%%%%%%%%%%%%%%%%%%%%%%%%%%%%%%%%%
\label{sec:results}
In this section we present a selection of representative configurations for each algorithm and compare them based on performance and signal loss. Implementations presented here assume \texttt{fp32} floating point numbers for the input. 

First we present signal loss for each algorithm and configuration for the idealized signal model (sec. \ref{sec:idealmodel}) using a rectangular pulse. We verify the correctness of our implementation of both algorithms using the idealized signal model by comparing the measured signal loss $\sll^m$ and $\slw^m$ ($\RSNRmax$, $\RSNRmin$) to predicted values of the signal loss $\sll$ and $\slw$. We also present the error in the detected width of the pulse and the error in the detected position in time relative to its true position (time localization), which are introduced by each algorithm.

Next we present RSNR and error in pulse width and time localization for the idealized signal model with Gaussian and double Gaussian pulse profiles. These pulse shapes are a better approximation of the pulses encountered in real observations. 

The recovered SNR and detected width for a rectangular, Gaussian and double Gaussian pulse shapes which are embedded in white noise are presented afterwards. For these results we have used pulses of $\mathrm{SNR}=\left\{8,5,3\right\}$. 

Following this we present the performance of our algorithms on two generations of NVIDIA GPUs, the P100 and the Titan V. The performance of the algorithm is measured in the number of dedispersion trials\footnote{A DM trial is a time-series which is the output of the de-dispersion transform algorithm, it corrects for the effects of interstellar medium.} (DM trials) which could be performed in real-time for sampling time $t_\mathrm{s}=64\mu s$. We also present the dependency of this number on the maximum width performed by the algorithm. 

Lastly we present a comparison of AstroAccelerate with Heimdall and its single pulse search algorithm. We also demonstrate how we calculate mean and standard deviation which is then used for the calculation of recovered SNR.

Results presented here are for three variants of the IGRID algorithm which we have described at beginning of section \ref{sec:IGRID_GPU_kernel}. To match these three IGRID configurations we chosen three configurations of the BoxDIT algorithm, each with a different ratio of sensitivity and performance. We can change the sensitivity of the BoxDIT algorithm by changing the value of the maximum boxcar width calculated in the boxcar filter step of the algorithm. These configurations are: BD-32 which calculates a maximum width of $\Lmax^i=32$ between decimations for any $i$; BD-16 with maximum width of $\Lmax^i=16$; and BD-8 which uses maximum width of $\Lmax^i=8$.

\subsection{Measurement of signal loss}
%%%%%%%%%%%%%%%%%%%%%%%% Measurements of signal loss %%%%%%%%%%%%%%%%%%%%%%%%
\label{sec:measurements}
To measure the signal loss of a given SPD algorithm we have to find the $\RSNR$ produced by the algorithm and then calculate signal loss using equation \eqref{eqa:signalloss}. The value of $\RSNR$ depends not only on the SNR of the injected pulse but also on its position within the time-series. This is because to localize the pulse and sum the power it contains, we use an incomplete set of boxcar filters. That is, the boxcar filters do not cover every possible pulse width at every possible time sample. Lower RSNR occurs when the injected pulse is not properly matched by these filters. 

Therefore, to accurately measure $\RSNRmm$ we have to create data where a pulse of given width will be present in all relevant positions in time. By sliding the pulse along its whole width in time we can evaluate all of these relevant positions for a pulse of a given width. The test data used for the measurement of the $\RSNRmm$ consists of a set of time-series where each pulse shift is stored in its own time-series.

The test data are then processed by the SPD algorithm. From each time-series we select the highest RSNR and from these we select the maximum, which is the $\RSNRmax$, and the minimum which is the $\RSNRmin$ for a given pulse width. 

To evaluate the value of the mean and standard deviation for longer boxcar filters we have used the white noise approximation as discussed in section \ref{sec:idealmodel} and shown in equation \eqref{eqa:whitenoise}. 

When we measure the $\RSNRmax$ and $\RSNRmin$ using our idealised signal model, the pulse will, as slides along its width, be detected by boxcar filters placed at different times. Since we select the first maximum (minimum) we introduce a bias into our measurement of time localization. Meaning that pulses are detected earlier than their true position in time.

When Gaussian noise is present in the test data we must modify how we measure $\RSNRmm$. Since we select the highest SNR for each pulse position, this would prefer pulses which are enhanced by the addition of the noise. Hence, when noise is present we use an averaged $\RSNR$ from multiple measurements of the pulse's $\RSNR$ at the same position. This way we can mitigate the effect of the noise to some degree. Measurement made with noise will always have a tendency to get higher values of $\RSNRmm$, because we select the maximum SNR. For the SNR calculation we have also used the white noise approximation \eqref{eqa:whitenoise}.

\subsection{Signal loss for idealized signal}
\begin{figure*}[htp]
	\begin{minipage}[t]{.49\textwidth}
		\centering
		\includegraphics[width=\linewidth]{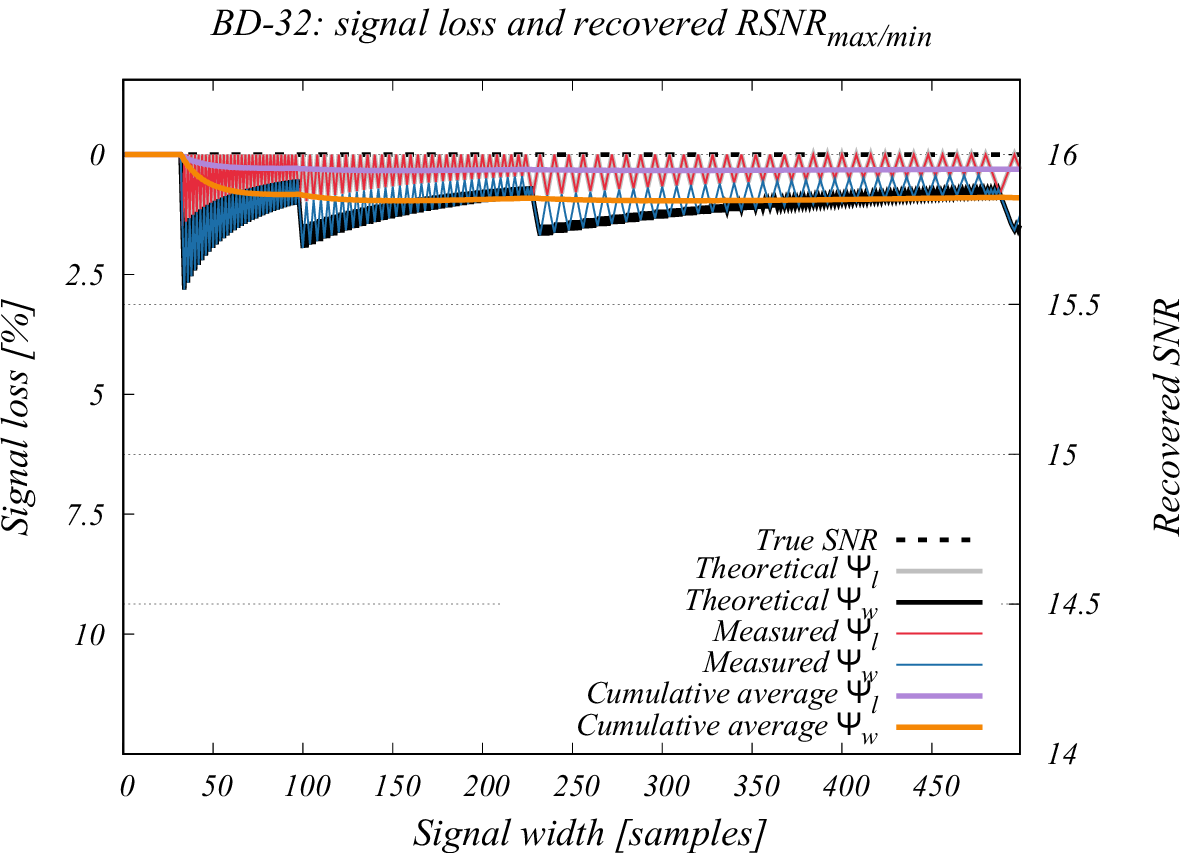}
	\end{minipage}
	\hfill%
	\begin{minipage}[t]{.49\textwidth}
		\centering
		\includegraphics[width=\linewidth]{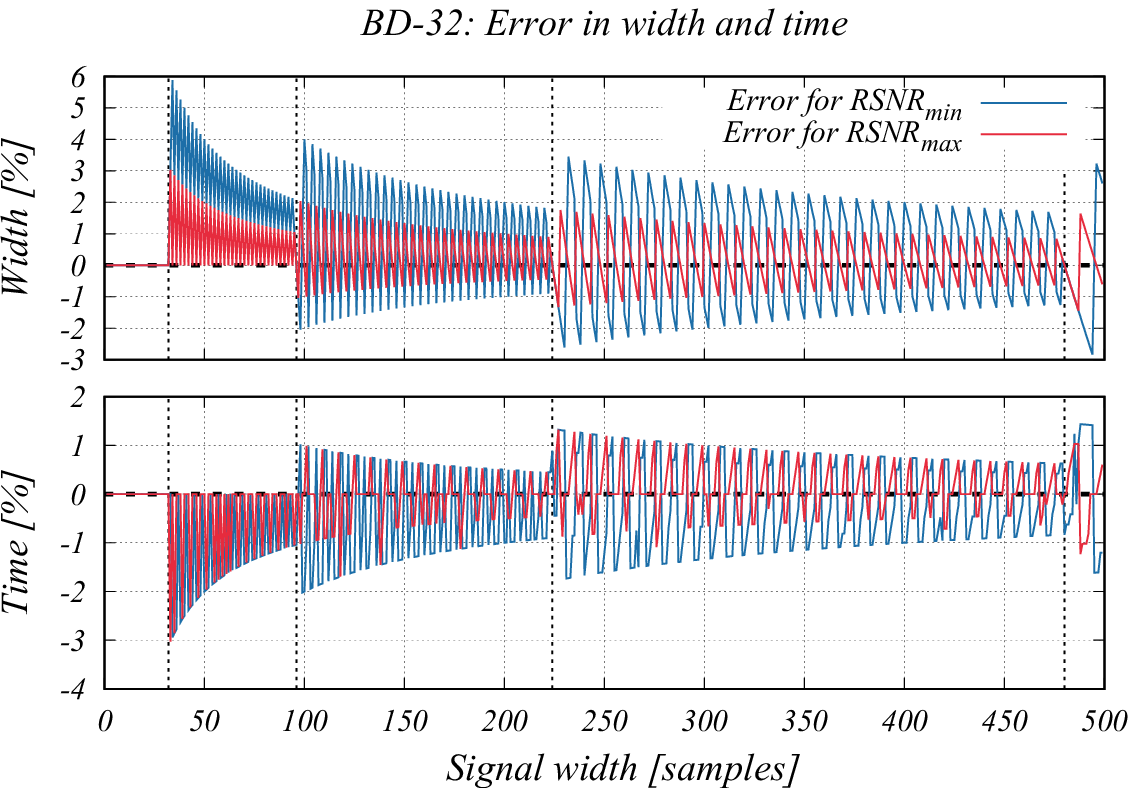}
	\end{minipage}\\
	%%%%%%%% row
	\begin{minipage}[t]{.49\textwidth}
		\centering
		\includegraphics[width=\linewidth]{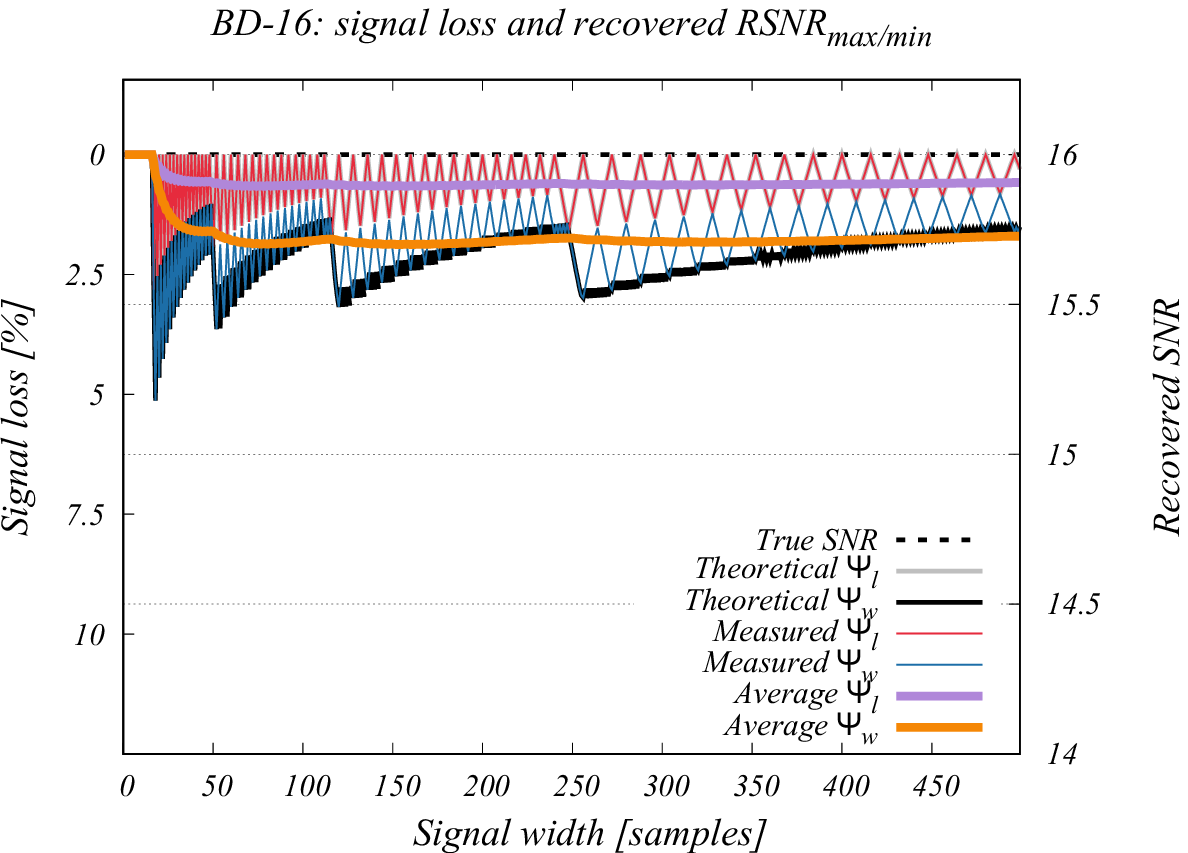}
	\end{minipage}
	\hfill%
	\begin{minipage}[t]{.49\textwidth}
		\centering
		\includegraphics[width=\linewidth]{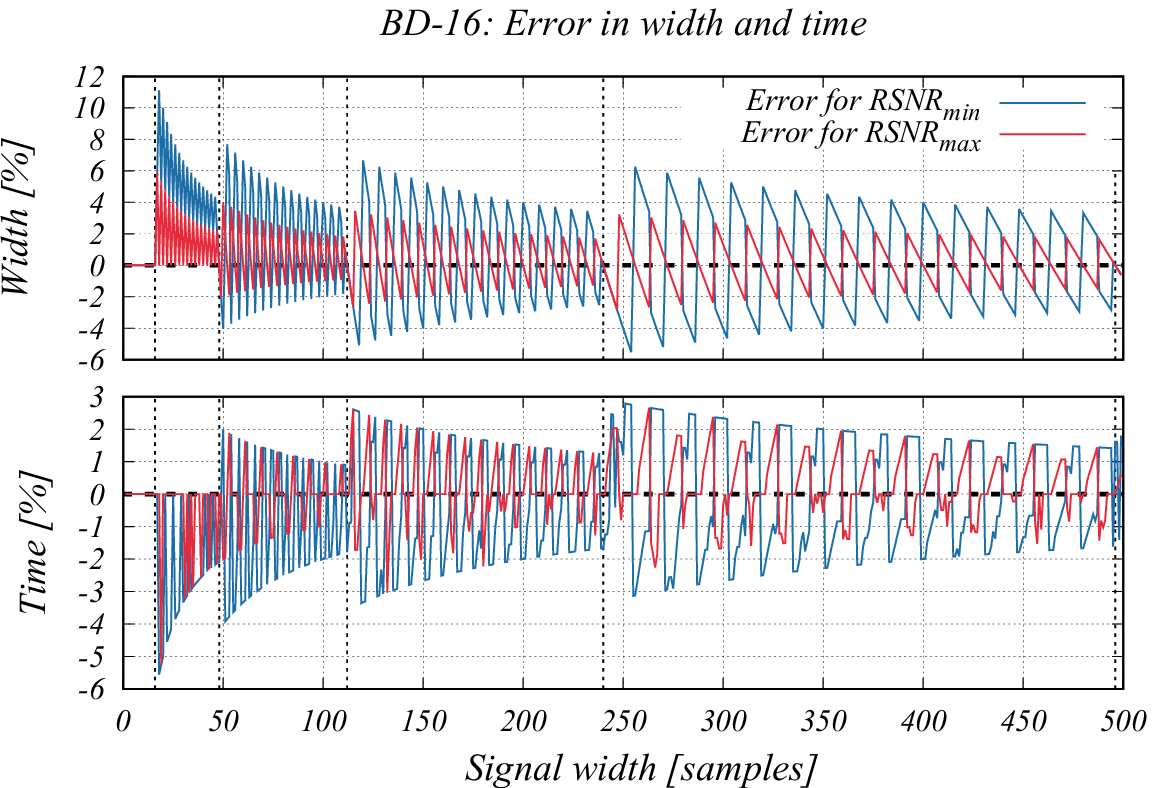}
	\end{minipage}\\
	%%%%%%%% row
	\begin{minipage}[t]{.49\textwidth}
		\centering
		\includegraphics[width=\linewidth]{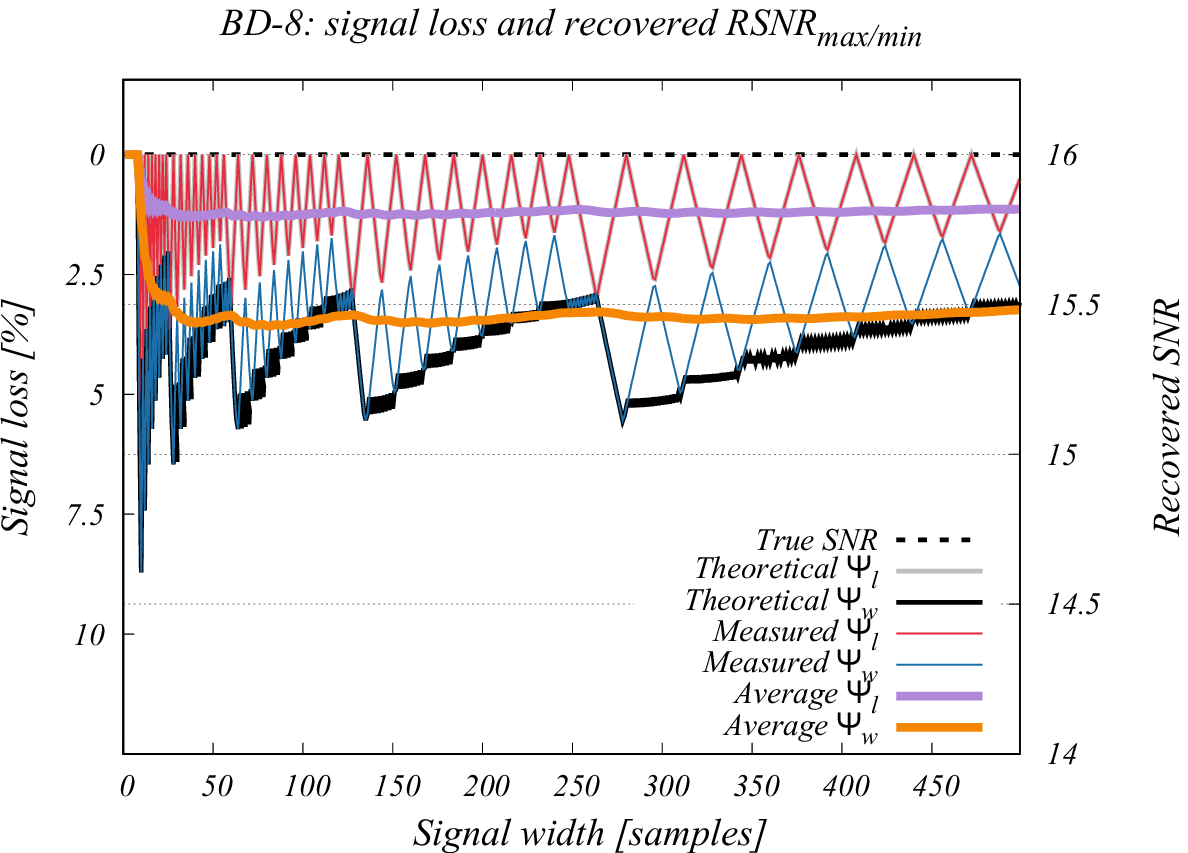}
	\end{minipage}
	\hfill%
	\begin{minipage}[t]{.49\textwidth}
		\centering
		\includegraphics[width=\linewidth]{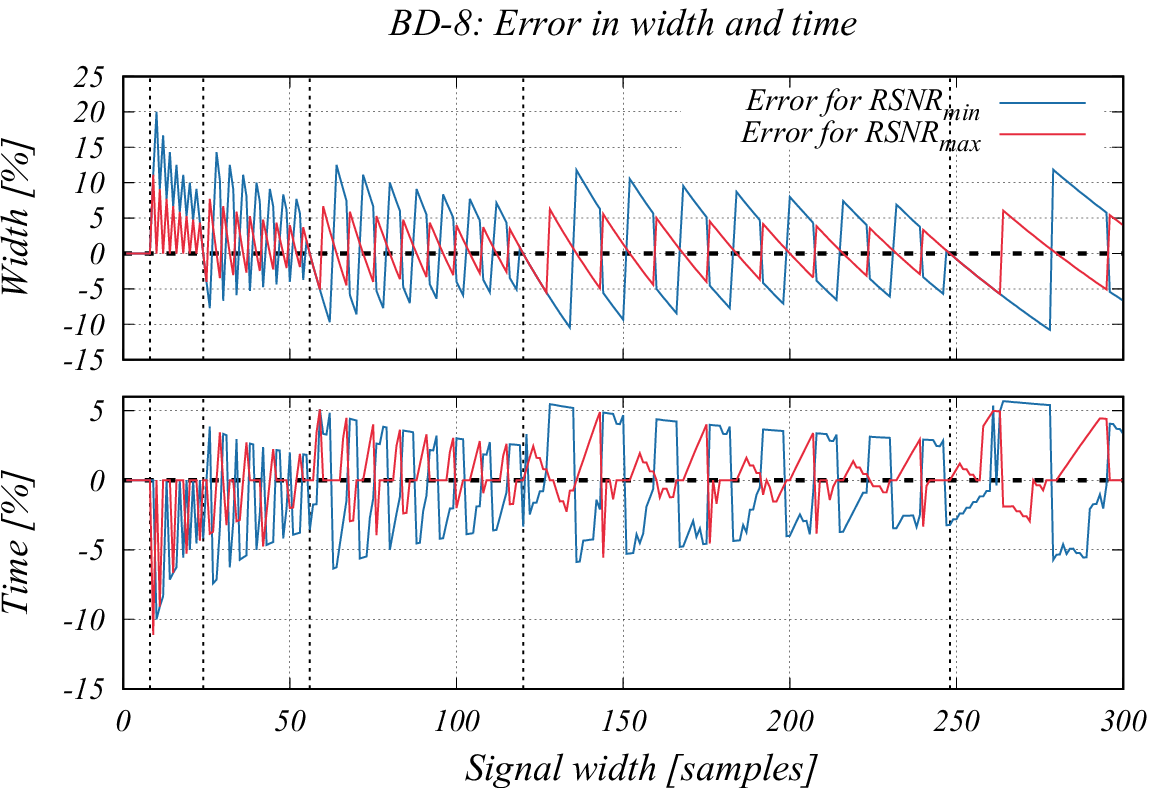}
	\end{minipage}

	\caption{BoxDIT: Measured signal loss $\sll$ and $\slw$ and associated $\RSNRmm$ (left) and error in measured width (as a percentage of the real pulse width), time difference relative to the true position of the pulse (as a percentage of the real pulse width) (right). These results are for our idealized signal model for three different configurations of our BOXDIT algorithm. A positive value of the error for the detected width indicates that the detected width is longer than the real width, a negative value indicates that it is shorter than the real width. For localization in time, negative values indicate that the pulse was detected at earlier time than the actual time at which the pulse occurred, a positive value indicates that the pulse was detected later. Together with measured signal loss (red and blue) we also show the predicted values of systematic signal loss $\sll$ in grey (located under red line) and worst signal loss $\slw$ in black. Lastly we also show the cumulative average signal loss in violet and orange.}
	\label{fig:SensitivityBoxDIT}
\end{figure*}

\begin{figure*}[htp]
	\begin{minipage}[t]{.49\textwidth}
		\centering
		\includegraphics[width=\linewidth]{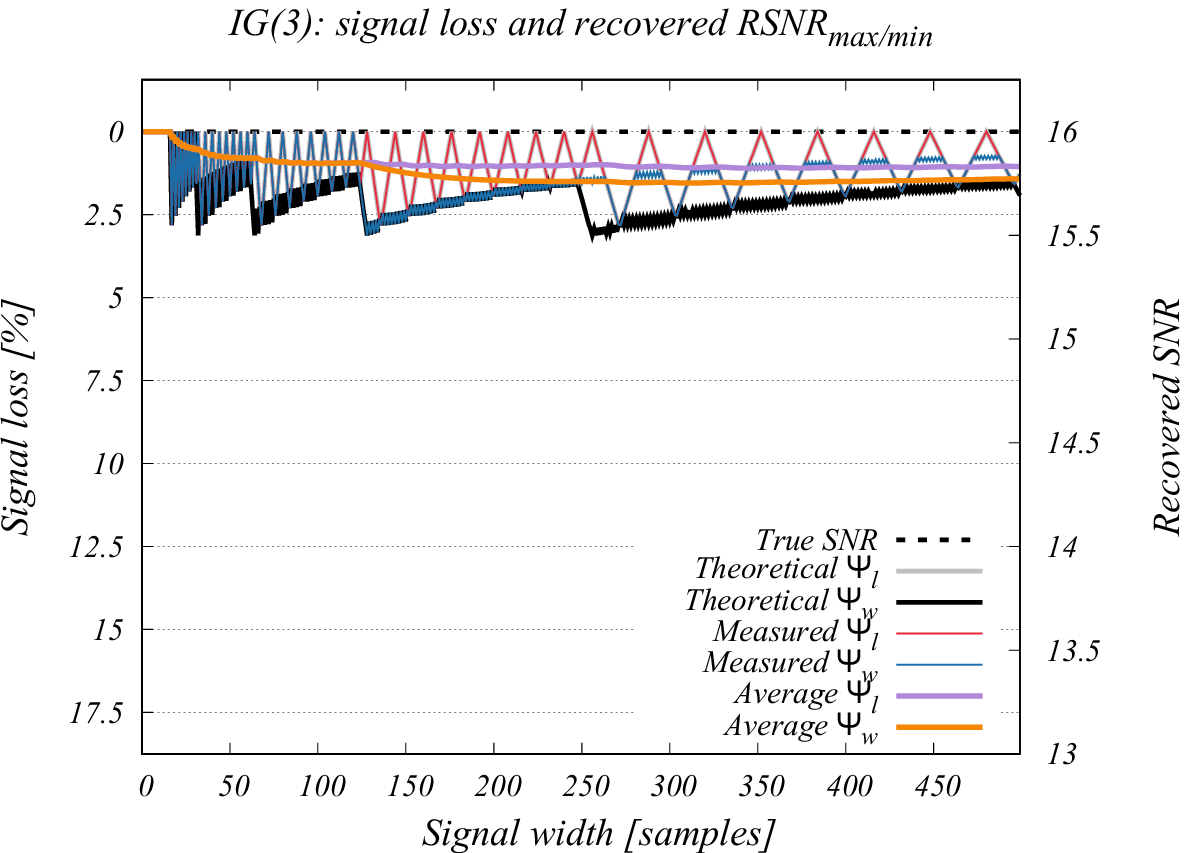}
	\end{minipage}
	\hfill
	\begin{minipage}[t]{.49\textwidth}
		\centering
		\includegraphics[width=\linewidth]{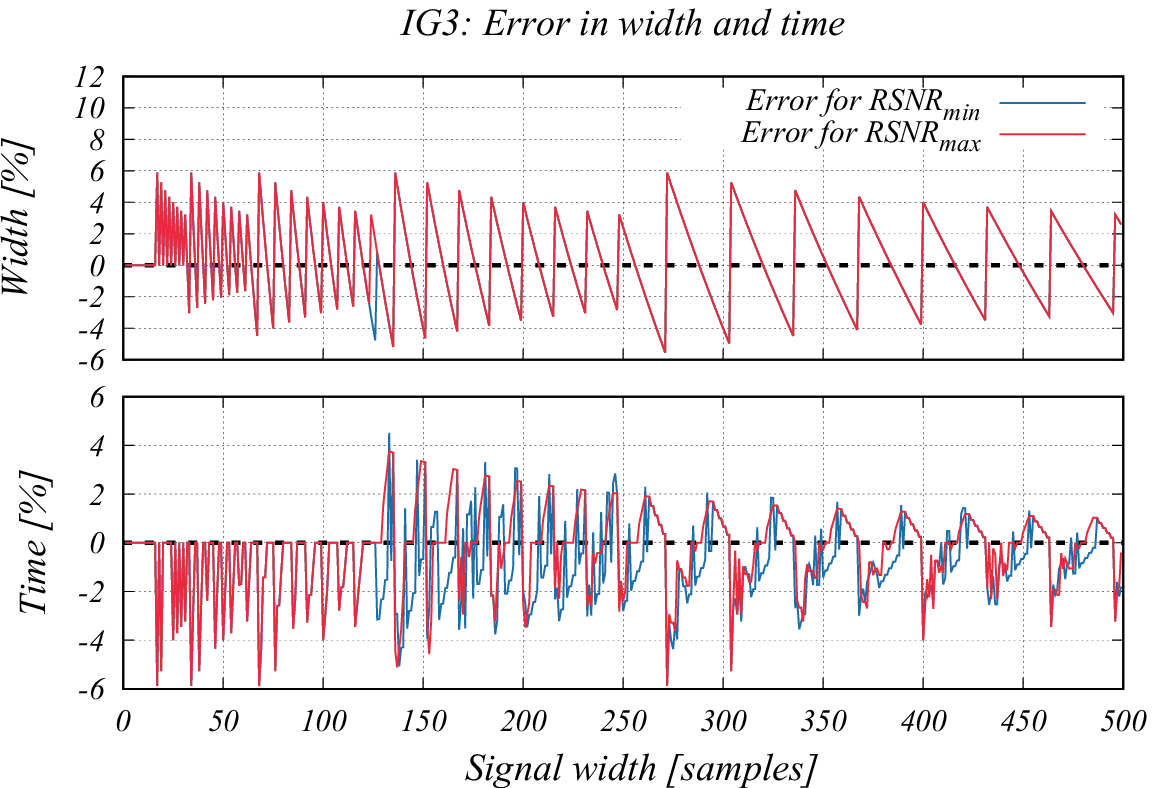}
	\end{minipage}\\
	%%%%%%%% row
	\begin{minipage}[t]{.49\textwidth}
		\centering
		\includegraphics[width=\linewidth]{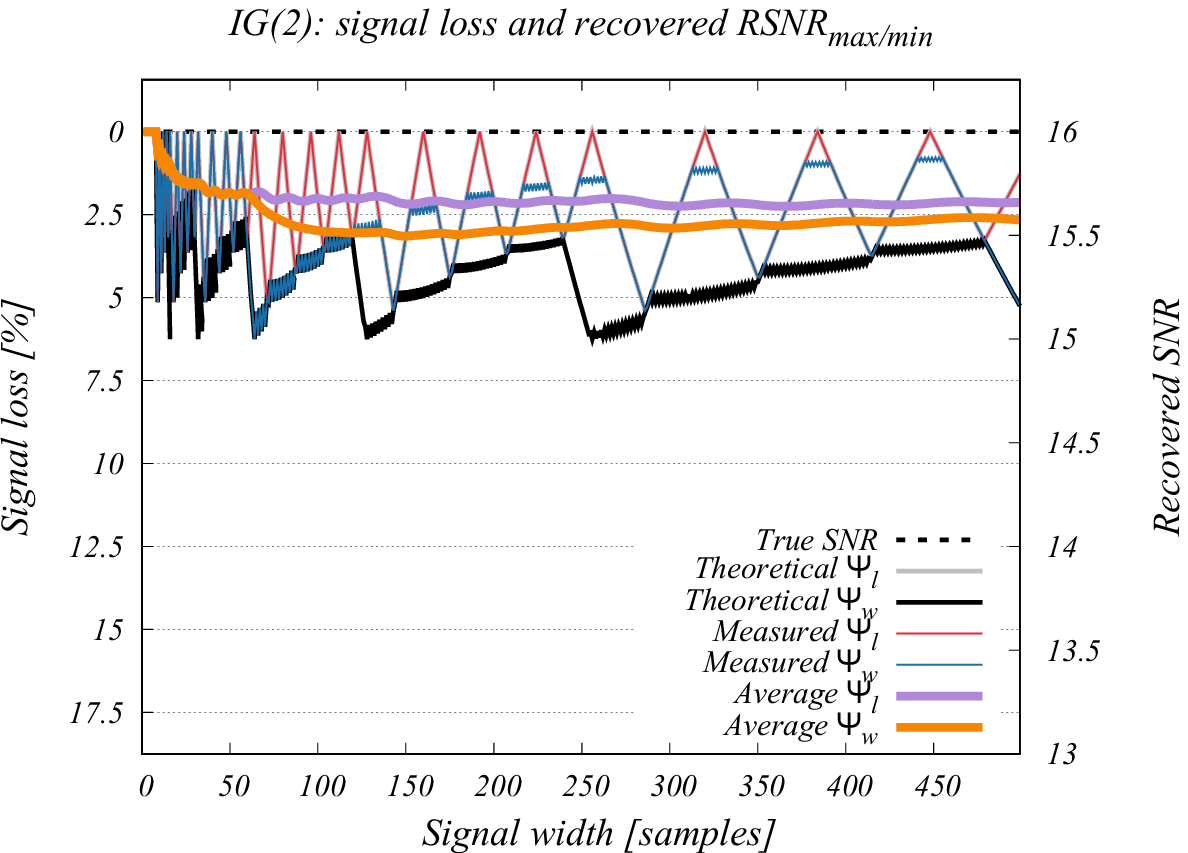}
	\end{minipage}
	\hfill
	\begin{minipage}[t]{.49\textwidth}
		\centering
		\includegraphics[width=\linewidth]{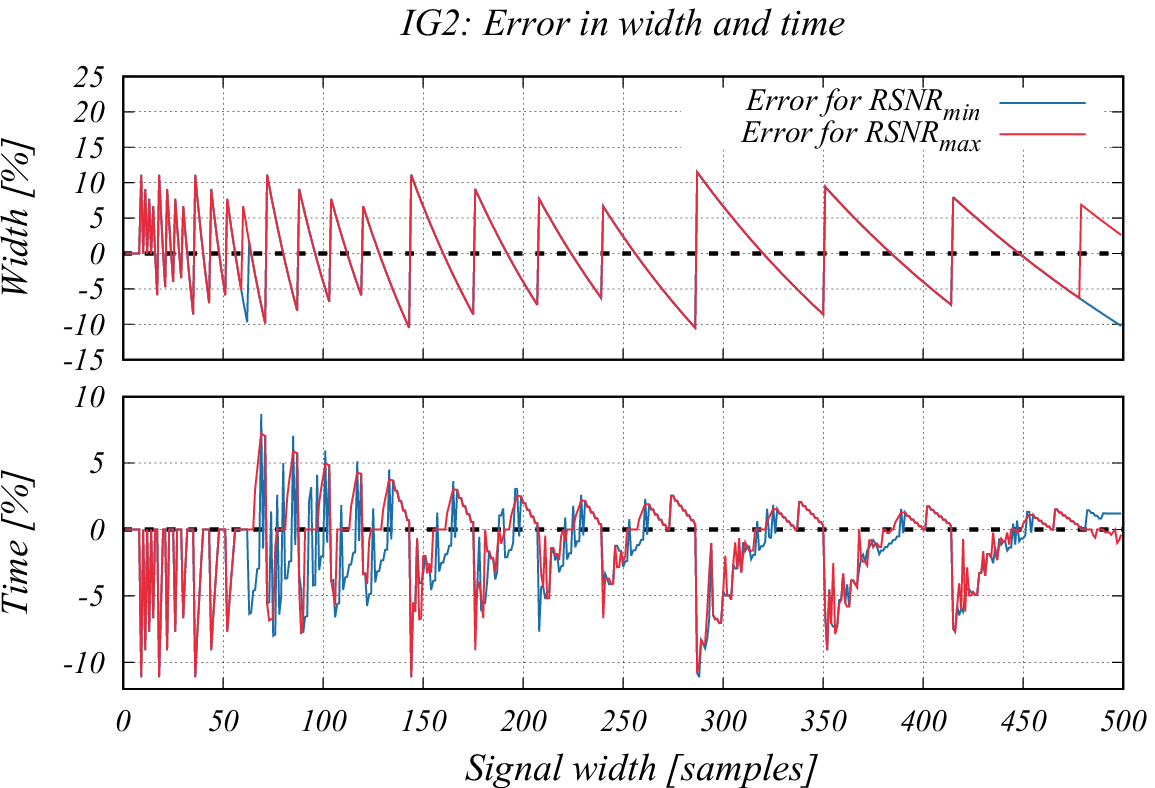}
	\end{minipage}\\
	%%%%%%%% row
	\begin{minipage}[t]{.49\textwidth}
		\centering
		\includegraphics[width=\linewidth]{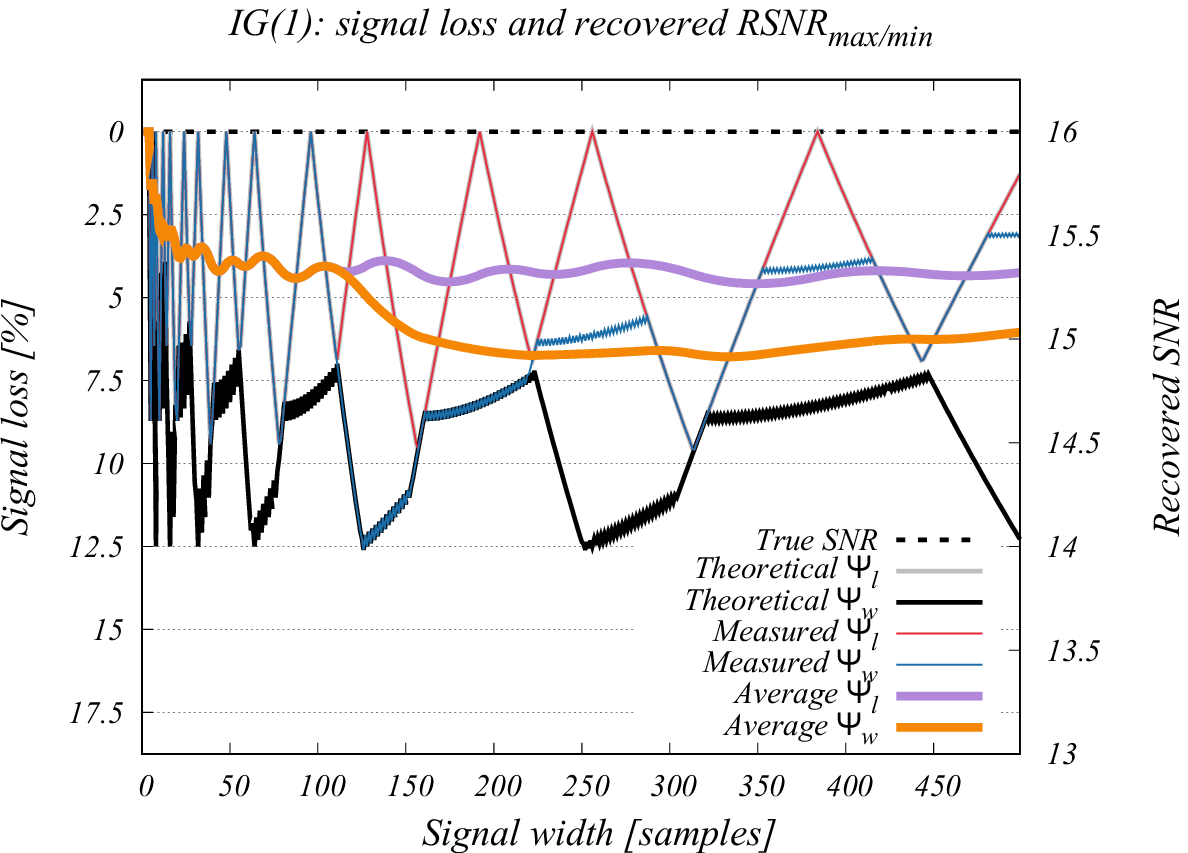}
	\end{minipage}
	\hfill
	\begin{minipage}[t]{.49\textwidth}
		\centering
		\includegraphics[width=\linewidth]{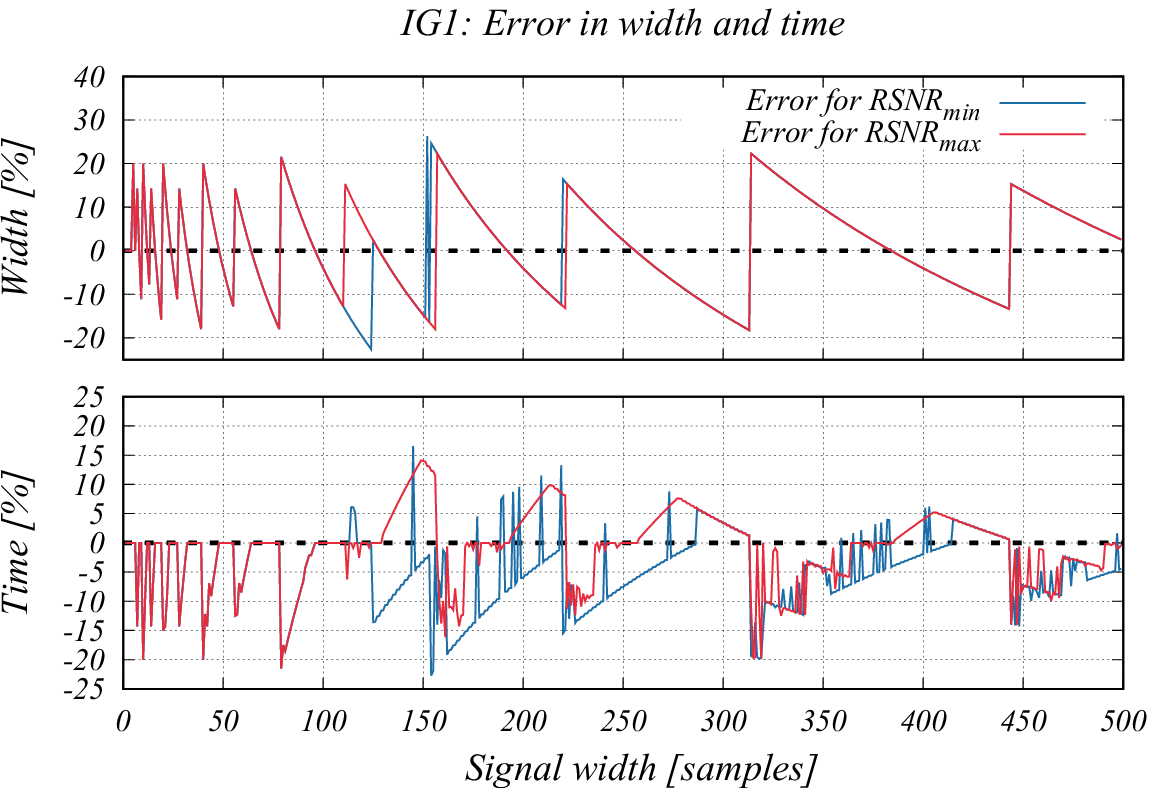}
	\end{minipage}

	\caption{IGRID: Measured signal loss $\sll$ and $\slw$ and associated $\RSNRmm$ (left) and error in measured width (as a percentage of the real pulse width), time difference relative to the true position of the pulse (as a percentage of the real pulse width) (right). These results are for our idealized signal model for three different configurations of our IGRID algorithm. A positive value of the error for the detected width indicates that the detected width is longer than the real width, a negative value indicates that it is shorter than the real width. For localization in time, negative values indicate that the pulse was detected at earlier time than the actual time at which the pulse occurred, a positive value indicates that the pulse was detected later. Together with measured signal loss (red and blue) we also show the predicted values of systematic signal loss $\sll$ in grey (located under red line) and worst signal loss $\slw$ in black. Lastly we also show the cumulative average signal loss in violet and orange.}
	\label{fig:SensitivityIGRID}
\end{figure*}

%%%%%%%%%%%%%%%%%%%%%%%% Signal loss for idealized signal %%%%%%%%%%%%%%%%%%%%%%%%
The measured $\RSNRmax$ and $\RSNRmin$ together with measured signal loss (eq. \eqref{eqa:signalloss}), error in detected width (eq. \eqref{eqa:errorpulsewidth}) and error in time localization (eq. \eqref{eqa:errorpulsetime}), of the rectangular pulse using our idealized signal model for our chosen configurations of our BoxDIT algorithm are shown in Figure \ref{fig:SensitivityBoxDIT} and for our IGRID algorithm in Figure \ref{fig:SensitivityIGRID}. Both errors are expressed as a percentage of the true pulse width.

The error in the time localization shown in Figures \ref{fig:SensitivityBoxDIT} and \ref{fig:SensitivityIGRID} are biased for earlier pulse detection as discussed above.

The comparison to predicted signal loss which is calculated by equations \eqref{eqa:RSNRmax} and \eqref{eqa:RSNRmin} are shown in figures \ref{fig:SensitivityBoxDIT} and \ref{fig:SensitivityIGRID} as gray and black lines. The measured systematic signal loss $\sll^m$ must be higher or equal to the predicted $\sll$, while measured worst signal loss $\slw^m$ must be lower or equal to the predicted worst signal loss $\slw$ for a given pulse width. In essence any measured value of the $\RSNR$ must lie be between values for $\sll$ and $\slw$.

Lastly, in figures \ref{fig:SensitivityBoxDIT} and \ref{fig:SensitivityIGRID}, we also present the cumulative average signal loss $\langle \Psi_X(S_0)\rangle$ up to a given pulse width $S_0$ which is calculated as
\begin{equation}
    \label{eqa:cumulativeaverage}
    \langle \Psi_X(S_0) \rangle = \frac{\sum_{s=1}^{S_0}\Psi_X(s)}{S_0-1}\,.
\end{equation}
This represents the mean signal loss $\langle \Psi_X(S_0)\rangle$ one can expect for all pulses with width $S<S_0$.

\subsubsection{Discussion}
The averaged signal loss for our chosen configurations of our BoxDIT algorithm (Fig. \ref{fig:SensitivityBoxDIT}) ranges from 1\% for configuration BD-32 to 3\% for configuration BD-8. For IGRID the averaged signal loss starts at 2\% for configuration IG(3) and increases to 12.5\% for IG(1). The sudden decreases in $\RSNR$ (increases in signal loss), which occur at different places for different configurations are caused by decimation in time. The decimation in time increases boxcar separation $\Ls$, which in the case of width and time localization determines the time resolution we are able to achieve. The boxcar separation represents the minimum increment in the measured width and the index of the time sample. Thus it is more significant for shorter pulse widths as it is a proportionally bigger part of the pulse's width. This is why the error in the width and time index decreases with pulse width.

Our IGRID algorithm behaves in a similar way (Figure \ref{fig:SensitivityIGRID}). Our IGRID algorithm has, in general, higher signal loss and a higher error in width and time localization than our BoxDIT algorithm. This is because IGRID places boxcar filters more sparsly than BoxDIT. 

A comparison of the measured signal loss $\sll^m$ and $\slw^m$ values to the predicted values of the signal loss $\sll$ and $\slw$ for both algorithms is shown on the left in Figure \ref{fig:SensitivityBoxDIT} for BoxDIT algorithm and in Figure \ref{fig:SensitivityIGRID} for IGRID algorithm. Both algorithms give the same values for $\sll$ as is predicted. However both algorithms have better worst signal loss $\slw$ than what is predicted by equation \eqref{eqa:RSNRmin}. This can be explained in each case. In the case of the IGRID algorithm this is because we calculate more layers than are necessary (section \ref{sec:IGRID_GPU_kernel}). This results in a smaller signal loss for some pulse widths. For the BoxDIT algorithm the higher $\RSNRmin$ is cause by boxcar filters of a shorter width which cover the pulse completely (best case) which has higher RSNR thus lower signal loss than boxcar filters considered in our sensitivity analysis for the worst case.

\subsection{Other pulse shapes}
%%%%%%%%%%%%%%%%%%% Other pulse shapes %%%%%%%%%%%%%%%%%
The profile of a real signal is usually far from rectangular, this is why we have included results from studies of two other pulse profiles; a Gaussian shape and combination of two Gaussians. We have measured $\RSNR$, error in detected width and error in time localization using the idealized signal model. For these alternative pulse shapes we cannot use the equation \eqref{eqa:signalloss} to calculate signal loss, because these pulses have a non-uniform distribution of power between samples of the pulse. Although the pulse is still normalized to the constant $\SNRT$ the non-uniform distribution means we can sum most of the pulse power with much shorter boxcar filters which overestimate SNR of the pulse, resulting in $\RSNR$ higher than $\SNRT$.
%Thus the assumption that a pulse has uniform distribution of power used in equation \eqref{eqa:RSNR} is no longer valid and as consequence we cannot determine $\SNRT$ which represent the highest RSNR value recoverable by any boxcar filter.

These pulse shapes are inserted in a similar fashion to the rectangular pulse shape. The input time-series is zeroed - all elements are set to zero. The pulse is then injected at the required position.

\begin{figure}[ht!]
	\centering
	\includegraphics[width=0.8\textwidth]{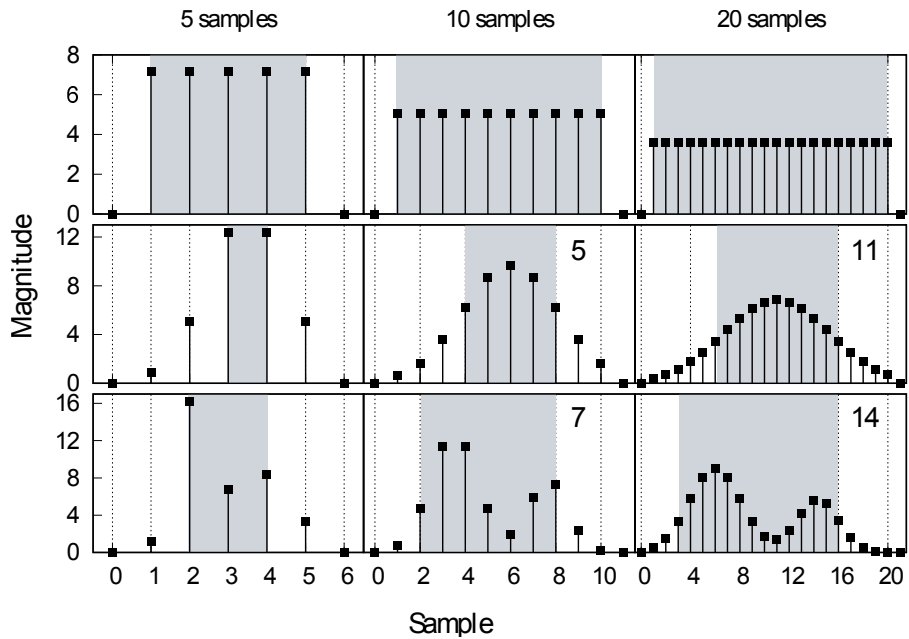}
	\caption{The three different pulse shapes used in this article. The pulses are shown with resolution of 5, 10 and 20 samples. The shaded region in each graph shows the samples of the pulse which were summed by the boxcar filter which produces the highest RSNR. The number of samples covered by the shaded area is at top right corner. \label{fig:pulseshapes}}
\end{figure}

To produce these pulses we have sampled the normal distribution given by 
\begin{equation}
f(x)=\frac{1}{\sqrt{2\pi\sigma^2}}\exp\left(-\frac{(x-\mu)^2}{2\sigma^2}\right)\,,
\end{equation}
where the mean is set to $\mu=S/2$ and the variance, $\sigma^2$, depends on the shape type and width of the pulse. For a Gaussian like shape we have used $\sigma^2=S/(8\ln2)$ \cite{BRANDT:2014:DataAnalysis} and for the double Gaussian profile we have used $\sigma^2=(S/4)/(8\ln2)$ and $\sigma^2=(S/5)/(8\ln2)$. The Gaussian profiles are then sampled at discrete intervals to produce values which are then injected into the idealized dataset. These pulses are shown in Figure \ref{fig:pulseshapes}. The signal is normalised such that when all its samples are summed together, i.e. a boxcar of signal width is applied to it, it will produce $\RSNR=C=16$.

\begin{figure*}[ht!]
	\centering
    \begin{minipage}[t]{.49\textwidth}
        \includegraphics[width=\textwidth]{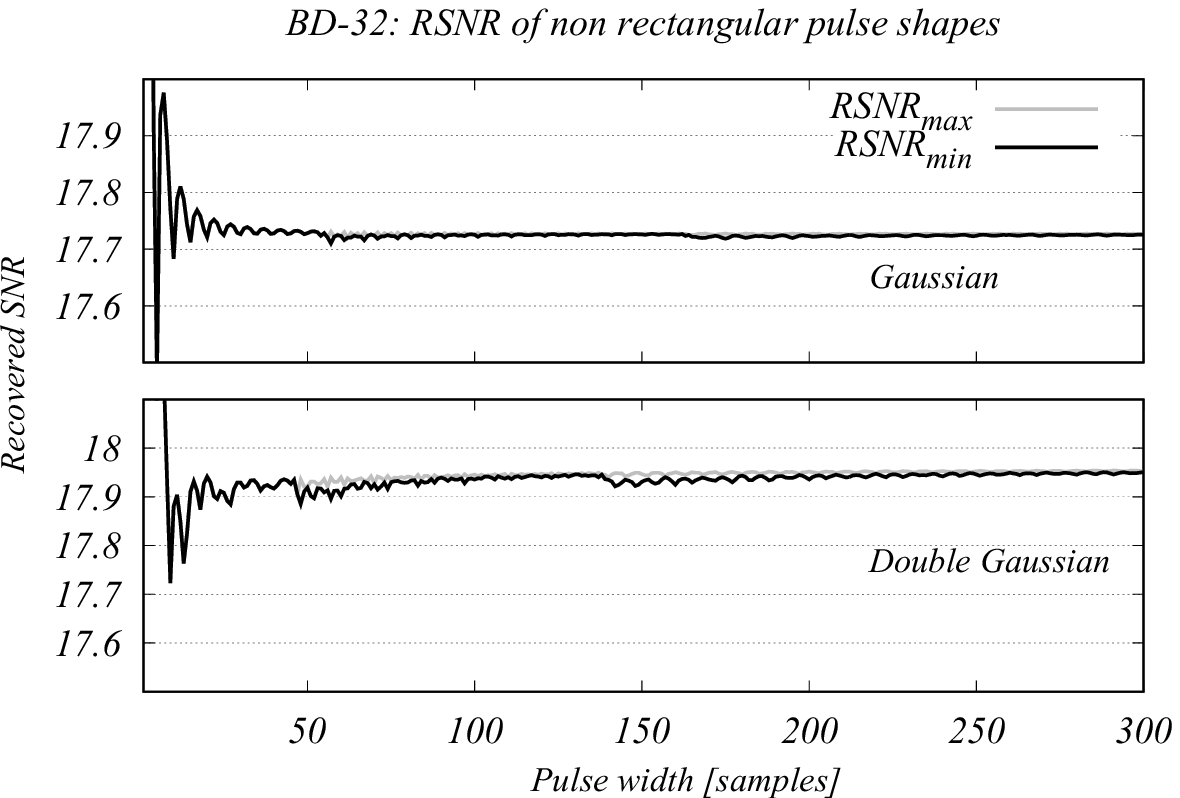}
    \end{minipage}
    \hfill
    \begin{minipage}[t]{.49\textwidth}
        \includegraphics[width=\textwidth]{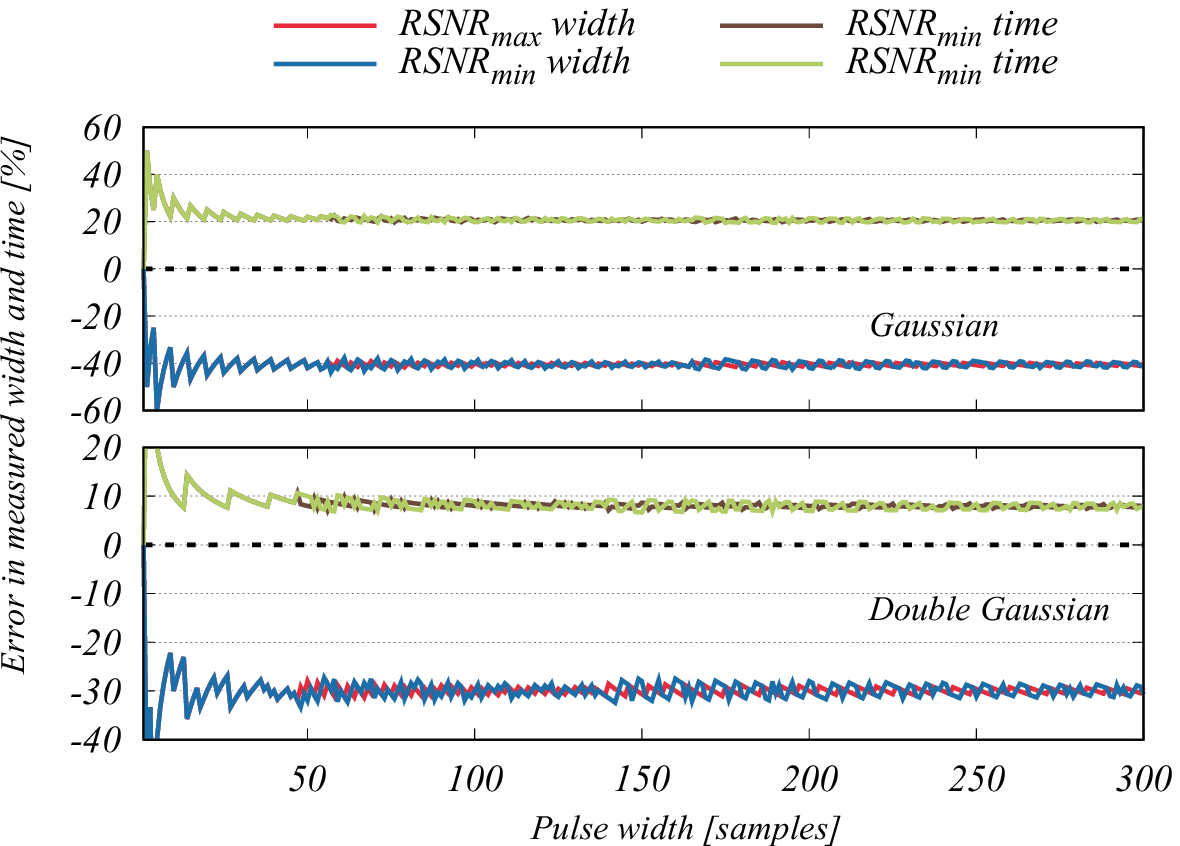}
    \end{minipage}\\
	\caption{Measured RSNR values and error in pulse width and time localization for different pulse shapes for BoxDIT BD-32 configuration. \label{fig:pulseshapes_SNR_boxdit}}
\end{figure*}

\begin{figure*}[ht!]
	\centering
    \begin{minipage}[t]{.49\textwidth}
        \includegraphics[width=\textwidth]{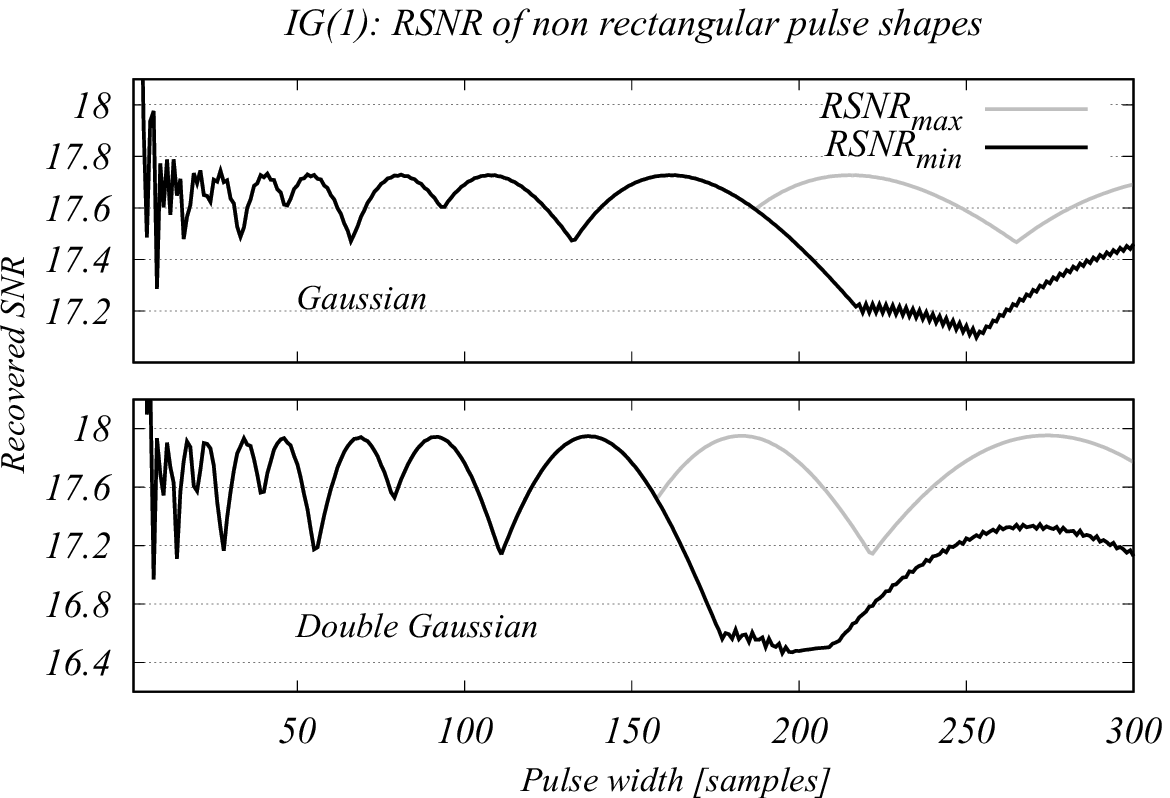}
    \end{minipage}
    \hfill
    \begin{minipage}[t]{.49\textwidth}
        \includegraphics[width=\textwidth]{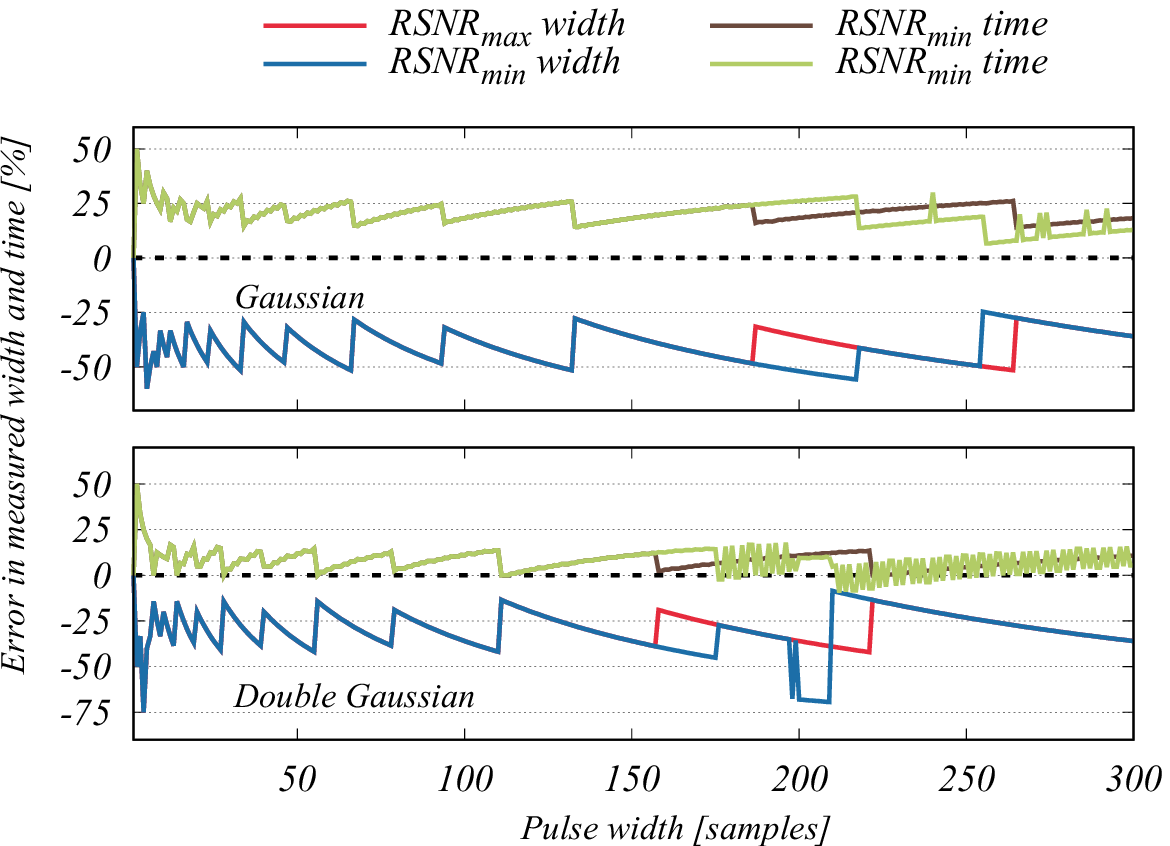}
    \end{minipage}\\
	\caption{Measured RSNR values and error in pulse width and time localization for different pulse shapes for IGRID IG(1) configuration. \label{fig:pulseshapes_SNR_igrid}}
\end{figure*}

\begin{figure*}[ht!]
	\centering
    \begin{minipage}[t]{.49\textwidth}
        \includegraphics[width=\textwidth]{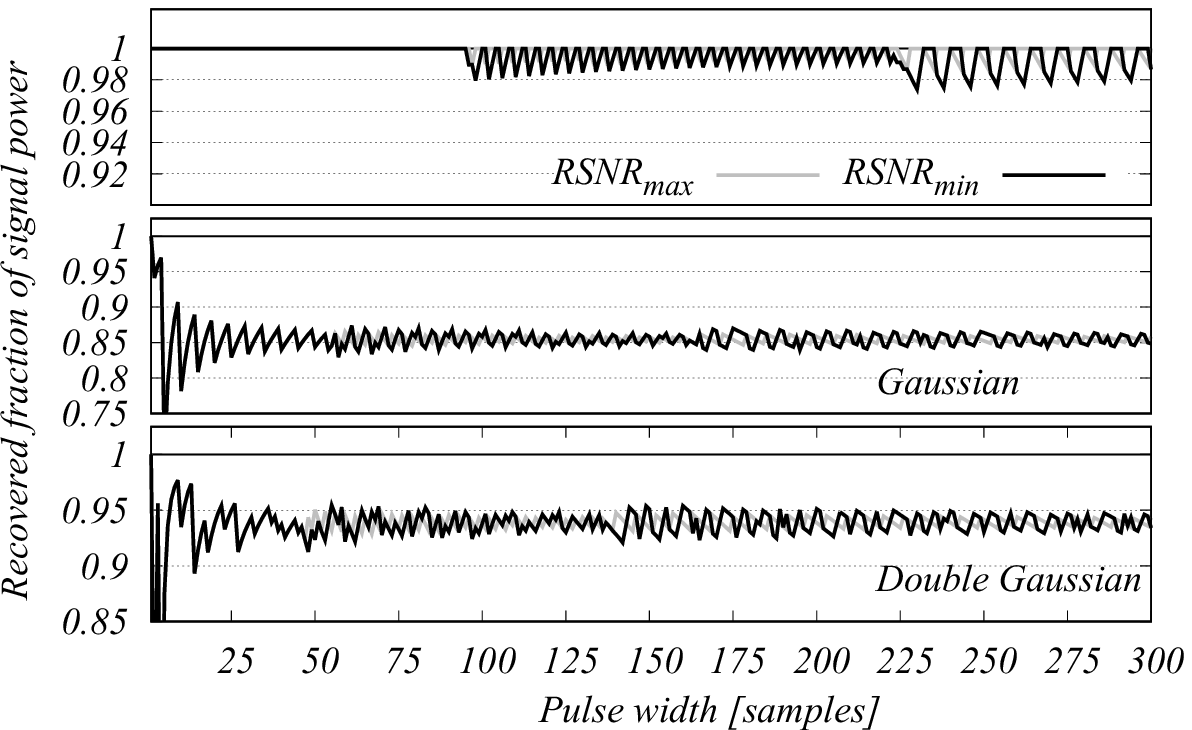}
    \end{minipage}
    \hfill
    \begin{minipage}[t]{.49\textwidth}
        \includegraphics[width=\textwidth]{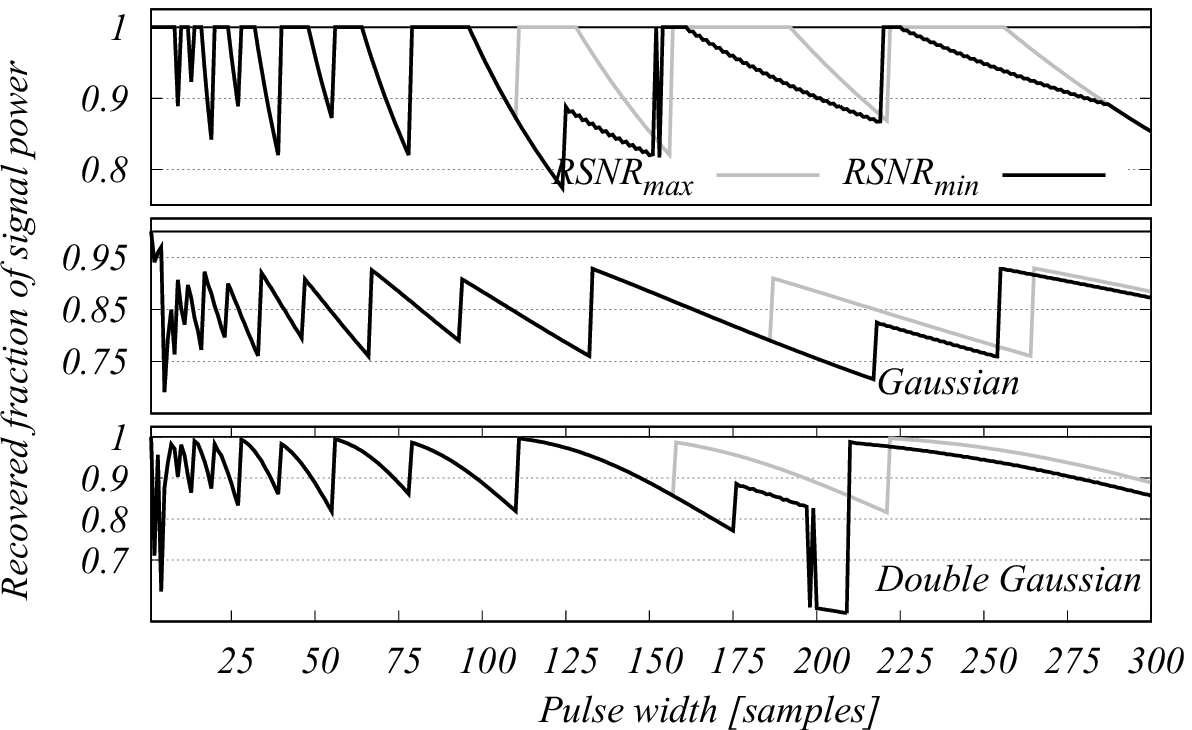}
    \end{minipage}\\
	\caption{Fraction of power recovered by the both algorithms. BoxDIT algorithm in configuration BD-32 (left) and IGRID algorithm in configuration IG(1) (right). \label{fig:pulseshapes_power}}
\end{figure*}

The results for the BoxDIT algorithm in the most sensitive configuration BD-32 are presented in Figure \ref{fig:pulseshapes_SNR_boxdit}. We present results for the fastest configuration of our IGRID algorithm (IG(1)) in Figure \ref{fig:pulseshapes_SNR_igrid}. We also show an alternative figure of merit to the signal loss in the form of the recovered fraction of a signal's power in Figure \ref{fig:pulseshapes_power} for both algorithms. 

\subsubsection{Discussion}
Figures \ref{fig:pulseshapes_SNR_boxdit} and \ref{fig:pulseshapes_SNR_igrid} show that the $\RSNR$ of pulse shapes other than rectangular are higher than the maximum for the rectangular pulse. This is due to the non-uniform distribution of power throughout these pulse shapes, where the power is concentrated into fewer samples. This means that a shorter boxcar filter can sum most of the power contained in the pulse which results in a higher RSNR. The fluctuations in $\RSNR$ for short widths is cause by the poor discreet representation of pulse shapes, shown in Figure \ref{fig:pulseshapes}.

The error in measured width and position in time is presented in figures \ref{fig:pulseshapes_SNR_boxdit} and \ref{fig:pulseshapes_SNR_igrid}. We see that boxcar filters systematically under-estimate pulse width when applied to a non-rectangular pulse. The error in the measured width is consistently about $40\%$ of the true pulse width for the single Gaussian and about $30\%$ for double Gaussian. These findings are consistent with the fact that Gaussian like pulse profiles tend to be detected with shorter boxcar filters than their true width due to the non-uniform distribution of power in the pulse.

The error in time localization is presented in figures \ref{fig:pulseshapes_SNR_boxdit} and \ref{fig:pulseshapes_SNR_igrid}. Time delay in detection is again expressed as a percentage of the true pulse width. Positive values indicate that the pulse is detected later in time, while negative values indicate that the pulse was detected earlier in time. These results show that Gaussian pulses are detected later than their true position in time. This is also shown in Figure \ref{fig:pulseshapes} which highlights the portion of the pulse which result in the highest RSNR detection. If we consider a Gaussian pulse that is 20 samples wide we see that it is detected by the boxcar filter with width of 11 samples, centered about the peak. Meaning that the pulse is detected with a time delay of 4 or 5 samples which represents $20\%$ or $25\%$ time delay compared to the true pulse width of 20 samples.

The recovered power by the BoxDIT (BD-32) and IGRID (IG(1)) algorithms for pulses of Gaussian like shapes are presented in Figure \ref{fig:pulseshapes_power}. We see that to produce the highest $\RSNR$ for Gaussian like pulses, the algorithm sums on average $85\%$ of the signal's power or $94\%$ for the double Gaussian pulse profile. These results show that boxcar filters which sum all the power contained within the pulse have lower $\RSNR$ and thus are not selected as candidates.

The noticeable increase in the width error visible in Figure \ref{fig:pulseshapes_SNR_igrid} for the double Gaussian pulse profile as well as the drop in the fraction of recovered power in Figure \ref{fig:pulseshapes_power} is because the IGRID algorithm (IG(1)) detects only the first Gaussian out of two that are present in the pulse. The error in time localization, which is unchanged indicates that it is the first Gaussian that is detected. For other configurations of our IGRID algorithm this was not observed. This is because these configurations have a denser set $\mathfrak{B}$ as well as shorter boxcar separation $\Ls$ and they are able to better match the entire pulse.  
%%%%%%%%%%%%%%%%%%% Other pulse shapes %%%%%%%%%%%%%%%%%

\subsection{Time-series with white noise}
\begin{figure}[htp]
    \begin{minipage}[t]{.49\textwidth}
        \centering
        \includegraphics[width=\linewidth]{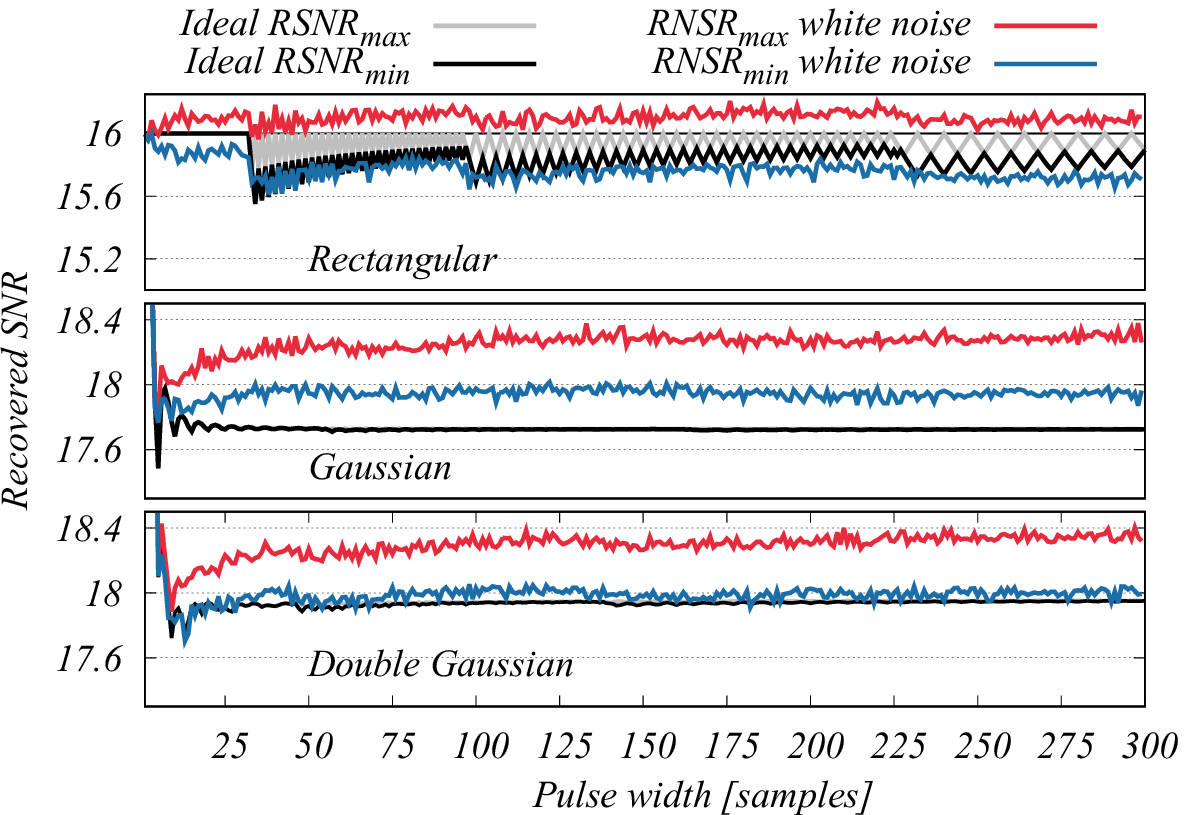}
    \end{minipage}
	\hfill
    \begin{minipage}[t]{.49\textwidth}
        \centering
        \includegraphics[width=\linewidth]{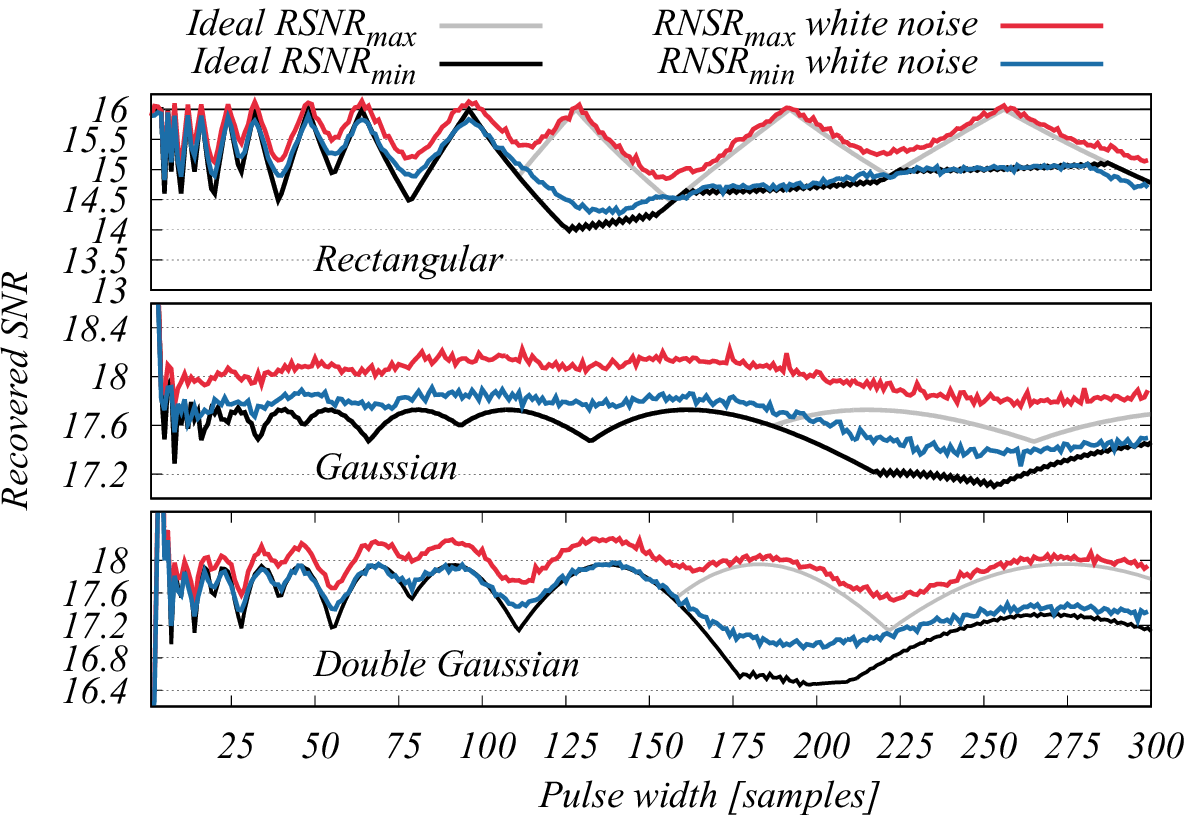}
    \end{minipage}\\
	\caption{Comparison of $\RSNRmm$ for our idealized signal without noise (gray and black) with $\RSNRmm$ for signals with Gaussian white noise (red and blue). \label{fig:SPDnoise}}	
\end{figure}

%%%%%%%%%%%%%%%%%%%%%% White noise %%%%%%%%%%%%%%%%%%%%%%
To verify the results of our SPD algorithms when there is noise present in the input data we have inserted Gaussian noise (mean $\mu=0$, standard deviation $\sigma=1$) into the idealized signal and measured $\RSNRmm$. To measure $\RSNRmm$ we have used the method described in section \ref{sec:measurements}. The measured values of $\RSNRmm$ for BoxDIT in configuration BD-32 and IGRID in configuration IG(1) are presented in Figure \ref{fig:SPDnoise}. The injected pulses have SNR=16. The effect of the noise on the measured $\RSNRmm$ is higher for pulses with lower SNR, as expected.

\begin{figure*}[ht!]
	\centering
    \begin{minipage}[t]{.49\textwidth}
        \includegraphics[width=\textwidth]{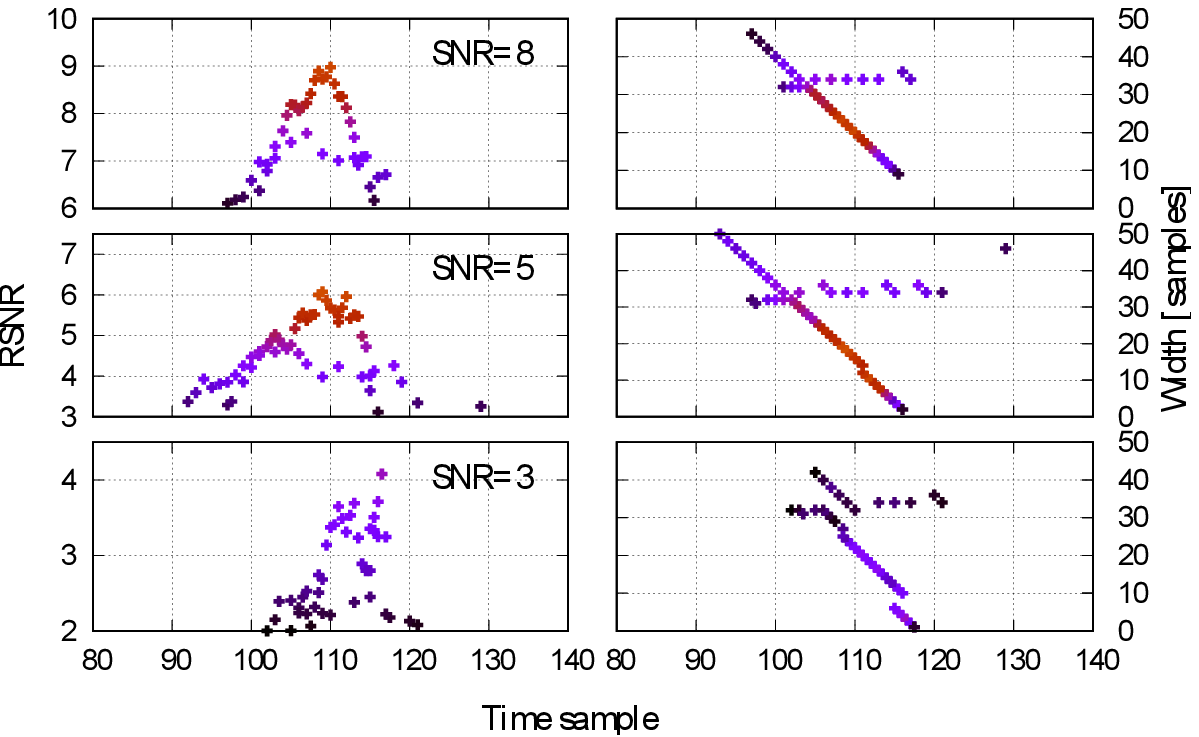}
    \end{minipage}
    \hfill
    \begin{minipage}[t]{.49\textwidth}
        \includegraphics[width=\textwidth]{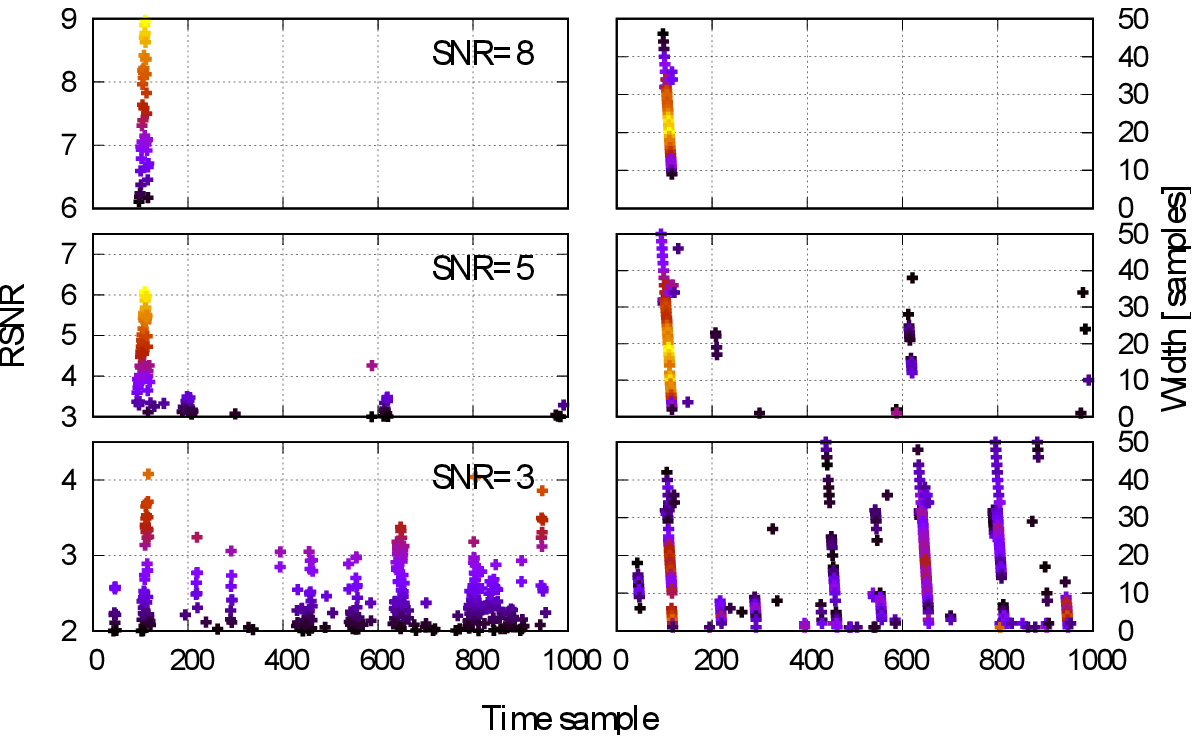}
    \end{minipage}\\
	\caption{Results for the rectangular pulse detections with different SNR embedded in white noise. The left graph shows a injected rectangular pulse of different initial SNR. It is split to show the RSNR values on the left and boxcar widths on the right. The points are colour coded by the RSNR. There might be more than one SNR value per sample because the SPD algorithm runs in multiple iterations and the results of all iterations are shown. Same results are shown in wider perspective in the graph on the right hand side. We see that the SNR=3 pulse would be hard to distinguish from background noise. This result is for BoxDIT algorithm with maximum search width $w = 8000$ samples. \label{fig:candidates_square}}
\end{figure*}

\begin{figure*}[ht!]
	\centering
    \begin{minipage}[t]{.49\textwidth}
        \includegraphics[width=\textwidth]{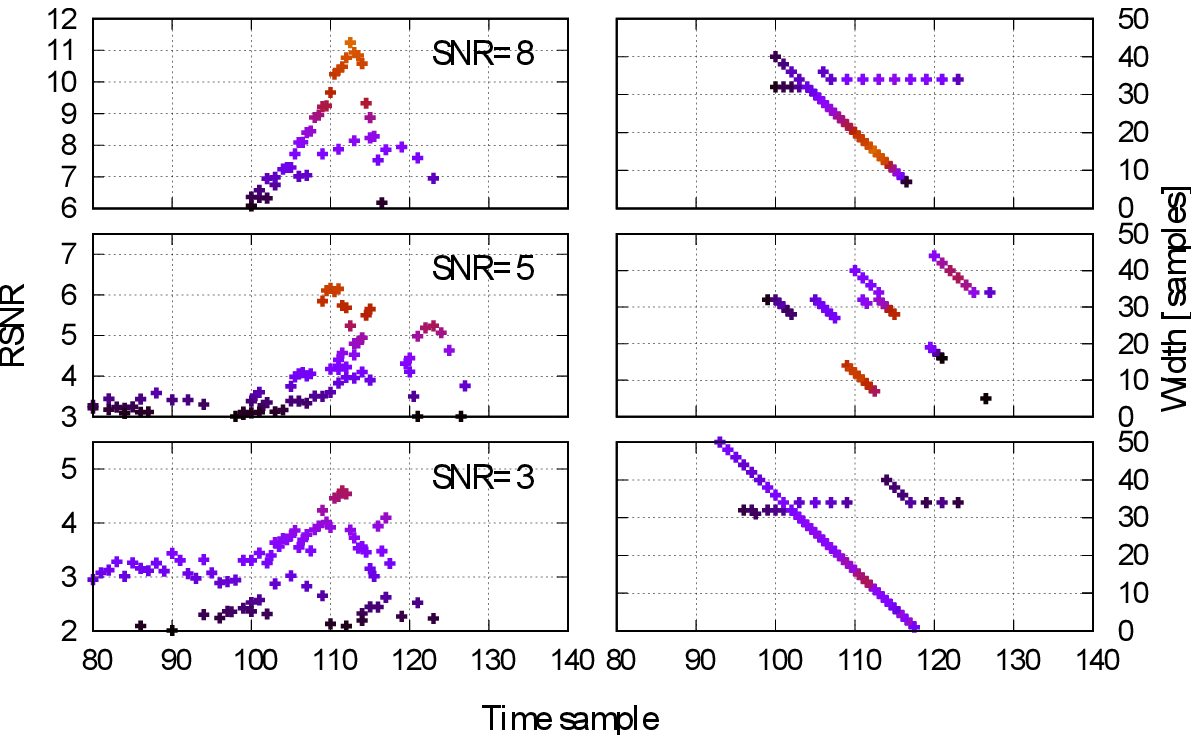}
    \end{minipage}
    \hfill
    \begin{minipage}[t]{.49\textwidth}
        \includegraphics[width=\textwidth]{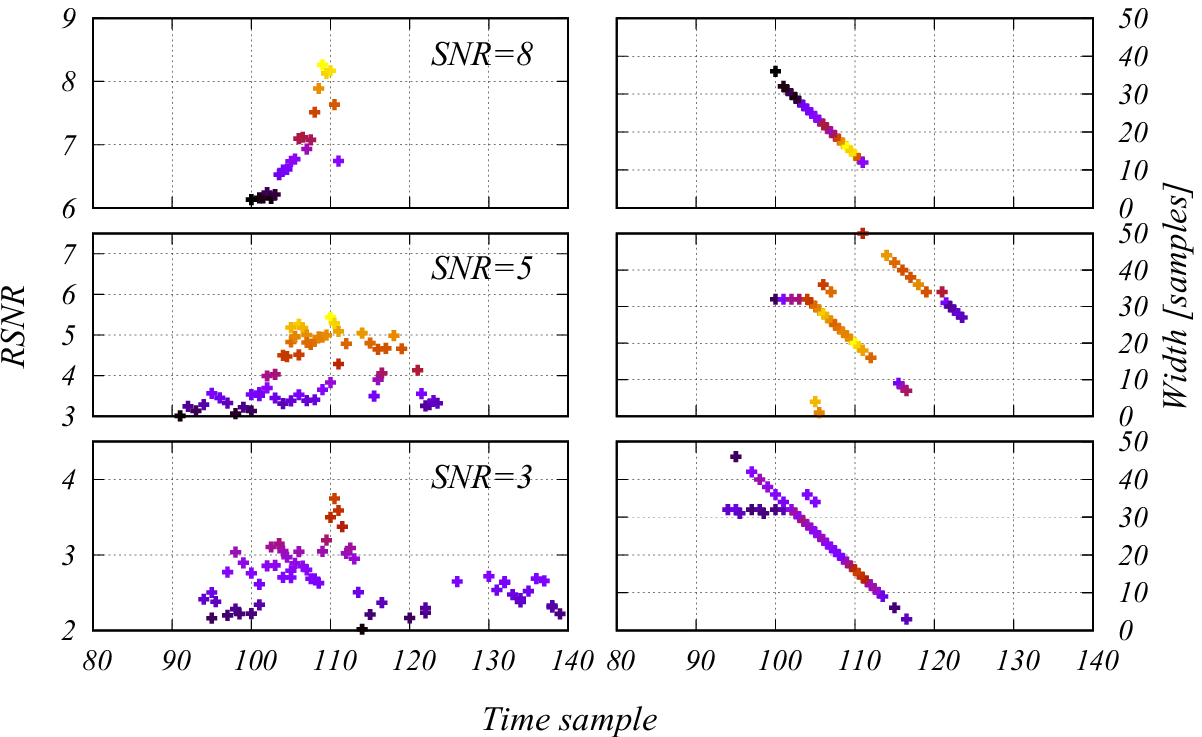}
    \end{minipage}\\
	\caption{Results for the detections for the Gaussian pulse with different initial SNR embedded in white noise is shown on the left set of graphs, and for double Gaussian on the right set of graphs. Each graph shows RSNR on the left and detected widths on the right. The points are colour coded by the RSNR. Results are for BoxDIT algorithm with maximum search width $w = 8000$ samples. \label{fig:candidates_gaussian}}
\end{figure*}

We have also used time-series with white noise to test SNR extraction. Examples for each pulse shape are presented in Figure \ref{fig:candidates_square} for the rectangular pulse, and in Figure \ref{fig:candidates_gaussian} for the Gaussian and double Gaussian pulse. The pulses have $\mathrm{SNR}=\{8,5,3\}$, width $S=20$ samples and they start at sample 100. For the rectangular pulse, the detection should occur at sample 110 by a boxcar filter of width 20 samples. For other pulse shapes the boxcar width is shorter as shown in Figure \ref{fig:pulseshapes}. The addition of the while noise means that single pulses shown in these figures do not represent overall sensitivity of our algorithms. 

\subsubsection{Disscusion}
The $\RSNR$ reported by our BoxDIT algorithm in configuration BD-32 and by our IGRID IG(1) algorithm are presented in Figure \ref{fig:SPDnoise}. The introduction of Gaussian white noise into our idealized model with rectangular pulse had only marginal effect on the detected RSNR by both algorithms. This is, for the most part, due to the initial SNR of the injected pulse which was $\SNRT=16$. We also see that $\RSNR$ produced by both SPD algorithms fluctuate about $\RSNRmin$ and often go below $\RSNRmin$. These variations are caused by fluctuations in the detection boxcar's width and it's position in time. 

Studying other pulse shapes we see that $\RSNR$ reported by BoxDIT is higher than in case without the noise. This is most visible for the Gaussian pulse shape. This again is due to the pulse shapes having a non-uniform distribution of power in their profile, hence they are detected with shorter boxcar filters than the actual pulse width. This makes it easier for the added noise to change the behavior of the SPD algorithm by changing the distribution of power throughout the pulse. For example, by decreasing the peak in the Gaussian pulse shape the boxcar which detects the pulse will be wider as it needs to accumulate power from more samples and in combination with our normalization this will decrease $\RSNR$ closer to 16. On the other hand by slightly increasing the peak in pulse profile, the pulse will be detected by a shorter boxcar producing higher $\RSNR$.

We see that in the case of the rectangular pulse (fig. \ref{fig:candidates_square}) the detection occurs at the correct position in time for SNR=8 and SNR=5. For pulse width SNR=3 the highest RSNR is detected at time sample 116 but the detected width underestimates the actual pulse width. This detection is mostly due to noise. The pulse itself is detected correctly at the time sample 110 with width 20, but it has lower RSNR. With decreasing pulse SNR we see that noise has stronger effect on the detected width and position of the pulse as would be expected. The Figure \ref{fig:candidates_square} also shows same data in a wider perspective with all candidates above a given threshold. For SNR=3 we see that the injected pulse is detected correctly, however it is hard to distinguish from background noise.
%%%%%%%%%%%%%%%%%%%%%% White noise %%%%%%%%%%%%%%%%%%%%%%

\subsection{Performance}
%%%%%%%%%%%%%%%%%%%%%% Performance %%%%%%%%%%%%%%%%%%%%%%
We have evaluated the performance of our algorithms and their implementation in different configurations by measuring the execution time. As a derived metric we have calculated the number DM trials that can be searched in real-time using sampling time $t_\mathrm{s}=64\mu s$. Processing data in real-time means performing all of the required operations on the data faster then we acquire the data. The number of DM trials processed in real-time $N_\mathrm{DM}$ is calculated as ratio of the number of processed time-samples per second by the SPD algorithm $N_\mathrm{P}$ and the number of acquired time-samples per second $N_\mathrm{S}=1/t_\mathrm{s}$. That is
\begin{equation}
    \label{eqa:DMtrialsinrealtime}
    N_\mathrm{DM}=\frac{N_\mathrm{P}}{N_\mathrm{S}}\,.
\end{equation}
The number of DM trials processed in real-time depends on the telescope sampling time and for a different sampling time other then $t_\mathrm{s}=64\mu s$, the number of DM trials searched in real-time will be different. For example for $t_\mathrm{s}=128\mu s$ the number of DM trials processed in real-time will be multiplied by factor of two $N^{(128)}_\mathrm{DM}=2N^{(64)}_\mathrm{DM}$.

The number of DM trials searched in real-time also depends on the maximum width of the boxcar filter $L_\mathrm{end}$ used for detection of the pulses. In the following results we have used $L_\mathrm{end}=8192$ samples, which for a sampling time $t_\mathrm{s}=64\mu s$ means a boxcar filter which is approximately half a second long. When searching for shorter pulses, the performance of the SPD algorithm can be increased by using shorter $L_\mathrm{end}$.

We present the performance of our implementations on two NVIDIA GPUs. The first is the TESLA P100 which is a scientific card from the Pascal generation. The second is the Titan V, which is a high end prosumer card from the Volta generation. We have not included any card from the current Turing generation since this generation is not designed for scientific tasks and workloads. The hardware specifications of these two GPUs are summarized in table \ref{tab:hardware}.

\begin{table*}[htbp]
	\caption{\label{tab:hardware} GPU card specifications. The shared memory bandwidth is calculated as $\mathrm{BW (bytes/s)} = \mathrm{(bank\, bandwidth\, (bytes))} \times \mathrm{(clock\, frequency\, (Hz))} \times \mathrm{(32\, banks)} \times \mathrm{(\#\, multiprocessors)}$.}
	\begin{center}
	\begin{tabular}{lrr}
	\hline
	\textbf{} & \textbf{P100} & \textbf{TITAN V} \\
	\hline
	Total CUDA Cores & 3584 & 5120 \\
	Streaming Mul. (SMs) & 56 & 80 \\
	Core Clock & 1303 MHz & 1455 MHz \\
	Memory Clock & 1406 MHz & 850 MHz \\
	Device m. bandwidth & 720 GB/s & 652 GB/s \\
	Shared m. bandwidth & 9121 GB/s & 14550 GB/s \\
	FP32 performance & 9.3 TFLOPS & 12.3 TFLOPS \\
	GPU memory size & 16 GB & 12 GB\\
	TDP & 250 W & 250 W\\
	CUDA version & 10.1 & 10.1\\
	Driver version & 418.39 & 415.27\\
	\hline
	\end{tabular}
	\end{center}
\end{table*}

\begin{figure*}[ht!]
	\centering
    \begin{minipage}[t]{.49\textwidth}
        \includegraphics[width=\textwidth]{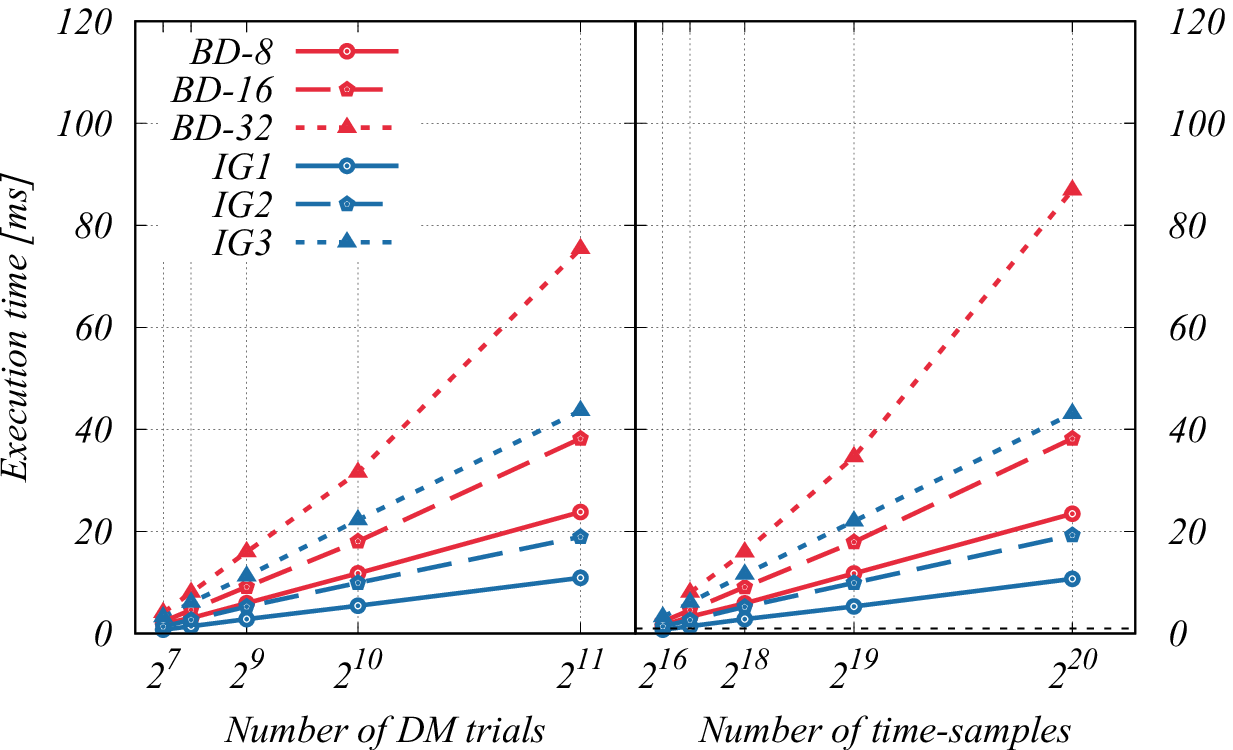}
    \end{minipage}
    \hfill
    \begin{minipage}[t]{.49\textwidth}
        \includegraphics[width=\textwidth]{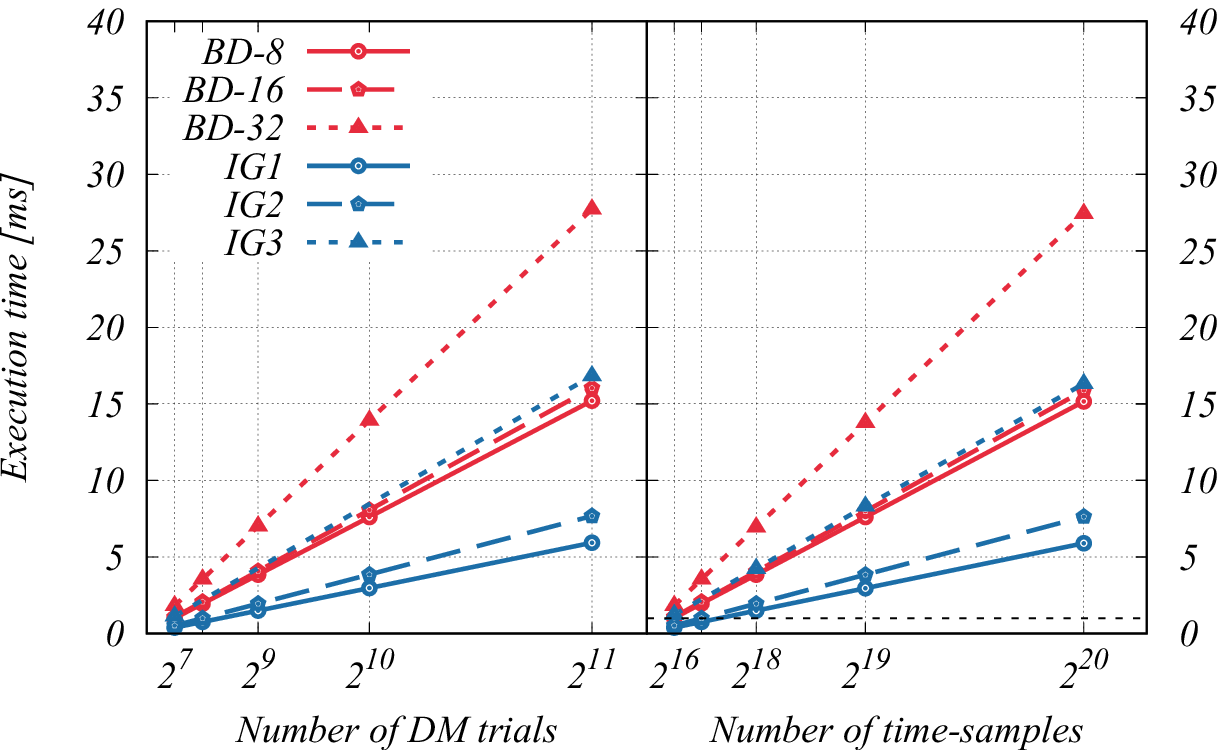}
    \end{minipage}\\
	\caption{Measured execution time (P100 and Titan V) for both BoxDIT and IGRID algorithms for increasing number of DM trials with 131072 time-samples per DM trial is on the left. Execution time for a varying number of time-samples per DM trial for 256 DM trials is shown on the right. All results are for a maximum boxcar width $L_\mathrm{end}=8192$ samples. In both cases our algorithms scale linearly. \label{fig:execution_time}}
\end{figure*}

\begin{figure}[ht!]
	\centering
	\includegraphics[]{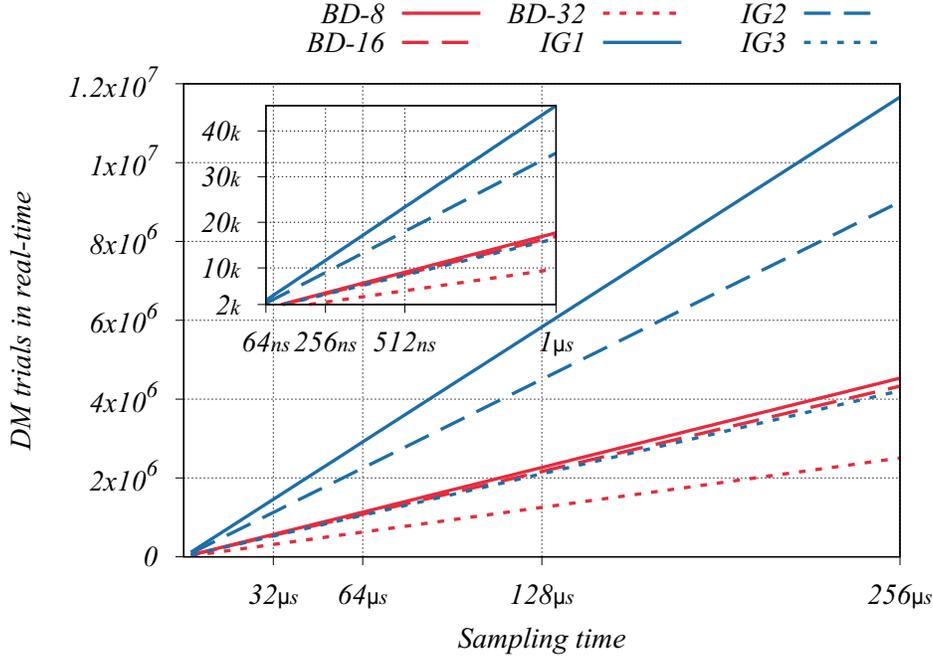}
	\caption{The number of DM-trials processed in real-time the dependency on sampling time. \label{fig:DM_trials}}
\end{figure}

Execution time and how it scales with increasing number of time-samples (per one DM trial) and execution time scaling with increasing number of DM-trials for the P100 and the Titan V are presented in Figure \ref{fig:execution_time}. The calculated number of DM-trials processed in real-time for different sampling times is presented in Figure \ref{fig:DM_trials}. 

\begin{figure*}
	\centering
    \begin{minipage}[t]{.49\textwidth}
        \includegraphics[width=\textwidth]{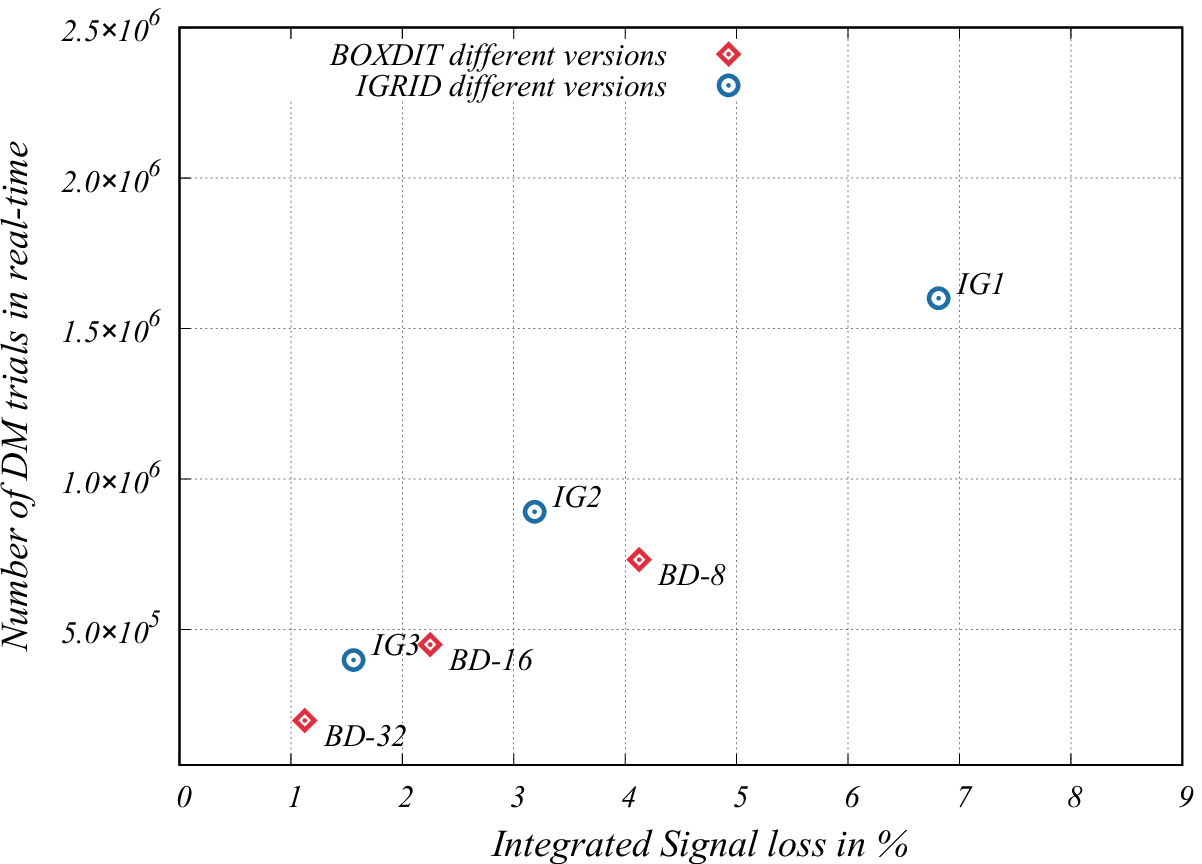}
    \end{minipage}
    \hfill
    \begin{minipage}[t]{.49\textwidth}
        \includegraphics[width=\textwidth]{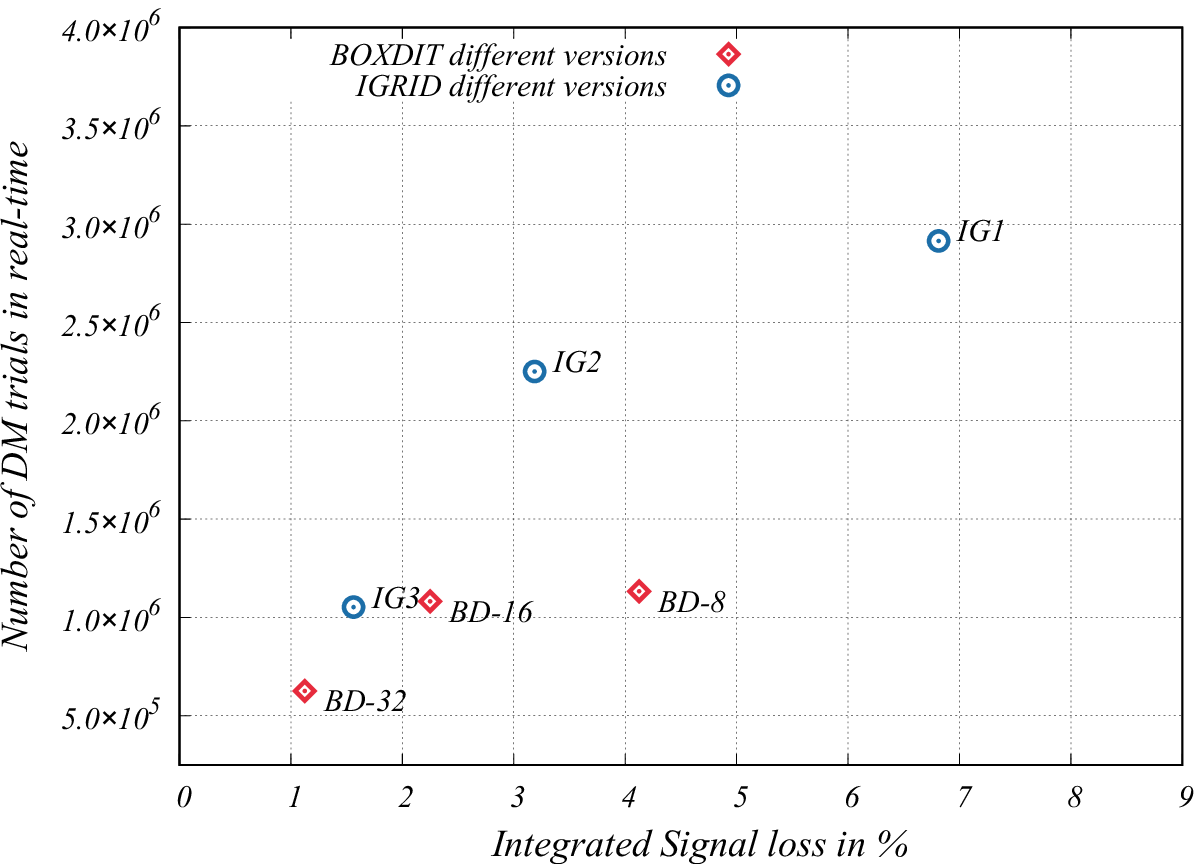}
    \end{minipage}\\
	\caption{Dependence of performance on the maximum width computed by the SPD algorithm used. Results are for P100 (left) and TITAN V (right). \label{fig:comparion_boxdit_IGRID}}
\end{figure*}

The comparison of the two algorithms with respect to their signal loss and achieved performance is presented in Figure \ref{fig:comparion_boxdit_IGRID}. This figure combines results presented in Figure \ref{fig:SensitivityBoxDIT} and Figure \ref{fig:SensitivityIGRID} with the number of DM trials processed in real-time using a sampling time $t_\mathrm{s}=64\mu s$. This figure allows us to compare the algorithms signal loss/performance ratio. 

\begin{figure*}
	\centering
    \begin{minipage}[t]{.49\textwidth}
        \includegraphics[width=\textwidth]{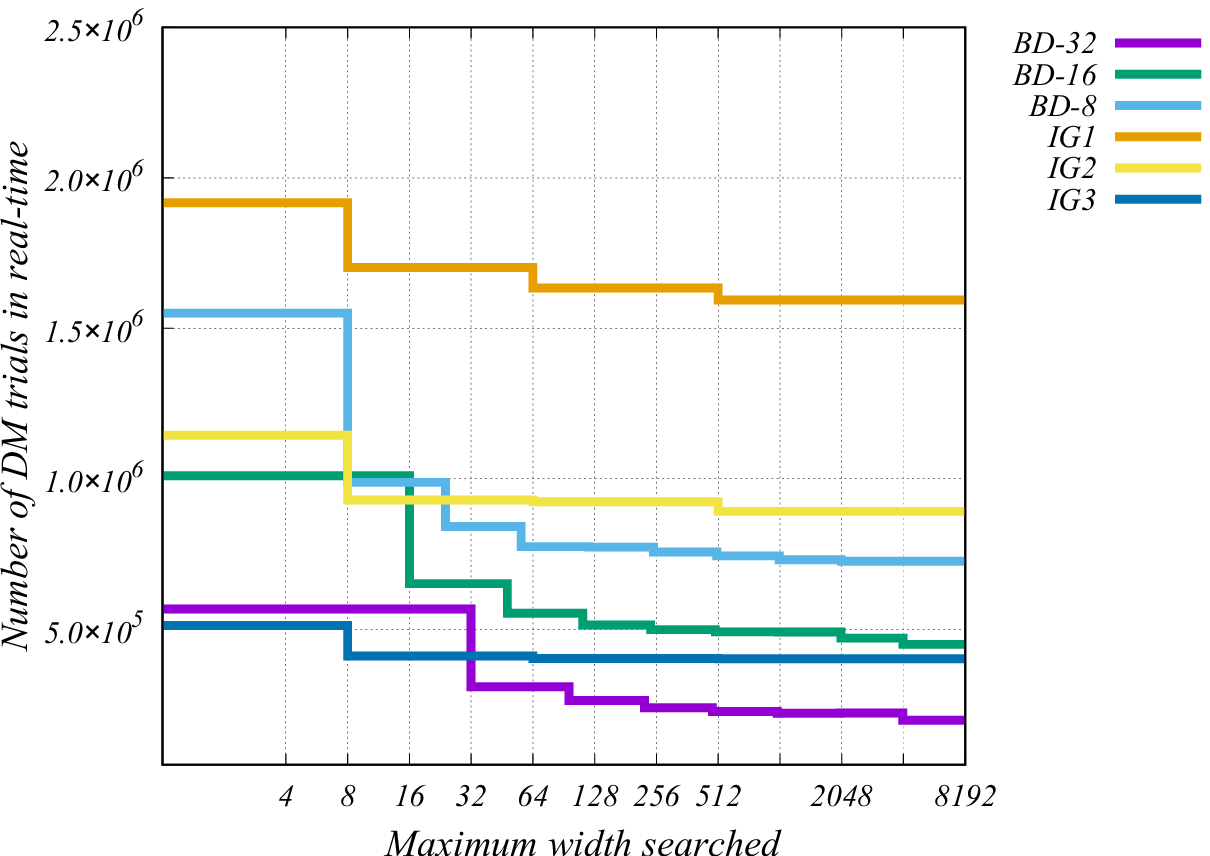}
    \end{minipage}
    \hfill
    \begin{minipage}[t]{.49\textwidth}
        \includegraphics[width=\textwidth]{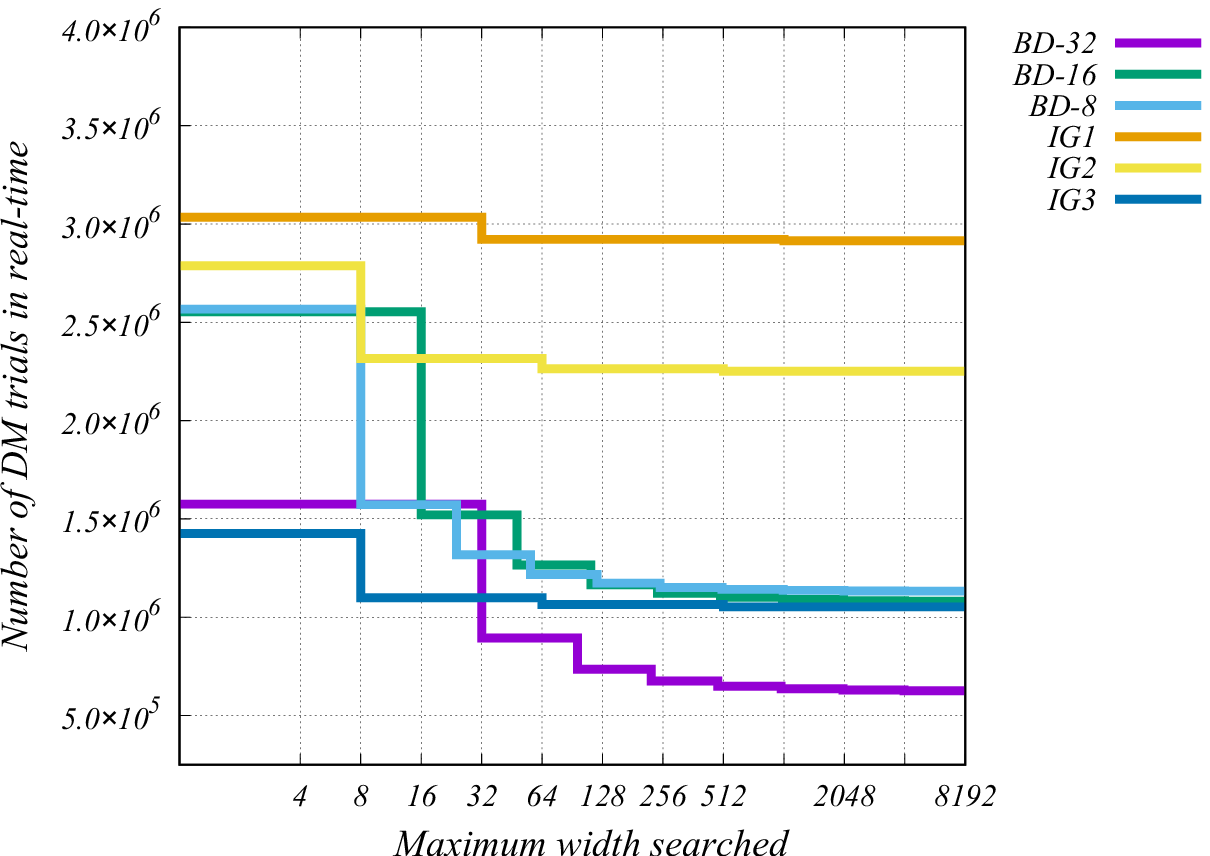}
    \end{minipage}\\
	\caption{Dependence of performance on the maximum width computed by the SPD algorithm used. Results are for P100 (left) and TITAN V (right). \label{fig:maxLperformance}}
\end{figure*}

Performance as a function of maximum boxcar width searched by the SPD algorithm is presented in Figure \ref{fig:maxLperformance}. In the case of IGRID algorithm, these results depend on how many IGRID steps we perform per GPU thread-block.

\subsubsection{Discussion}
The execution time for both algorithms is presented in Figure \ref{fig:execution_time}. Both algorithms scale linearly with the number of DM trials as well as with the number of time-samples. This is because the SPD problem is separable into independent parts, at the level of DM-trials as well as within a single DM-trial. Thus adding more time samples or DM-trials is equivalent.

The calculated number of DM-trials processed in real-time for the Titan V GPU is presented in Figure \ref{fig:DM_trials}. We see that for a sampling time of $t_\mathrm{s}=64\mu s$ the IG(1) algorithm is capable of processing three million DM-trials in real-time and the BD-32 algorithm is capable of processing almost six hundred thousand DM-trials in real-time. When considering radio-astronomy beyond the SKA era, for extremely short sampling times such as $t_s=256\mathrm{ns}$, the real-time performance of our IG(1) algorithm is about ten thousand DM-trials in real-time. This however has to be taken in the context of a whole pipeline in which the SPD algorithm is used.

The comparison of both SPD algorithms based on their real-time performance for $t_\mathrm{s}=64\mu s$ and their cumulative averaged signal loss is presented in Figure \ref{fig:comparion_boxdit_IGRID}. When considering the trade-off between signal loss and performance, both algorithms perform well. For roughly two times increase in signal loss the performance increases by roughly two times. There are some exeptions to this rule. For example BD-8 running on Titan V has almost the same performance as BD-16. Also note, IG(1) on Titan V offers a performance increase of only $1.4\times$ when compared to IG(2).

Figure \ref{fig:comparion_boxdit_IGRID} presents the different behaviour of the GPUs used and also differences between both algorithms. The GPU resource utilization as reported by the NVIDIA visual profiler for selected configurations of BoxDIT and IGRID algorithms is summarised in Table \ref{tab:GPUutilization}.

\begin{table}[htbp]
	\caption{\label{tab:GPUutilization} Compute/memory Utilization of used GPU's per configuration as reported by the NVIDIA visual profiler.}
	\begin{center}
		\begin{tabular}{lrr}
			\hline
			\textbf{} & \textbf{P100} & \textbf{TITAN V} \\
			\hline
			BoxDIT BD-32 & 90\%/15\% & 85\%/45\% \\
			BoxDIT BD-16 & 80\%/35\% & 80\%/85\% \\
			BoxDIT BD-8  & 80\%/55\% & 45\%/85\% \\
			IGRID IG(3)  & 75\%/25\% & 85\%/45\% \\
			IGRID IG(2)  & 85\%/35\% & 90\%/75\% \\
			IGRID IG(1)  & 85\%/55\% & 65\%/85\% \\
			\hline
		\end{tabular}
	\end{center}
\end{table}	

When we compare utilisation of both algorithms we see that they are similar on both GPUs. On the P100 GPU both algorithms are limited by compute performance. On the Titan V GPU they are limited by global memory bandwidth for higher signal loss configurations (BD-8, IG(1)), and by compute performance for lower signal loss configurations (BD-32, IG(3)). Algorithms with lower signal loss have to calculate a denser set of boxcar filters (more different widths) which are closer together. This requires more compute, but they use the same amount of input data. Despite similar GPU utilization IGRID is the better performing algorithm.

There are three main reasons why the IGRID algorithm is faster than BoxDIT. Firstly, the IGRID algorithm requires a smaller number of boxcar filters than the BoxDIT algorithm because IGRID decimates the input data after each IGRID step. If we compare BoxDIT BD-8 with the IGRID IG(2) algorithms the number of boxcar filters required by the IGRID IG(2) algorithm is half\footnote{The BD-8 performs 10 iterations with decimation after each step and in each step performs 8 boxcar filters. This gives $8\cdot 2\cdot N$ boxcar filters. The IGRID performs 14 IGRID steps of which the first 6 are in full resolution and all subsequent IGRID steps are decimated. This gives $6N + 2N = 8N$ boxcar filters.} compared to the number of boxcar filters required by the BoxDIT BD-8. On the P100 GPU where both algorithms are limited by compute we see that the IG(2) configuration has a lower signal loss but higher performance while compute utilization is almost the same for both algorithms. This shows that IGRID is more economical with the number of boxcar filters required to achieve a similar signal loss.

The IGRID algorithm also requires smaller input data, this is because in the IGRID algorithm the input time-series $x^i$ serves two functions, first it's values represent the values of the previously calculated boxcar filters which are used to built even longer boxcar filters. Secondly it serves as sample values, i.e. elements which are added to the existing boxcar filters to build up longer boxcars. In case of the BoxDIT algorithm we have to store these quantities in separate input arrays.

On Titan V both algorithms in their less sensitive variants are limited by device memory bandwidth. When we compare IGRID IG(2) with BoxDIT BD-8 we again see that IGRID is the better performing algorithm with lower signal loss while both algorithms are limited by device memory bandwidth.

The IGRID algorithm has some disadvantages as well. For high efficiency each configuration needs to be implemented individually and the complexity and size of the input required by IGRID increases with decreasing signal loss. Therefore extending IGRID to even higher sensitivities (lower signal loss) might be difficult. Furthermore, the signal loss of the IGRID algorithm depends on parameters of the implementation. Even though they do not decrease below the theoretically predicted worst signal loss $\slw$ the analysis of the data processed by the same configuration of IGRID algorithm (same depth) but with different parameters might cause issues. 

This problem is not shared by the BoxDIT algorithm. All presented configurations of our BoxDIT algorithm are served by a single implementation of the BoxDIT algorithm which only takes as an input, the number of boxcar filters to be performed (or maximum size of the boxcar filter) per iteration. Further more these might change from iteration to iteration which enables the user to decrease signal loss for certain pulse widths while increase it for others. 

Figure \ref{fig:maxLperformance} shows how performance depends on the maximum boxcar filter width $L_\mathrm{end}$ used. We see that IGRID algorithms are much less sensitive to increases in $L_\mathrm{end}$. This is due to lower number of GPU kernel execution needed by the IGRID algorithm to get to the desired boxcar width.

Lastly we present present fractions of real-time for SKA-like data. For SKA-like data we assume that the single pulse detection pipeline will process 6000 DM-trials per second with sampling time $t_s=64\mu s$. The BoxDIT algorithm in configuration BD-32 is capable of processing SKA-like data $33\times$ faster than real-time and $106\times$ faster than real-time on Titan V. The IGRID algorithm in IG(1) configuration is capable of processing SKA-like data $266\times$ faster than real-time on P100 GPU and $500\times$ faster than real-time on Titan V GPU. If the final SKA single pulse detection pipeline runs exactly in real-time with no spare processing time the IG(1) algorithm would take $0.4\%$ of processing time on P100 GPU and $0.2\%$ on Titan V GPU.

%%%%%%%%%%%%%%%%%%%%%%% Performance %%%%%%%%%%%%%%%%%%%

\subsection{Heimdall pipeline}
%%%%%%%%%%%%%%%%%%%%%%% Heimdal comparison %%%%%%%%%%%%%%%%%%%
In order to compare RSNR acquired by our SPD algorithms which are part of AstroAccelerate with another signal processing pipeline we have chosen to use Heimdall. Heimdall is a GPU accelerated transient detection pipeline developed by \citet{HeimdallSoft}, \citet{DDTR:2012:Barsdell}.

AstroAccelerate's single pulse search pipeline contains these steps: first data are de-dispersed; then the mean and standard deviation are calculated; followed by the SPD algorithm; lastly candidates are selected using a peak-finding algorithm. Since recovered/SNR by the pipeline depends also on other steps we will briefly describe them as well.

\subsubsection{De-dispersion transform}
%%%%%%%%%%%%%%%% Dedispersion %%%%%%%%%%%%%%%%%%%%%%%%
The de-dispersion transform used in AstroAccelerate was developed by \citet{DDTR:2012:Armour}. This implementation of the de--dispersion transform requires that the DM trials are calculated in groups with a constant step in DM. In contrast the de-dispersion transform used in the Heimdall pipeline has an adaptive step and so can have a varying step in DM for every DM-trial. This means that with AstroAccelerate we cannot fully match the de-dispersion scheme used by Heimdall (the Heimdall scheme should be more accurate). 

\begin{figure}[ht!]
	\centering
	\includegraphics[width=0.8\textwidth]{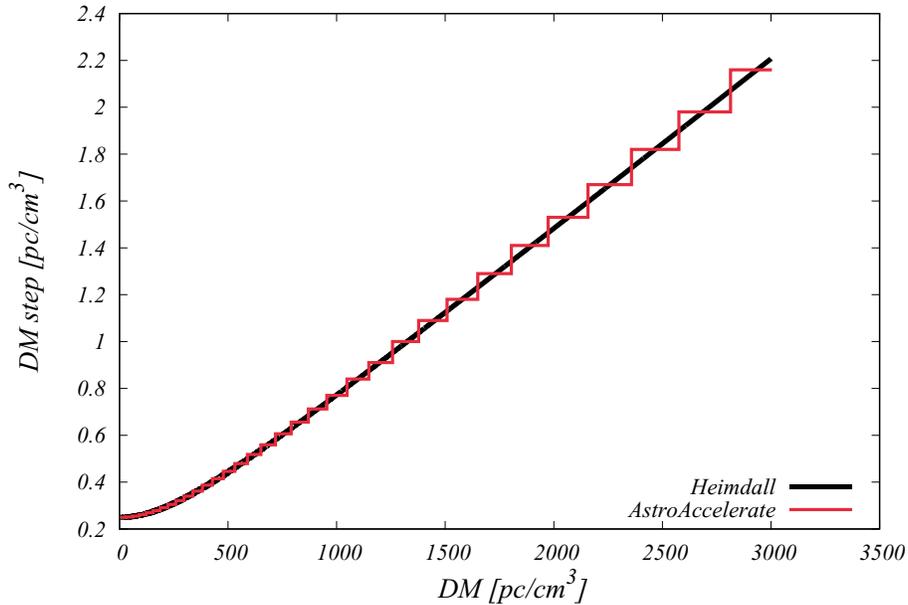}
	\caption{Comparison between DM steps used by Heimdall's and AstroAccelerate. \label{fig:DDTRplan}}
\end{figure}

We have based the AstroAccelerate de-dispersion scheme on the one used by Heimdall, a comparison of the two presented in Figure \ref{fig:DDTRplan}. Furthermore the input time series for each DM-trial can be decimated in time. This decimation step is matched for both codes apart from the first group of DM trials which, in the case of AstroAccelerate is performed at full time resolution. Heimdall in contrast, only performs the first DM trial at full time resolution.

%%%%%%%%%%%%%%%% Mean and standard deviation %%%%%%%%%%%%%%%%%%%%%%%%
\subsubsection{Calculation of mean and standard deviation}
\label{sec:MSD}
The correct calculation or estimation of the mean and the standard deviation of the base level noise in the input data is essential because these values are used to calculate the $\RSNR$. Only those detections with $\RSNR$ above a user defined number of standard deviations are considered significant and potential candidate detections. To ensure we have a fast, flexible and robust GPU code to calculate mean and standard deviation we have chosen to implement a streaming algorithm by \cite{Chan:1983:MSD} in CUDA for NVIDIA GPUs. This algorithm has several advantages. It is numerically stable and it is more precise than conventional methods. This algorithm is also very well suited to a parallel implementation. 

We estimate the true value of the standard deviation of the underlying base level noise by point-wise outlier rejection. Each point is compared with the current estimate of the mean and standard deviation and rejected if it is significantly different (the threshold is user defined). This is repeated until convergence.

\begin{figure}[ht!]
	\centering
	\includegraphics[width=0.8\textwidth]{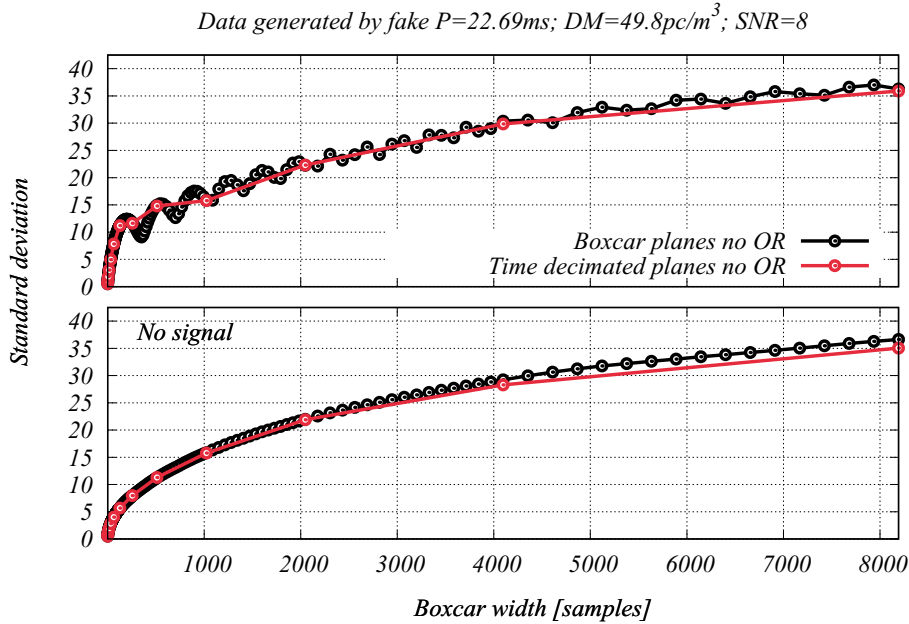}
	\caption{Difference between stdev values calculated from time-series after applying a boxcar of given width and standard deviations calculated from time-series after successive decimations in time. Data were generated by SIGPROC 'fake'. Top line: A time-series that contains a strong signal, Bottom line: A time-series without a signal present.  \label{fig:MSDexample}}
\end{figure}

Each application of a boxcar filter on the DM trial changes its mean and standard deviation. Since we cannot calculate the standard deviation after each boxcar filter, because it would be too computationally expensive, we need to approximate changes in the standard deviation introduced by application of boxcar filters. Therefore, we calculate values of the standard deviation only for time-series which are decimations of the input time-series in time, i.e. we calculate the standard deviation only for boxcar filters of width that are equal to a powers of two. We then interpolate values of standard deviation for all intermidiate boxcar widths. An example of the results produced by this method is presented in Figure \ref{fig:MSDexample} where we have used SIGPROC\footnote{\url{https://sourceforge.net/projects/sigproc/}} 'fake' to generate input data.

%%%%%%%%%%%%%%%% Heimdall %%%%%%%%%%%%%%%%%%%%%%%%
\subsubsection{Comparison to Heimdall pipeline}
We have compared the performance, reported SNR values and reported widths produced by AstroAccelerate to those produced by Heimdall. Both codes standard output provides SNR values and pulse widths. The execution time of Heimdall's SPD algorithm is measured using Heimdall's internal benchmark. To measure the execution time of the SPD algorithm within AstroAccelerate we have used the NVIDIA visual profiler.

The output of the Heimdall pipeline was compared to the AstroAccelerate SNR output for two different scenarios. The first scenario focused on the detection of single pulses with fixed initial SNR and increasing pulse width. The second scenario, on the response of both pipelines to pulses of fixed width but decreasing initial SNR of the pulses. For both scenarios, SIGPROC fake was used to generate an observation file containing a fake pulsar. AstroAccelerate's output contains many more data points per detection when compared with Heimdall's output. This is because we compare the full Heimdall pipeline with Astroaccelerate's basic candidate selection (peak-finding). In order to keep figures simple and readable, we do not display most of the points produced by AstroAccelerate and instead we show only the highest SNR candidates.

\begin{figure*}[htp]
	\begin{minipage}[t]{.2\textwidth}
		\centering
		\includegraphics[width=\linewidth]{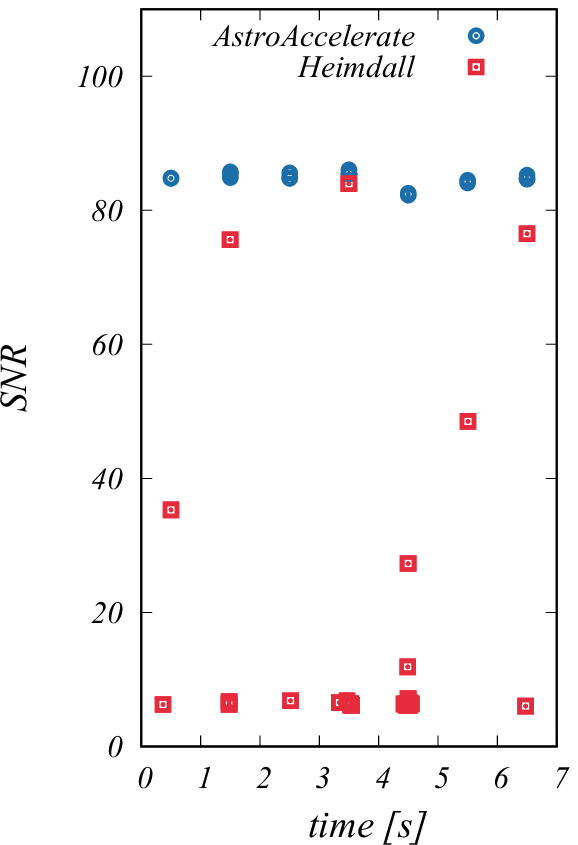}
	\end{minipage}%
	\hfill%
	\begin{minipage}[t]{.18\textwidth}
		\centering
		\includegraphics[width=\linewidth]{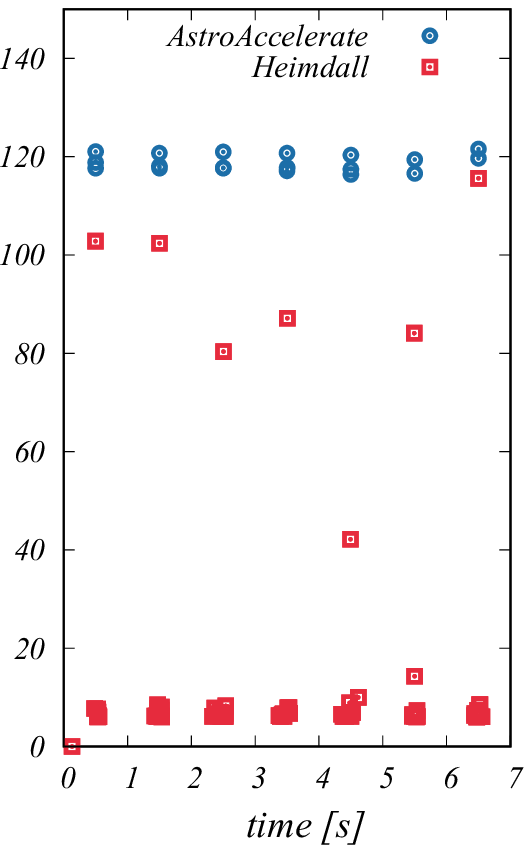}
	\end{minipage}%
	\hfill%
	\begin{minipage}[t]{.18\textwidth}
		\centering
		\includegraphics[width=\linewidth]{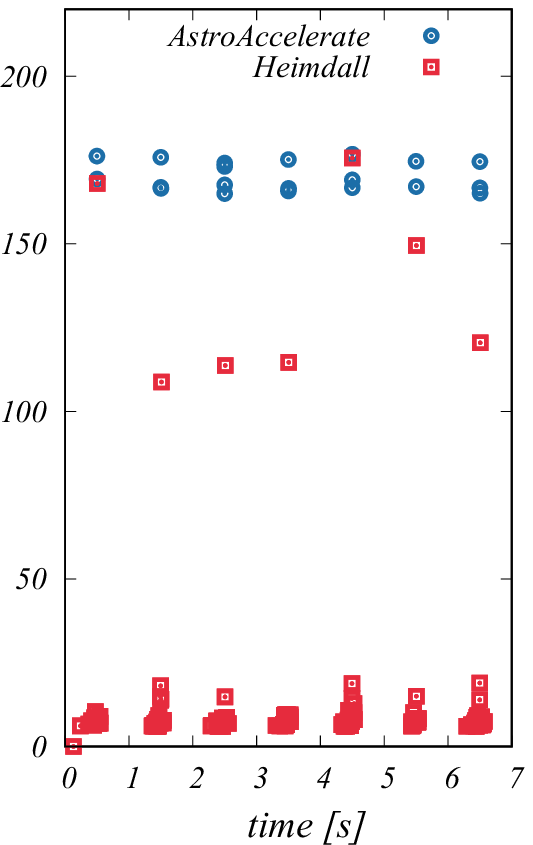}
	\end{minipage}%
	\hfill%
	\begin{minipage}[t]{.18\textwidth}
		\centering
		\includegraphics[width=\linewidth]{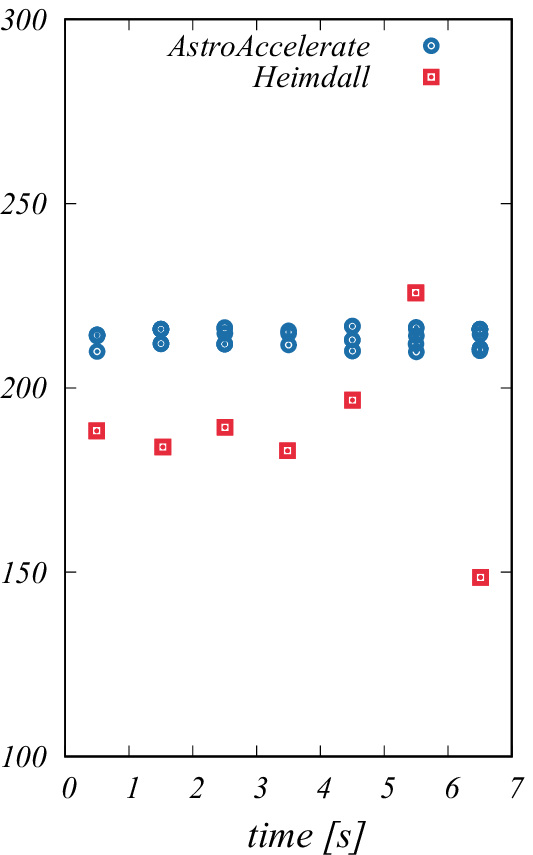}
	\end{minipage}\\
	%%%%%%%% row
	\caption{Comparison recovered SNRs by Heimdall and Astroaccelerate in single pulse case. Pulses are from left 100, 200, 400, and 800 samples wide. \label{fig:HvsAA_SNR}}
\end{figure*}

\begin{figure*}[htp]
	\begin{minipage}[t]{.2\textwidth}
		\centering
		\includegraphics[width=\linewidth]{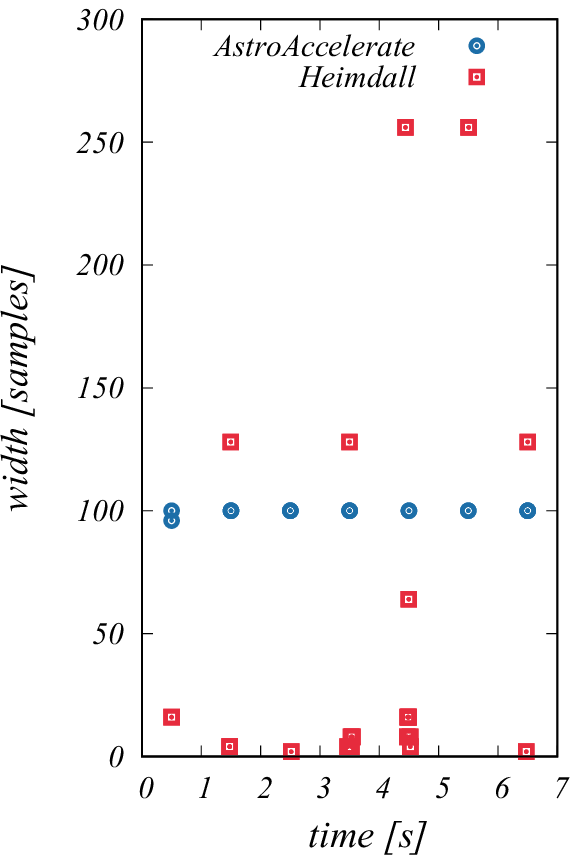}
	\end{minipage}%
	\hfill%
	\begin{minipage}[t]{.18\textwidth}
		\centering
		\includegraphics[width=\linewidth]{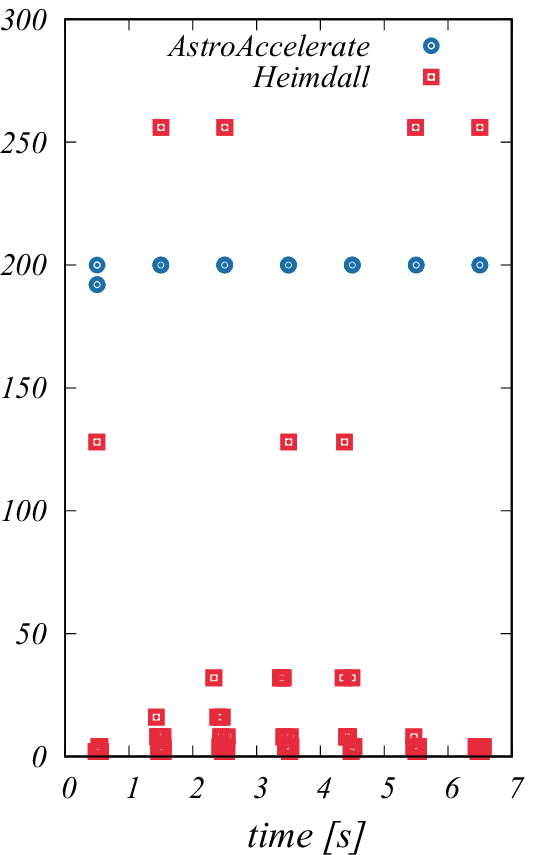}
	\end{minipage}%
	\hfill%
	\begin{minipage}[t]{.18\textwidth}
		\centering
		\includegraphics[width=\linewidth]{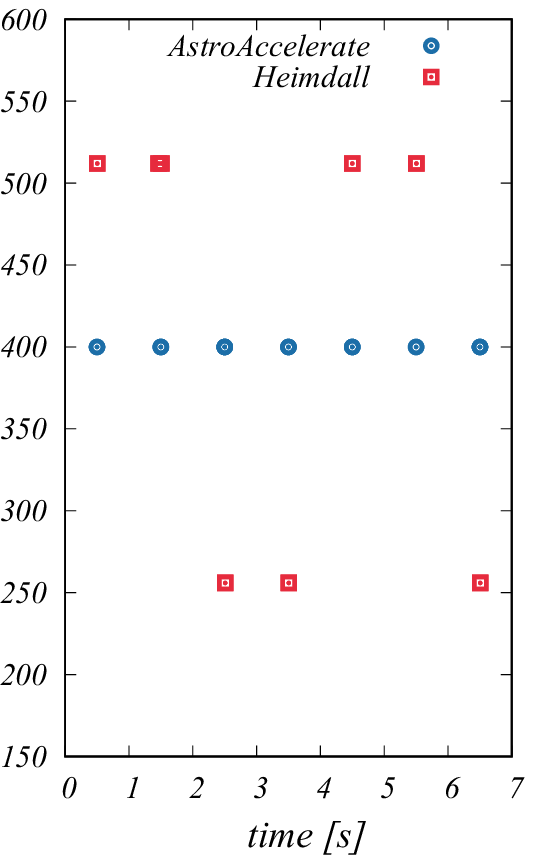}
	\end{minipage}%
	\hfill%
	\begin{minipage}[t]{.18\textwidth}
		\centering
		\includegraphics[width=\linewidth]{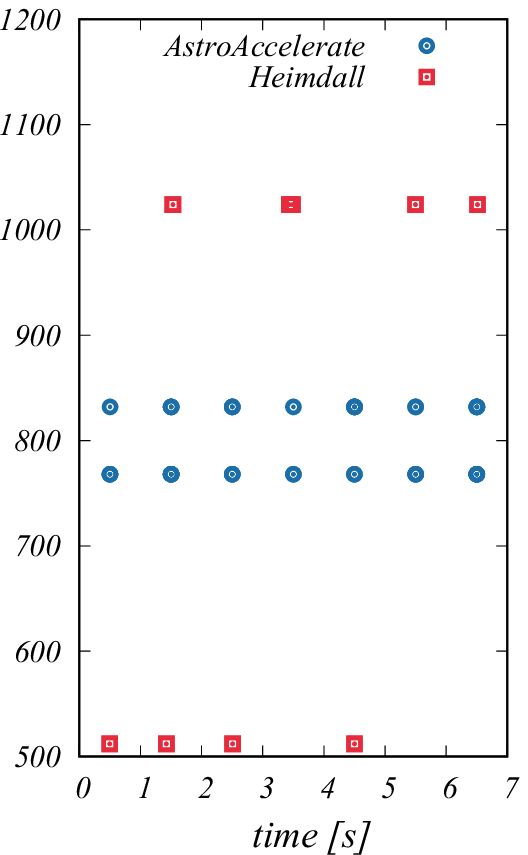}
	\end{minipage}\\
	%%%%%%%% row
	\caption{Comparison reported pulse width by Heimdall and Astroaccelerate in single pulse case. Pulses are from left 100, 200, 400, and 800 samples wide. \label{fig:HvsAA_width}}
\end{figure*}

In the first scenario, we have compared both codes using four different pulse widths (created by changing the duty cycle parameter in SIGPROC fake) with fixed initial SNR. The SIGPROC fake file used was generated using period $P=1000\mathrm{ms}$, $\mathrm{DM}=90\mathrm{pc/cm^3}$, $\mathrm{SNR}_\mathrm{init}=8$ and varying duty cycle. Recovered SNR for each case is presented in Figure \ref{fig:HvsAA_SNR} and the reported width of the pulses are shown in Figure \ref{fig:HvsAA_width}. Our implementation delivers SNR values which have a consistent SNR value for all peaks, independent of the position of the pulse within the input data. AstroAccelerate also reports more precisely the pulse width. We see that the Heimdall SPD algorithm, which uses a sequence of boxcar filter widths that are equal to powers of two (and hence cannot accurately represent all pulse widths), reports varying RSNR. This is because in some cases, the boxcar filter width is shorter(longer) that the actual width of the injected pulse or the injected pulses sit in-between adjacent boxcars used by Heimdall's SPD algorithm or both. This is shown in Figure \ref{fig:HvsAA_width} where Heimdall reports noticeably longer or shorter pulse widths. In both cases where the SPD algorithm does not sum all samples of the pulse (shorter boxcar filter width) or adds too much of the noise samples (longer boxcar filter width) the reported SNR will be lower than the true value. In the case where the position of the pulse in the time series straddles adjacent boxcar filters a similar effect to that above will occur, even if those boxcars have a width equal to the pulse.

\begin{figure*}[htp]
	\begin{minipage}[t]{.2\textwidth}
		\centering
		\includegraphics[width=\linewidth]{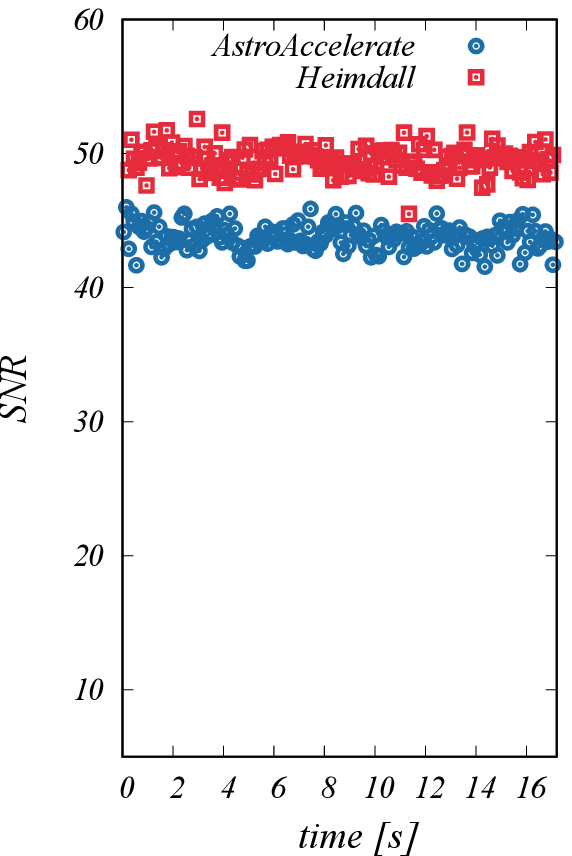}
	\end{minipage}%
	\hfill%
	\begin{minipage}[t]{.18\textwidth}
		\centering
		\includegraphics[width=\linewidth]{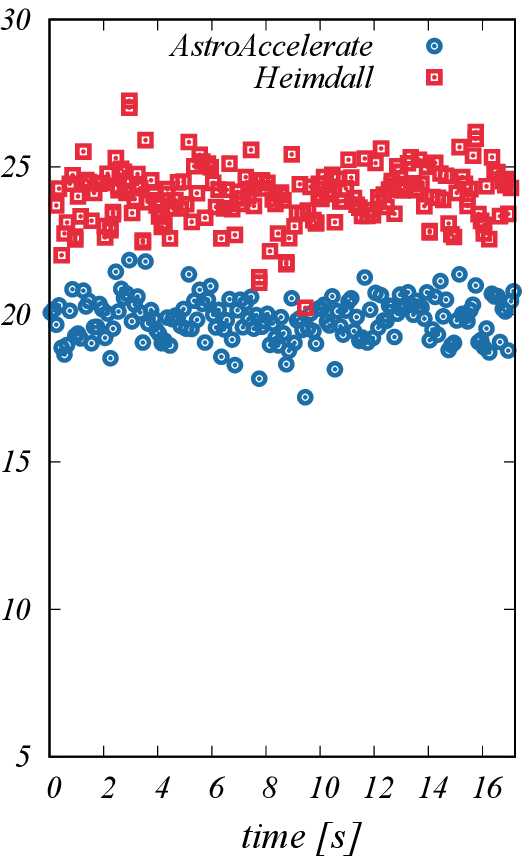}
	\end{minipage}%
	\hfill%
	\begin{minipage}[t]{.18\textwidth}
		\centering
		\includegraphics[width=\linewidth]{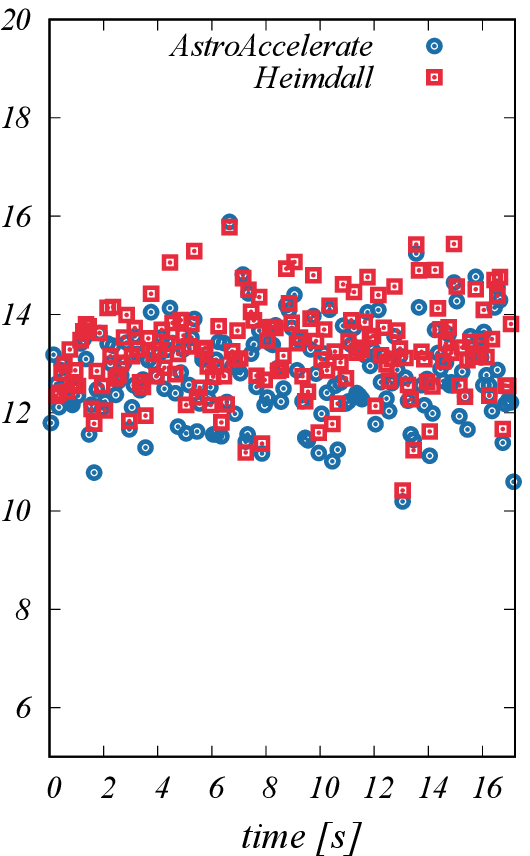}
	\end{minipage}%
	\hfill%
	\begin{minipage}[t]{.18\textwidth}
		\centering
		\includegraphics[width=\linewidth]{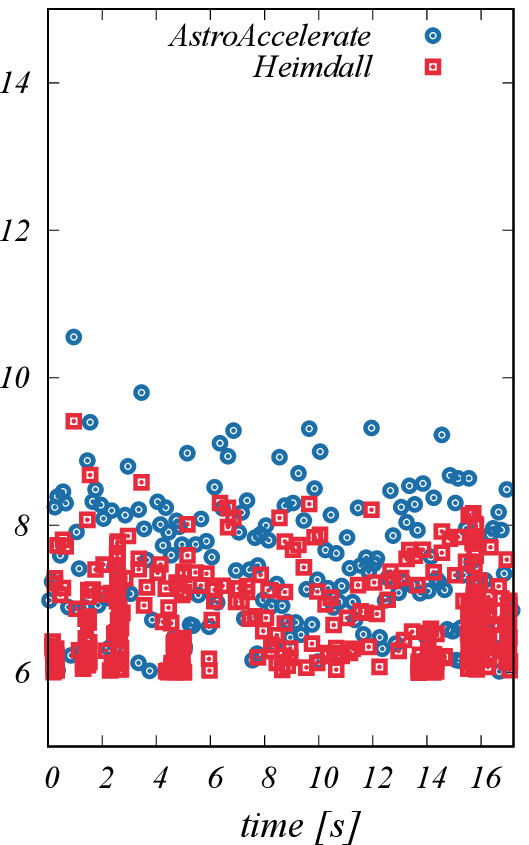}
	\end{minipage}\\
	%%%%%%%% row
	\caption{Comparison recovered SNRs by Heimdall and Astroaccelerate in pulsar case. Initial SNR of the pulsar is from left 8, 4, 2, and 1. \label{fig:HvsAA_pulsar}}
\end{figure*}

The second scenario compares the response of the SPD algorithm to different initial SNR values of the pulsar by processing a SIGPROC fake file which contained a pulsar with parameters $P=100\mathrm{ms}$, $\mathrm{DM}=50\mathrm{pc/cm^3}$ and a variable initial SNR. The pulse width was set to 32 samples which is an ideal width for the SPD algorithms in both pipelines. The results are presented in Figure \ref{fig:HvsAA_pulsar}. Our implementation delivers consistent results which are similar to the results from the Heimdall pipeline. The values of RSNR for initial pulsar $\mathrm{SNR}_\mathrm{init}=8$ and $\mathrm{SNR}_\mathrm{init}=4$ are higher for the Heimdall pipeline, but for lower initial pulsar SNR values the RSNR reported by both pipelines are similar. The values of RSNR depend on the correct estimation of the mean and the standard deviation of the underlying noise, where the two pipelines differ. To get these results we have used an approximation of the mean and standard deviation with outlier rejection where samples with $\sigma>2$ were rejected.

\begin{figure}[ht!]
	\centering
	\includegraphics[width=0.8\textwidth]{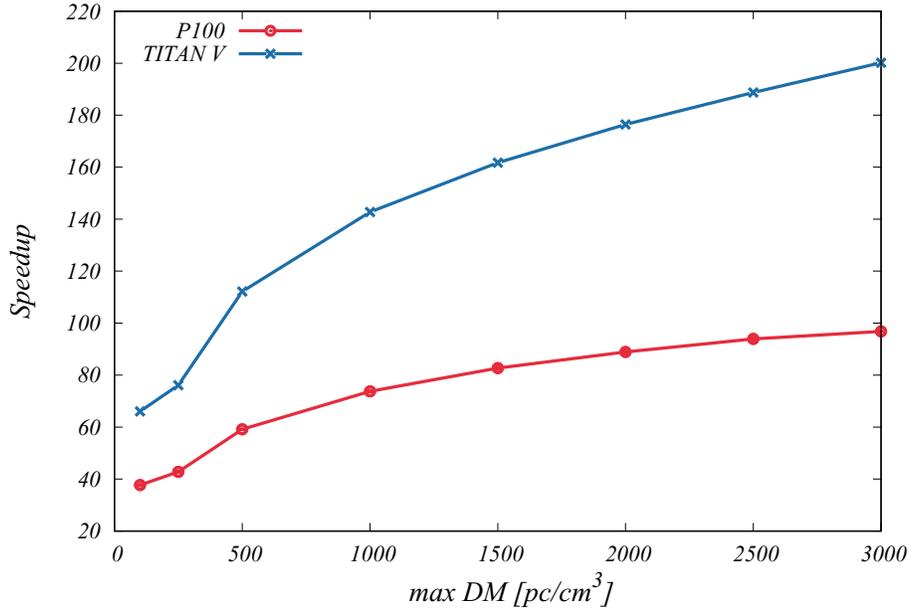}
	\caption{How performance depends on the maximum width computed by the SPD algorithm used. \label{fig:HvsAAspeedup}}
\end{figure}

For our performance comparison, we have used the same two NVIDIA GPU cards (P100 and Titan V). We present speed-up for different maximum DM values searched, i.e. we present ratio of Heimdall's SPD execution time over BoxDIT execution time in configuration BD-16. Our code is faster by an order of magnitude achieving almost $100\times$ speed-up for P100 and almost $200\times$ speed-up for Titan V. This is shown in Figure \ref{fig:HvsAAspeedup}. For lower values of maximum DM searched our speed-up is lower, because the first group of DM-trials in case of AstroAccelerate is processed in full time resolution in contrast to Heimdall which process only the first DM-trial in full time resolution.

\section{Conclusions}
\label{sec:conclusions}
In this work, we focused on single pulse detection using an incomplete set of boxcar filters. We have quantified properties of these single pulse detection (SPD) algorithms and quantified errors introduced by them. 

We have described the signal loss of the SPD algorithm in terms of the systematic signal loss and in terms of the worst case signal loss, these change with pulse width and parameters of the SPD algorithm. 
The systematic signal loss represent the loss in signal SNR which will occur even under best possible circumstances. This signal loss is different for different pulse widths and it is occurs because the SPD algorithm uses an incomplete set of boxcar filters. The worst case signal loss represent largest signal loss possible for a given pulse width and SPD algorithm. These two quantities represents upper and lower bounds on possible signal loss for given pulse width and SPD algorithm. 

We have also identified two governing parameters of the single pulse detection algorithm. These are the set of boxcar widths used for detection and the boxcar separation $L_s$, which is the distance between two neighbouring boxcar filters of the same width. The systematic signal loss depends only on how dense (lower signal loss) or sparse (higher signal loss) the set of boxcar filter widths are. The worst case signal loss depends strongly on boxcar separation $L_s$ and weakly on density of set of the boxcar filter widths.

Based on the importance of these two parameters we have designed two distinct single pulse detection algorithms. We have also presented our parallel implementation of these algorithms on many-core architectures, namely NVIDIA GPUs. We have verified the correctness of these implementation and discussed their performance. 

Our first algorithm is called BoxDIT, the BoxDIT algorithm calculates boxcar filters at every point of the input time-series which is progressively decimated. Decimation steps increase performance, but also increase the signal loss. This algorithm is designed to be flexible and to minimise systematic signal loss $\sll$. It is more suited for situations were high sensitivity (low signal loss) is required where it offers good performance. This makes it optimal for processing large amounts of archival data. The disadvantage of the algorithm is that with increasing signal loss performance does not increase accordingly.

The second algorithm is called IGRID and it is based on a distribution of boxcar filters using a binary tree structure. This algorithm has higher performance than our BoxDIT algorithm at the cost of higher signal loss, but for low signal loss this algorithm is slower. IGRID is best suited for real-time search scenarios where performance is critical. The disadvantage of the algorithm is that decreasing signal loss increases the memory footprint of the algorithm which leads to lower performance.

Using IGRID in its less sensitive configuration we are able to process SKA-MID like data 266$\times$ faster then real-time on P100 GPU and 500$\times$ faster then real-time on Titan V GPU. Even when considering the beyond SKA era of ultra high time resolution radio-astronomy, decreasing sampling time to nanosecond scales ($t_s=512\mathrm{ns}$) the IGRID algorithm on Titan V GPU would be able to process 22000 DM-trials in real-time.

BoxDIT algorithm is part of AstroAccelerate software package \citet{AstroAccelerate_2019_2556573} and AstroAccelerate with BoxDIT algorithm was used by \citet{2018ATel11606....1M}. 

\section*{Acknowledgements}
This work has received support from an STFC Grant (ST/R000557/1). The authors would like to acknowledge the use of the University of Oxford Advanced Research Computing (ARC) facility in carrying out this work. The authors would also like to express thanks to Aris Karastergiou for valuable discussions and input. 

\bibliography{APJS_journal_SPDT}

%%%%%%%%%%%%%%%%%%%%%%%%%%%%%%%%%%%%%%%%%%%%%%%%%%%%%%%%%%%%%%%%%%%
%%%%%%%%%%%%%%%%%%%%%%%%%%%%%%%%%%%%%%%%%%%%%%%%%%%%%%%%%%%%%%%%%%%

\appendix

\section{Sensitivity analysis}
    \label{app:sensitivity}
    Here we present a detailed description of our derivation of sensitivity analysis for a given SPD algorithm. We use signal-to-noise ratio (SNR) as defined in equations \eqref{eqa:SNR} and \eqref{eqa:SNRlong} as a figure of merit for pulse detection by a given SPD algorithm. To distinguish SNR produced by the SPD algorithm from the true SNR of the injected pulse we use recovered SNR $\RSNR$. To simplify the sensitivity analysis have used an idealized signal model as defined in section \ref{sec:idealmodel} which includes the normalization of the injected rectangular pulse as given by equation \eqref{eqa:amplitudenormalization}. Furthermore, we assume that the SPD algorithms return only the highest RSNR for a given sample, that is if multiple boxcar filters of different widths are applied only the boxcar filter which has produced the highest RSNR is returned by the SPD algorithm.
    
    Using the idealized signal we can simplify the SNR calculation \eqref{eqa:SNRlong} and define RSNR of the boxcar filter which acts on a rectangular pulse. When a boxcar filter encounters the pulse it will give a value $X_L=d_{s}A$, where $d_s$ is the portion of the pulse (in the number of time samples) that has been summed by the boxcar filter or \textit{pulse coverage}. The pulse coverage is an integer value between zero and the pulse width, $0~\leq~d_s~\leq~S$. Using the formula for SNR \eqref{eqa:SNRlong} together with white noise approximation \eqref{eqa:whitenoise}, mean ($\mu(x)=0$) and standard deviation ($\sigma(x)=1$) for the idealized signal, and with normalization of the pulse's amplitude $A(S)$ \eqref{eqa:amplitudenormalization}, we can write the \textit{recovered SNR} ($\RSNR$) by the boxcar filter of width $L$ as
    \begin{equation}
    \label{eqa:RSNR}
    \RSNR(S,L,d_s)=\frac{d_s A(S)}{\sqrt{L}}=\frac{d_{s}C}{\sqrt{L}\sqrt{S}}\,.
    \end{equation}
    The recovered SNR by the boxcar filter in general depends on signal width $S$, boxcar width $L$, and on the portion of the pulsed covered by the boxcar filter $d_s$. The dependence on the pulse coverage $d_s$ is another way of saying that the RSNR depends on the pulse's position.

    A boxcar filter which does not intersect with the injected rectangular pulse will sum up to zero ($X_L=0$ from eq. \eqref{eqa:boxcar}), that is $d_s=0$ thus $\RSNR(S,L,d_s)=0$ as well.

    To quantify $\RSNRmax(S)$ and $\RSNRmin(S)$ we first have to define the worst case detection event and the best case detection event.
\subsubsection{Best detection case}
    %%%%%%%%%%%%%%%%%%%%% Best detection case %%%%%%%%%%%%%%%%%%%%%
    By investigating the best detection case and the signal loss of the SPD algorithm under best possible circumstances we are also investigating the minimal signal loss which is introduced by the SPD algorithm.

\begin{figure}[ht!]
		\centering
		\includegraphics[width=0.8\textwidth]{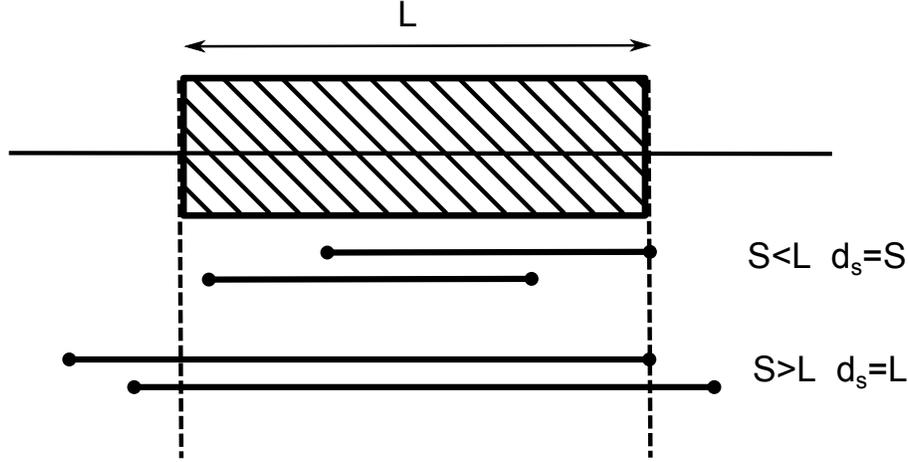}
        \caption{The best case detection event. The pulse is shown as lines with circles at the ends and the boxcar is shown as a hatched square.\label{fig:bestcaseboxcarlayout}}
    \end{figure}

    The maximum $\RSNR((S,L)$ for a fixed $S$ and $L$ is when a boxcar filter covers the whole pulse or as much of the pulse as is possible if the pulse is wider than the boxcar detecting it. There are three possible configurations, depending on the size of the pulse $S$ with respect to the size of the boxcar filter $L$. These are $S<L$, $S>L$ and $S=L$. The situation is summarized in Figure \ref{fig:bestcaseboxcarlayout}, which only shows the first two cases.

    The values of pulse coverage $d_s$ and $\RSNR$, using equation \eqref{eqa:RSNR} for all three possible cases are:
    \begin{itemize}
	    \item{$S<L$ then $d_s=S$ and $\RSNR_{S<L}=(\sqrt{S}C)/\sqrt{L}$}
	    \item{$S=L$ then $d_s=S=L$ and $\RSNR_L=C$}
	    \item{$S>L$ then $d_s=L$ and $\RSNR_{S>L}=(\sqrt{L}C)/\sqrt{S}$}
    \end{itemize}

\begin{figure}[ht!]
		\centering
		\includegraphics[width=0.8\textwidth]{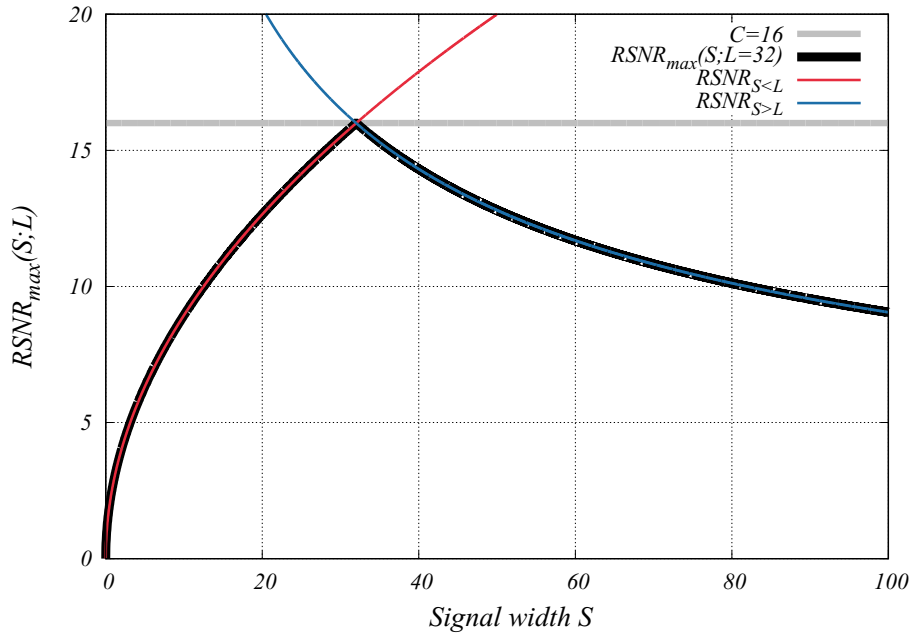}
        \caption{The response of the boxcar filter of with $L=32$ to different pulse widths, that is the behaviour of $\bRSNRmax(S;L)$. The peak is located at $L=S$, where the boxcar filter adds up all of the pulse power.\label{fig:bRSNRmax}}
    \end{figure}

    As the SPD algorithm may consist of many different boxcar filter widths, we first determine the response $\bRSNRmax(S;L)$ of a single boxcar filter of width $L$ to a different pulse width $S$. We define $\bRSNRmax(S;L)$, a function of $S$ and parametrized by $L$, as a combination of the cases we have discussed above, that is 
    \begin{equation}
        \bRSNRmax(S;L) = \left\{
            \begin{array}{lr}
                \frac{\sqrt{S}}{\sqrt{L}}C & : S < L\\
                \frac{\sqrt{L}}{\sqrt{S}}C & : S \geq L
            \end{array}
        \right.
    \end{equation}
    This quantity gives us the highest $\RSNR$ possible for a given boxcar of width $L$ for any pulse of width $S$. The behavior of $\bRSNRmax(S;L)$ for a fixed boxcar filter of width $L$ is shown in Figure \ref{fig:bRSNRmax}.

    To find the response of the whole SPD algorithm to a pulse of width $S$ we need to find the maximum value of $\bRSNRmax(S;L)$ produced by boxcar filters of every boxcar width $L$ used by the SPD algorithm. That is
    \begin{equation}
        \RSNRmax(S)=\max_{L \in B}\left( \bRSNRmax(S;L)\right)\,.
        \label{eqa:RSNRmax}
    \end{equation}

\begin{figure}[ht!]
		\centering
		\includegraphics[width=0.8\textwidth]{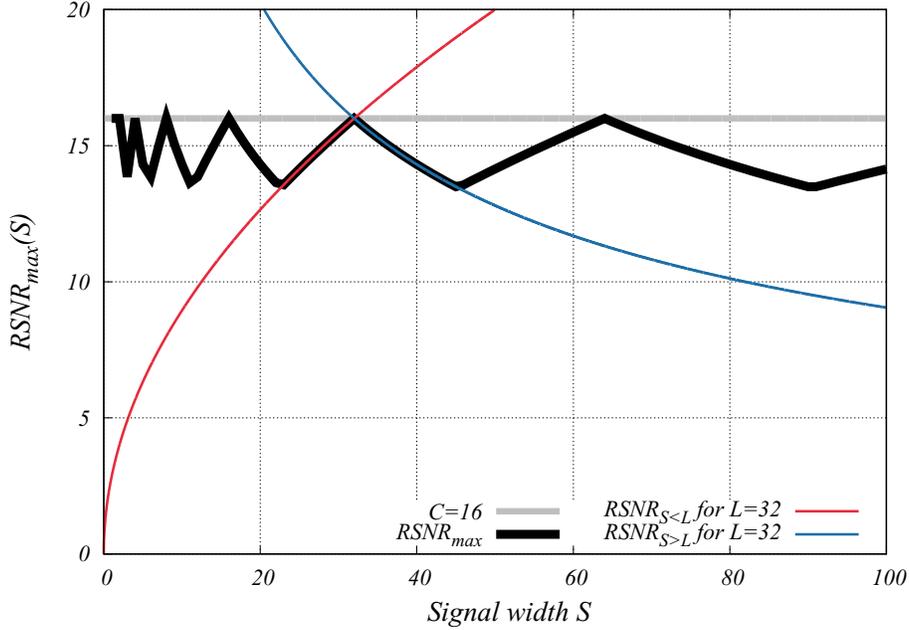}
        \caption{Response (behaviour of $\RSNRmax(S)$) of an SPD algorithm, where widths of boxcar filters are powers of two, that is $\mathfrak{B}=\{1,2,\ldots,2^t\}$, and boxcar separation is $L_s=L$. The $\bRSNRmax(S;L)$ for $L=32$ is emphasized to illustrate how $\RSNRmax(S)$ is constructed. \label{fig:RSNRmax}}
    \end{figure}

    An example of the behavior of $\RSNRmax(S)$ is shown in Figure \ref{fig:RSNRmax}. We see that the form of $\RSNRmax(S)$ depends on the number of boxcar widths we combine together into the SPD algorithm. Thus in order to increase (decrease) $\RSNRmax(S)$ we need to add (remove) more boxcar filters of different widths to (from) set $\mathfrak{B}$. 
    
    The associated systematic signal loss $\sll$ is calculated using equation \eqref{eqa:signalloss}. This represents minimal signal loss which is always introduced by the SPD algorithm for given pulse width $S$.

    %%%%%%%%%%%%%%%%%%%%% Best detection case %%%%%%%%%%%%%%%%%%%%%

\subsubsection{Worst detection case}
    %%%%%%%%%%%%%%%%%%%%% Worst detection case %%%%%%%%%%%%%%%%%%%%%
    To find the worst detection case we need to find a position of the pulse which minimizes $\RSNR(S,L,d_s)$. Let's suppose that we have a rectangular pulse of arbitrary width $S$, which is detected by a boxcar filter of fixed width $L$. Let's also suppose that the starting time of our rectangular pulse is increasing, that is the pulse is sliding forward in time. We assume that the boxcar filters of given width are distributed evenly throughout the time-series with boxcar separation $L_s$. As the pulse slides forward in time it crosses the beginning of the next boxcar filter covering the time-series every $L_s$ steps. When the signal crosses the beginning of the next boxcar filter the situation is the same as it was $L_s$ steps before. Which means that the only shifts that matter happens between two consecutive boxcar filters on the span of $L_s$ samples. Thus for the investigation of the worst case detection we need to consider two consecutive boxcar filters.

\begin{figure}[ht!]
		\centering
		\includegraphics[width=0.8\textwidth]{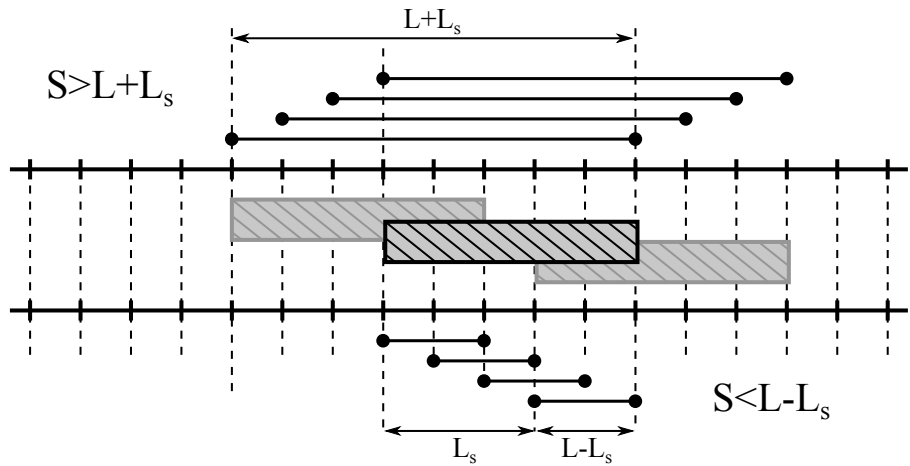}
        \caption{Pulses of width $S \leq L-\Ls$ and $S \geq L+\Ls$ are shown as lines ending with circles and boxcar filters are shown as grey hatched rectangles. The pulses shift in time from one boxcar to the next. The boxcar which is always covered by the pulse for $S \leq L+\Ls$ or always covered by the pulse width $S \geq L-\Ls$ is emphasized in black. Parameters are $L=5$, $\Ls=3$ with pulses of width $S=8$ and $S=2$. \label{fig:worstcaseallcases}}
    \end{figure}

\begin{figure}[ht!]
		\centering
		\includegraphics[width=0.8\textwidth]{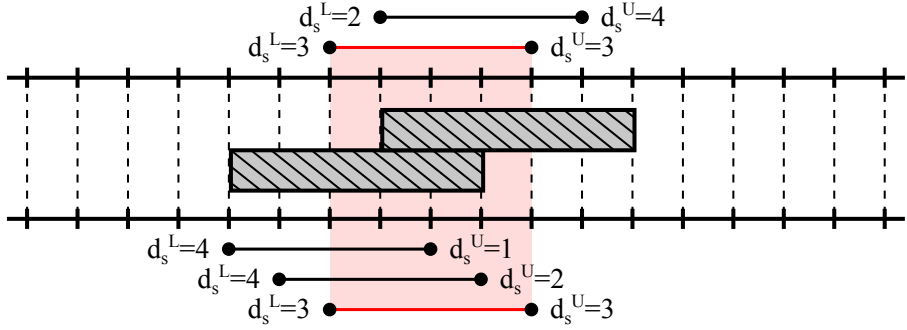}
        \caption{Pulses are shown as lines ending with circles and boxcar filters are shown as hatched rectangles. A pulse of width $L-L_s<S<L+L_s$ starts at the beginning of the lower boxcar and as we shift the pulse in time it moves to the starting point of the upper boxcar. The worst case is emphasized in red. It also shows pulse coverage for both lower and upper boxcars. Parameters are $L=5$, $L_s=3$, $S=4$. \label{fig:worstcaseboxcarlayout}}
    \end{figure}

    Depending on the pulse width we can have three cases:
    \begin{description}
        \item [$S \geq L+L_s$] the pulse is wider than the extent of two consecutive boxcar filters. There is no way to shift the pulse, so the pulse would not cover at least one boxcar filter. The pulse coverage is $d_s=L$.
        \item [$S \leq L-L_s$] the pulse is shorter than the overlap of two consecutive boxcar filters. Wherever the pulse is, the pulse will always be covered by at least one boxcar filter with pulse coverage $d_s=S$.
        \item [$L-L_s < S < L+L_s$] the pulse lies between two consecutive boxcar filters.
    \end{description}
    The first two cases are shown in Figure \ref{fig:worstcaseallcases} and the last case is shown in Figure \ref{fig:worstcaseboxcarlayout}.

    We see that from these three cases, the worst case can only occur when $L-L_s<S<L+L_s$. Our goal is then to find the position of the pulse which minimizes $\RSNR$.

    Let's denote the boxcar which starts earlier in time as the \textit{lower boxcar} with pulse coverage $d_s^L$ and boxcar next to it as the \textit{upper boxcar} with $d_s^U$. To continue our example with the pulse shifting in time, let's inject our rectangular pulse so that its beginning coincides with the beginning of the lower boxcar. As the pulse shifts in time, the pulse coverage of the lower boxcar $d_s^L$ will decrease while pulse coverage of the upper boxcar $d_s^U$ will increase. Let's now suppose that the shift is continuous instead of discrete. In the continuous case, there will be a time when $d_s^L=d_s^U$, this is shown in Figure \ref{fig:worstcaseboxcarlayout}. If we shift the pulse in either direction one of the pulse coverages will increase at the expense of the other. When performing single pulse detection we are interested only in the highest RSNR, for fixed $S$ and $L$, this means the highest pulse coverage $d_s$. In the worst case scenario we are looking for the lowest maximum produced by the SPD algorithm. This is achieved when both the lower and upper boxcar produce the same $\RSNR$. Therefore the condition for the worst case detection is 
    \begin{equation}
        \label{eqa:worstcondition}
        d_s^L=d_s^U\,.
    \end{equation}

    For clarity it is useful to split the range $L-L_s < S < L+L_s$ into two further cases $L-L_s < S < L$ and $L \leq S < L+L_s$. For range $L-L_s < S < L$ we can express pulse coverage for both boxcars as
    \begin{equation}
        \label{eqa:worstcase1}
        \begin{aligned}
            d_s^L&=S-\alpha L_s\,,\\
            d_s^U&=L-L_s+\alpha L_s\,,
        \end{aligned}
    \end{equation}
    where $0 \leq \alpha \leq 1$ is a shift of the pulse in time. Using condition \eqref{eqa:worstcondition} together with equations \eqref{eqa:worstcase1} we can get the expression for the parameter $\alpha$. Substituting $\alpha$ into one of the equations \eqref{eqa:worstcase1} will give pulse coverage
    \begin{equation}
        \label{eqa:ds_wc1}
        d_s^<=S-\left\lfloor \frac{S-L+L_s}{2} \right\rfloor,
    \end{equation}
    where $\left\lfloor \, \right\rfloor$ represents the floor function which rounds the argument down to nearest lower integer. Rounding is necessary because we are working with discrete data. This is depicted in Figure \ref{fig:worstcaseboxcarlayout}.

    For the second case where $L \leq S < L+L_s$, we can express pulse coverage for both boxcars as
    \begin{equation}
        \label{eqa:worstcase2}
        \begin{aligned}
            d_s^L&=L-\alpha L_s\,,\\
            d_s^U&=S-L_s+\alpha L_s\,.
        \end{aligned}
    \end{equation}
    Solving equation \eqref{eqa:worstcondition} using \eqref{eqa:worstcase2} for $\alpha$ and then substituting into $d_s^L$ gives
    \begin{equation}
        \label{eqa:ds_wc2}
        d_s^>=L-\left\lfloor \frac{L+L_s-S}{2} \right\rfloor.
    \end{equation}
    We can get the same expression using $d_s^U$.

    To summarize we have the following cases:
    \begin{itemize}
	    \item{$S \leq L-L_s$ then $d_s=S$ and $\RSNR=(\sqrt{S}C)/\sqrt{L}$}
	    \item{$L-L_s < S < L$ then $d_s^<=S-\left\lfloor (S-L+L_s)/2 \right\rfloor$, and $\RSNR=d_{s}C/\sqrt{LS}$}
	    \item{$L \leq S < L+L_s$ then $d_s^>=L-\left\lfloor (L+L_s-S)/2 \right\rfloor$, and $\RSNR=d_{s}C/\sqrt{LS}$}
	    \item{$S>L+L_s$ then $d_s=L$ and $\RSNR=(\sqrt{L} C)/\sqrt{S}$}
    \end{itemize}

    After the investigation into all possible cases we can define the worst case response of the boxcar filter of width $L$ to a rectangular pulse of different widths $\bRSNRmin(S;L)$ as
    \begin{equation}
        \bRSNRmin(S;L) = \left\{
        \begin{array}{lr}
            \frac{\sqrt{S}}{\sqrt{L}}C & : S \leq L-L_s\\
            d_s^<\frac{C}{\sqrt{LS}}   & : L-L_s < S < L\\
            d_s^>\frac{C}{\sqrt{LS}}   & : L \leq S < L+L_s\\
            \frac{\sqrt{L}}{\sqrt{S}}C & : S \geq L+L_s
        \end{array}
        \right.
    \end{equation}
    where $d_s^<$ and $d_s^>$ are given by equations \eqref{eqa:ds_wc1} and \eqref{eqa:ds_wc2}.

    The worst case response of the whole SPD algorithm $\RSNRmin(S)$ is then given by
    \begin{equation}
        \RSNRmin(S)=\max_{L \in \mathfrak{B}} \left(\bRSNRmin(S;L)\right)\,.
        \label{eqa:RSNRmin}
    \end{equation}

\begin{figure*}
	\centering
    \begin{minipage}[t]{.49\textwidth}
        \includegraphics[width=\textwidth]{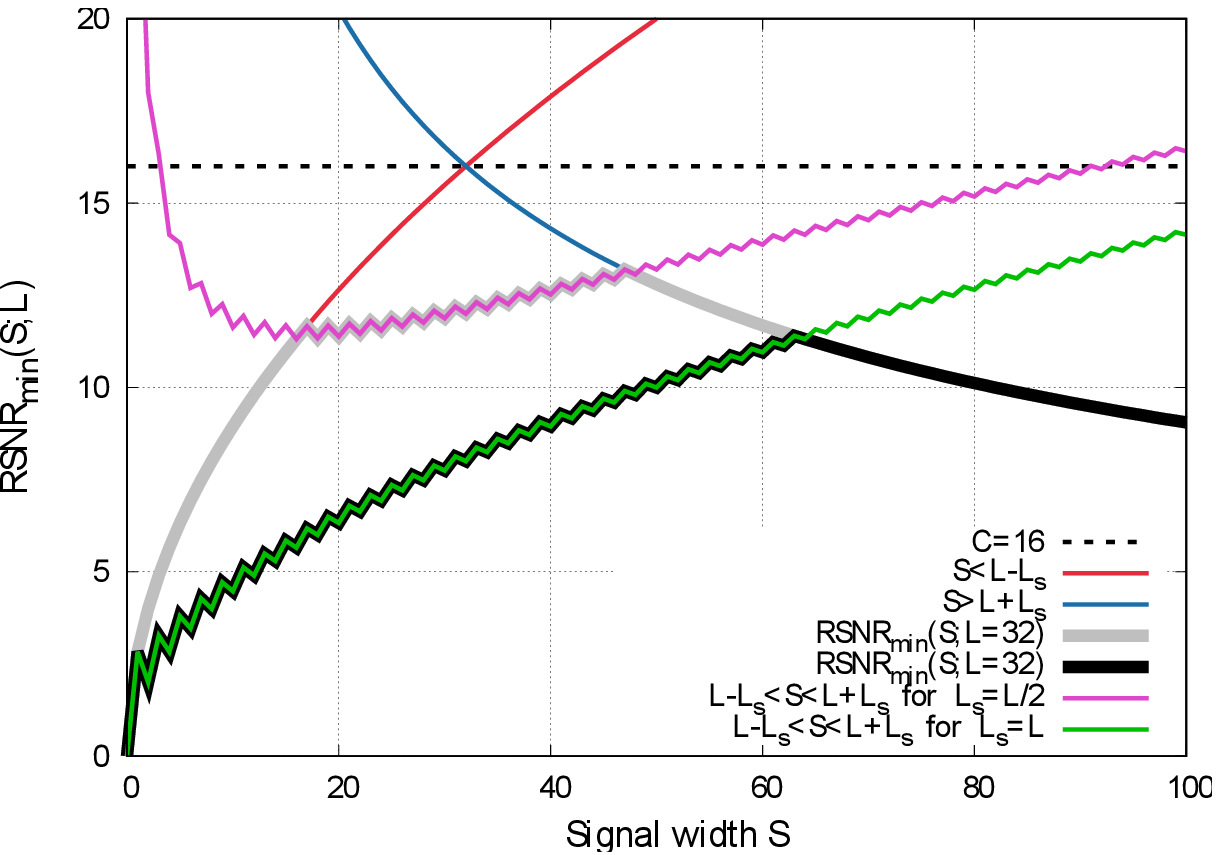}
    \end{minipage}
    \hfill
    \begin{minipage}[t]{.49\textwidth}
        \includegraphics[width=\textwidth]{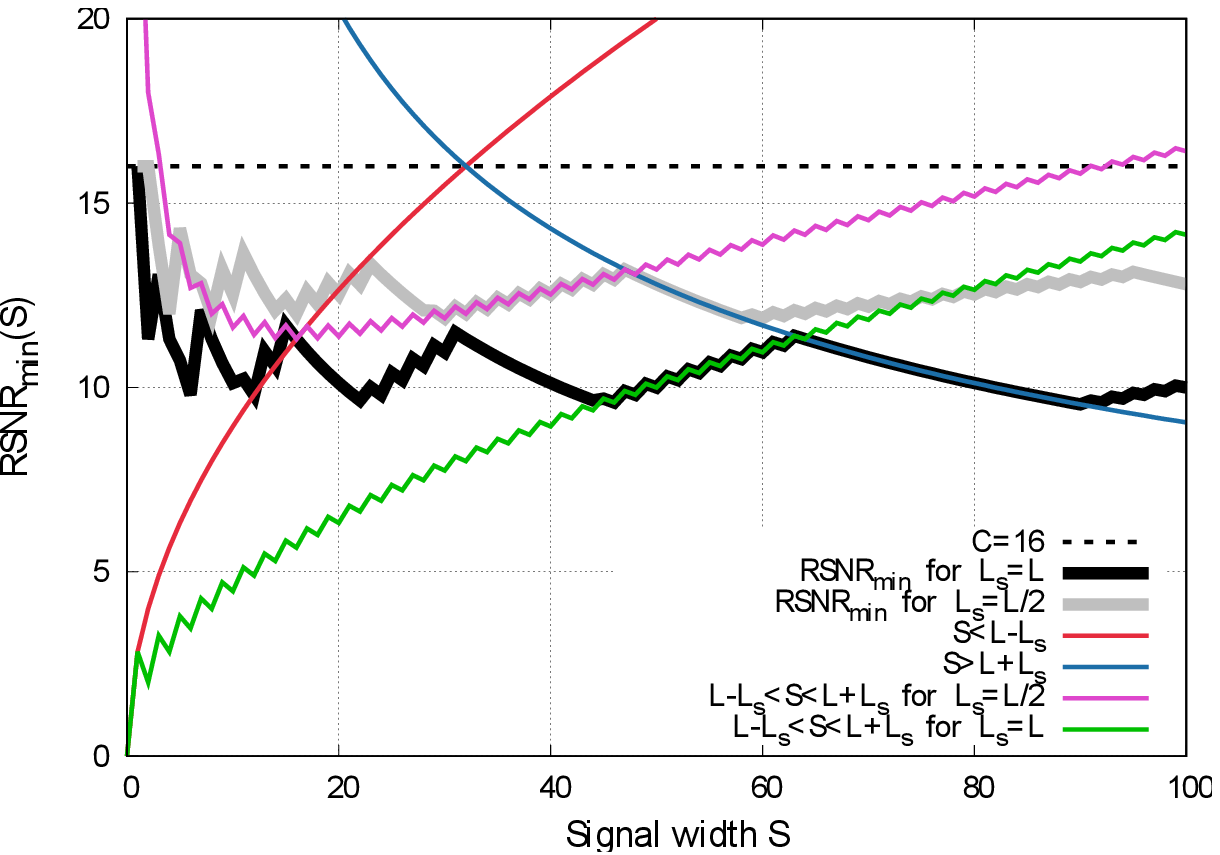}
    \end{minipage}\\
    \caption{Left: The worst case response $\bRSNRmin(S;L)$ of the boxcar filter of width $L=32$ for two values of $L_s=L$ (black) and $L_s=L/2$ (grey). We see that a change in $L_s$ has a noticeable effect on the value and behaviour of $\bRSNRmin(S;L)$, thus to increase (decrease) $\bRSNRmin(S;L)$ we need to decrease (increase) the value of $L_s$. Right: The response of the whole SPD algorithm $\RSNRmin(S)$, where boxcar widths are powers of two $B=\{1,2,\ldots,2^t\}$, and $L_s=L$ is in black and $L_s=L/2$ is in grey. \label{fig:RSNRmin}}
    \end{figure*}

    The behavior of $\bRSNRmin(S;L)$ and $\RSNRmin(S)$ for two different values of $L_s$ is shown in Figure \ref{fig:RSNRmin}. We see that $\RSNRmin(S)$ is also influenced by how many different widths of boxcar filters are used. Furthermore, we see that it is perhaps even more influenced by the distance between boxcar filters $L_s$. 

    %%%%%%%%%%%%%%%%%%%%% Worst detection case %%%%%%%%%%%%%%%%%%%%%

\end{document}